\documentclass[twocolumn,showkeys,prd,nofootinbib,floatfix,preprintnumbers]{revtex4-1}

\usepackage[utf8]{inputenc}
\usepackage{multirow}
\usepackage{amsfonts,amsmath,amssymb} 
\usepackage{graphicx,graphics,color}
\usepackage{gensymb}
\usepackage{longtable}
\usepackage{bbding}
\usepackage{subfigure}

\usepackage{hyperref}
\usepackage[normalem]{ulem}
\hypersetup{
    colorlinks=true,
    linkcolor=blue}

  \def\apjl{ApJ} 
\def\apss{Ap\&SS}   
  
  \def\prd{Phys. Rev. D}
 \def\prl{Phys. Rev. Lett.}

\begin{document}

\title{New Non-Abelian Reissner-Nordstr\"{o}m Black Hole Solutions in the Generalized SU(2) Proca Theory And Some Astrophysical Implications}

\author{Gabriel G\'omez}
\email{gabriel.gomez.d@usach.cl}
\affiliation{Departamento de F\'isica, Universidad de Santiago de Chile,\\Avenida V\'ictor Jara 3493, Estaci\'on Central, 9170124, Santiago, Chile}

\author{Jos\'e F. Rodr\'iguez}
\email{jose.rodriguez2@correo.uis.edu.co}
\affiliation{Escuela de F\'{\i}sica, Universidad Industrial de Santander,  Ciudad Universitaria, Bucaramanga 680002, Colombia\\}
\affiliation{ICRANet, Piazza della Repubblica 10, 65122, Pescara PE, Italy}

\begin{abstract}
The Generalized SU(2) Proca theory is a vector-tensor theory of gravity whose action is invariant under global transformations of the SU(2) group and includes second-order derivative self-interactions of the vector field beyond the massive Yang-Mills theory. We find, in particular, that the presence of two Lagrangian pieces consisting of four gauge fields minimally coupled to gravity gives rise to an exact Reissner-Nordstr\"{o}m black hole solution endowed with two different non-Abelian effective charges that depend on the specific combination, $\chi = 2\chi_1 + \chi_2$, of the respective coupling constants. After studying the spacetime structure of the black hole,  which allows us to characterize the parameter space that preserves the weak cosmic censorship conjecture, some astrophysical implications of the black hole solutions are investigated. First, joint analysis of observations of the EHT's first images of Sagittarius A$^{\star}$ of our Galaxy and the Keck telescope set the first serious constraint on the free parameters of the theory beyond the theoretical bounds found. Second, we investigate the accretion properties of spherical steady flows around this class of non-Abelian Reissner-Nordstr\"{o}m black hole. Specifically, we examine the general conditions under which transonic flow is allowed. An analytical solution for critical accretion is found in terms of the coupling constant. In addition, we explore the effect of changing $\chi$ on the radial velocity and mass density numerically and show how the extremal Reissner-Nordstr\"{o}m and the standard Schwarzschild solutions as limit cases are achieved. Finally, working in the fully relativistic regime, an analytical expression for the critical mass accretion rate of a polytropic fluid onto a black hole is derived. As a main result, we find that the critical
accretion rate efficiency can be noticeably improved
compared to the Schwarzschild case for a specific region of the parameter space where the non-Abelian charge becomes imaginary.

\end{abstract}

\maketitle

\section{Introduction}

Black holes (BHs) are among the most fascinating objects in the Universe, resulting from the gravitational collapse of massive objects, as predicted by general relativity (GR), see e.g. \cite{Poisson:2009pwt}. Apart from the fundamental conceptions and interesting properties they harbor, 
BHs are ideal laboratories, due to their intense gravitational fields, to study high-energy astrophysical processes occurring in their vicinity \cite{Shapiro:1983du}. Furthermore, observations of BHs in the strong field regime provide a unique opportunity to study the properties of spacetime and understand the nature of gravity in extreme environments. This is the primary program of theories beyond GR, which aim to predict deviations from Einstein's theory.
Motivated by these concerns, BHs have been the central target of current astrophysical experiments, including the Event Horizon Telescope (EHT) and the Very Large Telescope global networks \cite{EventHorizonTelescope:2019dse,EventHorizonTelescope:2022wkp}, GRAVITY collaboration \cite{GRAVITY:2020gka}, and the LIGO-Virgo collaboration \cite{LIGOScientific:2017ync,LIGOScientific:2017vwq} among others.
While the predictions of GR are well consistent with all available observational data within the current uncertainties \cite{Will:2014kxa}, some theories beyond Einstein's theory can also explain the observed phenomena \cite{LIGOScientific:2020tif,Cardoso:2019rvt,EventHorizonTelescope:2020qrl,Vagnozzi:2022moj}). Therefore, current and future measurements of BHs at the event-horizon scale pose a challenge for theories beyond GR. %Although the results derived from these observations are well consistent with GR, they are not conclusive in the sense that they can be reproduced by non-trivial spacetime metrics (see, e.g., \cite{LIGOScientific:2020tif,Cardoso:2019rvt,EventHorizonTelescope:2020qrl,Vagnozzi:2022moj}).
Such observations, however, do provide strong evidence about the existence of BHs. 

BHs can possess electric and magnetic charges as described by the Reissner-Nordstr\" {o}m (RN) solution in the Einstein-Maxwell theory \cite{1916AnP...355..106R,1918KNAB...20.1238N}. However, it is widely believed that astrophysical BHs are electrically neutral due to charge neutralization by astrophysical plasma, among other suitable physical mechanisms. Alternatively, it is possible for BHs to carry, for instance, U(1) charge instead of electromagnetic charge due to mechanisms in the early Universe within the dark (hidden) sector with no coupling to Standard-Model particles (see, e.g., \cite{DeRujula:1989fe,Cardoso:2016olt}). See also \cite{Zajacek:2019kla} for a discussion from an astrophysical point of view. Regardless of the underlying physical process behind the charge mechanism, this remains an open issue that has recently received significant attention after measurements made by the EHT of the supermassive BH M87$^\star$ shadow size, and the detection of gravitational waves from compact object binaries \cite{EventHorizonTelescope:2021dqv,Bozzola:2020mjx,Liu:2020vsy,Christiansen:2020pnv,Wang:2021vmi,Benavides-Gallego:2022dpn}. These discoveries demand a careful examination of BH charges beyond academic considerations. See also \cite{Zakharov:2014lqa,Zajacek:2018ycb} for observational limits on the charge of the Galactic Center BH.

Moreover, several observations have confirmed that BHs must rotate to account for various astrophysical phenomena, such as X-rays streaming off material near BHs due to the formation of an accretion disk (see e.g., Ref.~\cite{Done:2007nc}). On the other hand, the RN solution provides a useful first approximation for studying realistic and complex phenomena in the presence of electric charge, such as accretion and shadow phenomena. The charge of the RN black hole plays a significant role in the event horizon structure, similar to the role of spin. However, it is important to note that RN and Kerr black holes are fundamentally different scenarios. 

The electric and magnetic charges of the RN black hole can have similar effects as the spin parameter of a rotating Kerr black hole, which has been observed in the magnetar J1745-2900 orbiting around the supermassive BH Sagittarius A$^\star$ \cite{Juraeva:2021gwb}. In addition, recent observations on the motion of S-stars have constrained the spin of the central object in the Milky Way to be relatively small, with $a/M \lesssim 0.1$ \cite{2020ApJ...901L..32F}. This suggests that even though BHs are expected to have significant spin, \emph{there exist objects with small angular momentum} that can be described by a stationary and spherically symmetric spacetime \cite{2017grav.book.....M}.

After the discovery of (purely magnetic) static spherically symmetric non-Abelian BH solutions in the Einstein SU(2) Yang-Mills (EYM) model \cite{Bizon:1990sr,Volkov:1989fi,Kuenzle:1990is}, it was soon demonstrated that they are perturbatively unstable \cite{Lee:1991qs}. To address this problem, higher-order curvature terms of the gauge field \cite{Radu:2011ip,Mazharimousavi:2009mb}, as well as non-trivial combinations with other theories, have been introduced into the gravitational sector (see, e.g., Ref.~\cite{Volkov:1998cc} for previous proposals). Furthermore, some BH solutions with non-Abelian hairs have been found in theories beyond the canonical Yang-Mills theory but still within the framework of GR \cite{Herdeiro:2017oxy,Radu:2011ip,Mazharimousavi:2009mb}. In addition, higher curvature terms of the metric tensor, such as $f(R)$ gravity coupled to the Yang-Mills field, also admit BH solutions with single or double horizons \cite{Mazharimousavi:2011nc}.

The EYM case is interesting as there exists a RN solution \cite{1980JMP....21.2236H,1988PhRvL..61..141B}. However, unlike the Einstein-Maxwell case, the EYM RN solution is unstable, indicating that although both models share the same spacetime configuration, they are perturbatively different. The same is true in the Einstein-Yang-Mills-Higgs case, where a RN solution exists but can be stabilized by the Higgs mechanism \cite{1995NuPhB.442..126B}.

The quest to construct classical theories beyond Einstein's theory is an ongoing and active area of research, fueled by the need to address longstanding issues related to singularities  \cite{Penrose:1964wq,Hawking:1970zqf} and renormalization \cite{Deser:1974hg} in GR. While Einstein's theory of gravity provides an effective description of the gravitational interaction, it is only valid up to a certain cutoff scale before it loses its regime of validity \cite{,Donoghue:1994dn,Burgess:2003jk}. If this breakdown occurs, for instance, in the strong gravity regime, then modified gravity theories may play a significant role in describing the behavior of compact objects such as BHs and neutron stars. More importantly, these theories have the potential to provide a more complete and accurate description of the gravitational interaction at these scales, which is crucial for a deeper understanding of the nature of gravity and the behavior of astrophysical objects. 

One promising approach to modify Einstein's theory is to introduce new gravitational degrees of freedom (see e.g., \cite{Heisenberg:2018vsk}). The simplest example is the Horndeski theory \cite{Horndeski:1974wa}, which introduces a scalar field and yields field equations that are, at most, of second order. This is crucial for avoiding the Ostrogradski ghost \cite{Ostrogradsky:1850fid}. The Generalized Proca theory \cite{Heisenberg:2014rta,Allys:2015sht,Allys:2016jaq,GallegoCadavid:2019zke}, is a vector-tensor version of the Horndeski theory, where the internal gauge symmetry of the vector field has been abandoned to allow for the existence of extra terms \cite{Allys:2016kbq,Gomez:2019tbj,GallegoCadavid:2020dho,GallegoCadavid:2022uzn}. Adding a global SU(2) internal symmetry to the Generalized Proca leads to what we called the Generalized SU(2) Proca (GSU2P) theory \cite{Allys:2016kbq,Gomez:2019tbj,GallegoCadavid:2020dho,GallegoCadavid:2022uzn}, whose action is 
invariant under diffeomorphisms  and globally invariant under the SU(2) group transformations. However, the equivalence between these two vector-tensor theories is not straightforward due to the non-Abelian nature of the GSU2P theory, leading to the presence or absence of new terms, which can potentially result in new phenomenology \cite{Garnica:2021fuu}. Although the GSU2P theory was first formulated in \cite{Allys:2016kbq} by imposing a primary constraint-enforcing relation to eliminate the non-physical degree of freedom from the vector field, a secondary constraint-enforcing relation was required to close the constraint algebra \cite{GallegoCadavid:2020dho}. Therefore, all Lagrangian building blocks in the GSU2P theory were constructed to ensure the propagation of the correct number of physical degrees of freedom, thereby avoiding the Ostrogradski instability.

The aim of this paper is to explore the astrophysical implications of the GSU2P theory, which provides a useful framework for investigating the theory at that relevant scale. The field equations in the GSU2P theory are highly non-trivial, which makes it challenging to obtain analytical solutions for the entire theory. Therefore, our focus in this study is to examine the phenomenological aspects of individual Lagrangian pieces, building upon our previous work on particle-like solutions \cite{Martinez:2022wsy}. Specifically, we delve into the analysis of two Lagrangian terms that involve quartic order non-derivative self-interactions of the gauge field. We choose these Lagrangian terms because they enable the derivation of exact analytical solutions for BH, unlike the other Lagrangian terms of the theory. Moreover, these solutions are distinguished by possessing an effective global charge, as was unveiled in our recent paper \cite{Martinez:2022wsy}. Interestingly, the solutions we obtained correspond to a Reissner-Nordstr\"{o}m solution with a non-Abelian magnetic charge. This magnetic charge adds an intriguing aspect to the traditional Reissner-Nordstr\"{o}m solution, highlighting the presence of non-Abelian gauge fields and their impact on the properties of BHs. We plan to report on numerical BH solutions for the derivative self-interaction terms in a separate work. 

In this paper, we report an analytical exact Reissner Nordstr\"{o}m solution within the GSU2P theory. The solutions are characterized by two non-Abelian effective charges that depend on the corresponding coupling constants of the Lagrangian pieces. While this solution shares the same RN spacetime structure as the standard solution, there is a crucial difference: the charge in our solution is not of electromagnetic origin and can even be imaginary, resulting in negative energy density. This leads to interesting astrophysical implications that we uncover in this work. It is worthwhile mentioning that our findings are rooted in modified theories of gravity, although they were particularly derived from Lagrangian pieces minimally coupled to gravity. 

%On the other hand, it is possible to construct a healthy theory that includes higher derivative self-interactions of the SU(2) gauge field while still propagating the correct number of degrees of freedom. This theory is known as the Generalized SU(2) Proca (GSU2P) theory which is the non-Abelian version of the Generalized Proca theory \cite{Heisenberg:2014rta,Allys:2015sht,Allys:2016jaq,GallegoCadavid:2019zke} and belongs to a class of vector-tensor theories that is inspired by Horndeski's theory \cite{Horndeski:1974wa}. Considering particularly some Lagrangian pieces that involve four gauge fields minimally coupled to gravity, which arise from a systematic construction in the full theory, gives place to BH solutions with two different non-Abelian effective charges that depend on the coupling constants. It is worthwhile mentioning that our findings are rooted in modified theories of gravity although they were particularly derived from these Lagrangian pieces.
Accretion processes of ideal and polytropic fluids onto black holes have been an area of intense study in astrophysics (see e.g \cite{Bondi:1952ni,1972Ap&SS..15..153M,Richards:2021zbr,Aguayo-Ortiz:2021jzv}), serving as a probe of concept in both the context of GR and more general frameworks. In particular, accretion flows in an arbitrary spacetime have been extensively studied as valuable astrophysical probes for detecting any deviations from GR and testing alternative theories of gravity.\cite{Bauer:2021atk,Salahshoor:2018plr,Feng:2022bst,Zuluaga:2021vjc,Ditta:2020jud,Uniyal:2022vdu,Chakhchi:2022fls,John:2019was,Perez:2017spz,Liu:2021yev,Heydari-Fard:2021ljh,Stashko:2021lad,Shaikh:2019hbm,VanAelst:2021uem}.

The paper is structured as follows. In Section \ref{sec:2}, we introduce the model and derive in detail an exact non-Abelian RN BH solution in terms of the coupling constants of the theory. Some properties of the BH solutions, such as the event horizon, photon sphere, and shadow, are studied. In particular, observational data of the EHT's first images of Sagittarius A$^{\star}$ are used to infer the first constraints on the effective coupling constant. In Section \ref{sec:3}, we present a general description of the hydrodynamics equations of the accretion flow, and in Section \ref{sec:4}, we calculate the critical accretion rate for isothermal and polytropic fluids using analytical and numerical computations. In Section \ref{sec:5}, we discuss the main findings and possible extensions of the work, along with further observational constraints on the theory that can be used in the future. Throughout the manuscript, Latin indices are internal SU(2) group indices and run from 1 to 3, while Greek indices stand for spacetime indices and run from 0 to 3. We use geometrized units with $c = G = 1$.

%%%%%%%%%%%%%%%%%%%%%%%%%%%%
\section{Reissner Nordstrom black hole with non-Abelian charge}\label{sec:2}
The action of the model, which corresponds to some Lagrangian pieces of the GSU2P theory \cite{GallegoCadavid:2020dho}, includes quartic order self-interactions of the vector field\footnote{As the inclusion of a mass term $\mu^{2}B_{a\alpha}B^{a\alpha}$ spoils the existence of the solution, it has been taken away from the model. This result is similar to the classical massive vector field, where the mass needs to vanish to guarantee regularity of the solution and to allow, therefore, a vector hair to exist \cite{Bekenstein:1971hc}.},
\begin{multline}
    S = \frac{1}{16\pi}\int \sqrt{-g}\, d^4x[R - F_{a\mu\nu}F^{a\mu\nu}\\
    +\chi_1 B_{a \mu}B^{a\mu}B_{b\nu}B^{b\nu} + \chi_2 B_{a\mu}B^{a}{}_{\nu}B_{b}{}^{\mu} B^{b \nu}],\label{eqn:action}
\end{multline}
where $R$ is the Ricci scalar, $B_{a\mu}$ represents the vector fields, $F_{a\mu\nu}= \partial_\mu B_{a\nu}-\partial_\nu B_{a\mu}+ \tilde{g} \epsilon_{abc}B^{b}{}_{\mu} B^{c}{}_{\nu}$ is the field strength, $\tilde{g}$ is the gauge coupling constant and $\epsilon_{abc}$ is the structure constant tensor of the SU(2) group. In geometrized units $\tilde{g}$ has units of inverse length, and the free parameters $\chi_1$ and $\chi_2$ have units of inverse square length.

The line element in a stationary and spherical symmetric spacetime has the following form,
\begin{align}
   ds^2 &= g_{tt}(r) dt^2 + g_{rr}(r)dr^2 + r^2d\Omega^2 \nonumber\\ 
   &= - e^{-2\delta}Ndt^2 + N^{-1} dr^2 + r^2d\Omega^2\label{eqn:ds2-0}, 
\end{align}
where $N = 1 -2m/r$, $\delta$ and $m$ are functions of the coordinate $r$, and $d\Omega$ is the  solid angle element.
Regarding the vector fields we chose the Wu-Yang monopole, given by,
\begin{equation}
    \mathbf{B} =   (w/v + 1)\, \mathbf{t}_{\phi} d\theta + (v-w)\sin\theta\, \label{eqn:wuyang}\mathbf{t}_{\theta} d\phi 
\end{equation}
%\begin{equation}
%    \mathbf{B}_t = \mathbf{B}_r=0
%\end{equation}
%\begin{equation}
%    \mathbf{B}_{\theta} = (w/v + 1)\, %\mathbf{t}_{\phi}
%\end{equation}
%\begin{equation}
%    \mathbf{B}_{\phi} = (v-w)\sin\theta\, %\mathbf{t}_{\theta},
%\end{equation}
where,
\begin{align}
    \mathbf{t}_{\theta} &= \cos\theta\cos\phi \,\mathbf{t}_1 + \cos\theta\sin\phi\,\mathbf{t}_2 - \sin\theta\, \mathbf{t}_3,\\
    \mathbf{t}_{\phi} &=-\sin\phi\, \mathbf{t}_1 +\cos\phi\, \mathbf{t}_2,
\end{align}
in which $\mathbf{t}_i = -i\sigma_i/2$ correspond to the vector basis of the SU(2) algebra with $\sigma_i$ being the Pauli matrices, $w$ is constant and $v$ is an integer denoting the azimuthal winding number.
We use the coupling constant $\tilde{g}$ to define the normalized variables, $\hat{r}=r\tilde{g}$, $\hat{m}=m\tilde{g}$, $\hat{\chi}_1 = \chi_1/\tilde{g}^2$ and $\hat{\chi}_2=\chi_2/\tilde{g}^2$. The form of the equations in the normalized variables can be obtained effectively by setting $\tilde{g}=1$. Hereafter, all the equations are normalized, but we drop the hat to ease the notation.

The field equation obtained after varying the action with respect to $B_{a\mu}$ is,
\begin{equation}
    (v+w) \left[\left(v^2 \chi _1+v^2 \chi _2+\chi _1\right) (v+w)^2+w (v-w)\right]=0.\label{eqn:weq}
\end{equation}
The solutions of this last equation are, 
\begin{equation}
    w_{\rm schw} = -v \label{eqn:schwsol},
\end{equation}
\begin{widetext}
\begin{equation}
    w_{\rm I,II} = \frac{v+2 v \chi _1 + 2 v^3 \left(\chi _1+\chi _2\right)\pm\sqrt{v^2 \left[8 v^2 \chi _2+8 \left(v^2+1\right) \chi _1+1\right]}}{2 \left(1- v^2 \chi _1-v^2 \chi _2-\chi _1\right)}.\label{eqn:wI-II}
\end{equation}
\end{widetext}
The first solution \eqref{eqn:schwsol} is the trivial solution with vanishing vector field which corresponds to the Schwarzschild spacetime\footnote{To fully understand this solution, we need to consider
Eq.~(\ref{eqn:dmdr}). If we set $w_{\rm sch}=-v$ we find $m^\prime=0$, implying that the mass function takes the form of the Schwarzschild solution, where $m=M$ and $\delta=0$. Here $M$ is the total gravitational mass. This, in turn, is consistent with an asymptotically flat solution.}. Instead, the latter solutions \eqref{eqn:wI-II}, with two branches I and II, allow the existence of a non-trivial vector field we shall focus on, and constitutes, therefore, an important outcome of this work, as will be described in detail below.

%In this last case the Eqs \eqref{eqn:d2deltadr2} and \eqref{eqn:weq} coincide. 

The field equations obtained by varying the action \eqref{eqn:action} with respect to metric are given by,
\begin{multline}
   m'-\frac{ \left(v^3-v w^2\right)^2}{2 r^2 v^4} \\+\frac{(v+w)^4 \left[v^4 \chi _2+\left(v^2+1\right)^2 \chi _1+\chi _2\right]}{{4 r^2 v^4}}=0 \label{eqn:dmdr}
\end{multline}
\begin{equation}
    \delta'=0 \label{eqn:ddeltadr}
\end{equation}
\begin{multline}
    m''+\frac{ v^2 (v-w) (v+3 w) (v+w)^2}{ r^3 v^4}\\
    +\frac{\left[\left(v^4+5\right) \chi _2+\left(v^4+6 v^2+5\right) \chi _1\right] (v+w)^4}{2 r^3 v^4}=0.\label{eqn:d2deltadr2}
\end{multline}
We look for asymptotically flat solutions, i.e. the components of the metric have the following behavior $g_{\mu\nu}\to \eta_{\mu\nu} + \mathcal{O}(r^{-1})$ when $r\to \infty$, where $\eta_{\mu \nu}$ are the components of the Minkowski metric (see e.g. \cite{1984ucp..book.....W}). This implies that the functions tend asymptotically to $\delta \to 0$, $m\to M \equiv {\rm finite}$ \cite{1984ucp..book.....W}. 
The Arnowitt-Desser-Misner mass, which coincides with the Komar mass in this spherically symmetric and stationary case, is given by the asymptotic value of $m(r)$, thus $M$ corresponds the total gravitational mass \cite{1979JMP....20..793A}. 
Under these conditions Eq. \eqref{eqn:ddeltadr} is easily solved as $\delta=0$. 
On the other hand, the solution of mass function $m$ for the cases given by \eqref{eqn:wI-II} has the following Reissner-Nordstr\"{o}m solution,
\begin{equation}
    m = M - \frac{Q_{\rm NA}^2}{2 r}, \label{eqn:metricsol}
\end{equation}
where is a constant representing the effective charge and depends on the free parameters of the theory.
%as follows,
%\begin{multline}
%    Q^2_{\rm NA} = \frac{\left(v^2-w_{\rm I,II}^2\right)^2}{v^2}\\
%    -\frac{(v+w_{\rm I,II})^4 \left[\left(v^4+1\right) \chi _2+\left(v^2+1\right)^2 \chi _1\right]}{2 v^4}.\label{eqn:charge} 
%\end{multline}

Eqs. \eqref{eqn:dmdr} and \eqref{eqn:d2deltadr2} must be consistent, which implies an additional constraint between $\chi_1, \chi_2 $ and $v$ given by, 
\begin{widetext}
\begin{equation}
    \frac{3 \left(v^4-1\right) \left(\chi _1+\chi _2\right) \left\{4 v^3 \chi _2+3 \sqrt{v^2 \left[8 v^2 \chi _2+8 \left(v^2+1\right) \chi _1+1\right]}+4 \left(v^3+v\right) \chi _1+5 v\right\}^2}{ \left[v^2 \chi _2+\left(v^2+1\right) \chi _1-1\right]{}^4} = 0 \label{eqn:constraintFE}.
\end{equation}
\end{widetext}
One solution fixes the winding number as, 
\begin{equation}
    v = \pm 1, \label{eqn:1branch}
\end{equation}
with the parameters $\chi_1$ and $\chi_2$ being independent.
The other solution gives a relation between the free parameters of the action model,
\begin{equation}
    \chi_2 = -\chi_1 \label{eqn:2branch},
\end{equation}
with the winding number now unconstrained\footnote{It seems that there is another solution given by the vanishing of the expression between braces in the numerator of Eq. \eqref{eqn:constraintFE}. Nevertheless, this does not constitute a solution because it makes also the denominator to vanish, inducing a divergence.}.

Despite the fact that the solution set by \eqref{eqn:1branch} gives two possible solutions, it represents only one since changing the sign of the winding number interchanges the solutions $w_{\rm I}$ and $w_{\rm II}$ as can be verified in Eq.~\eqref{eqn:wI-II}. In the solution given by  \eqref{eqn:1branch} the value of $w$ depends on the combination $\chi = 2\chi_1 + \chi_2$,
\begin{equation}
    w_{\rm I,II} = \frac{1 + 2\chi\pm\sqrt{1+8\chi}}{2 -  2\chi} \label{eqn:sol_wc2}\,.
\end{equation}
In the other solution corresponding to \eqref{eqn:2branch}, the value of $w/v$ has the same functional form of \eqref{eqn:sol_wc2} after making  $\chi\mapsto \chi_1$. Therefore, the physical behavior of both cases can be analyzed by means of \eqref{eqn:sol_wc2}. Hence, all subsequent analysis will be carried out in terms of the new effective coupling constant $\chi$. Notice that once the value of $\chi$ is determined, the value of $w$ is fully specified. This is consistent with the Wu-Yang monopole Ansatz, which assumes that $w$ is a constant, as stated in Eq. (\ref{eqn:wuyang}).

\begin{figure*}
    \centering
    \includegraphics[width=0.45\textwidth]{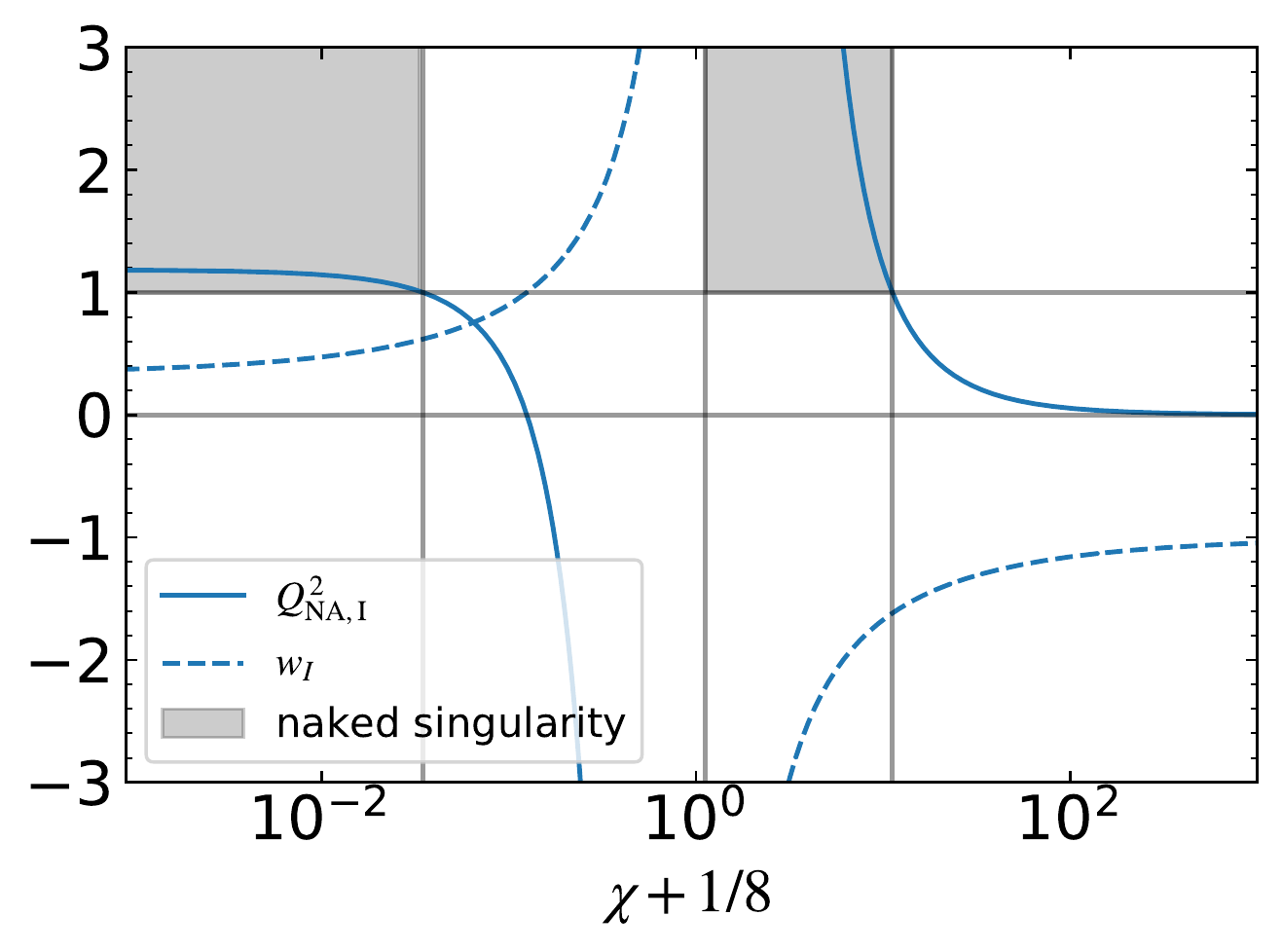}
    \includegraphics[width=0.45\textwidth]{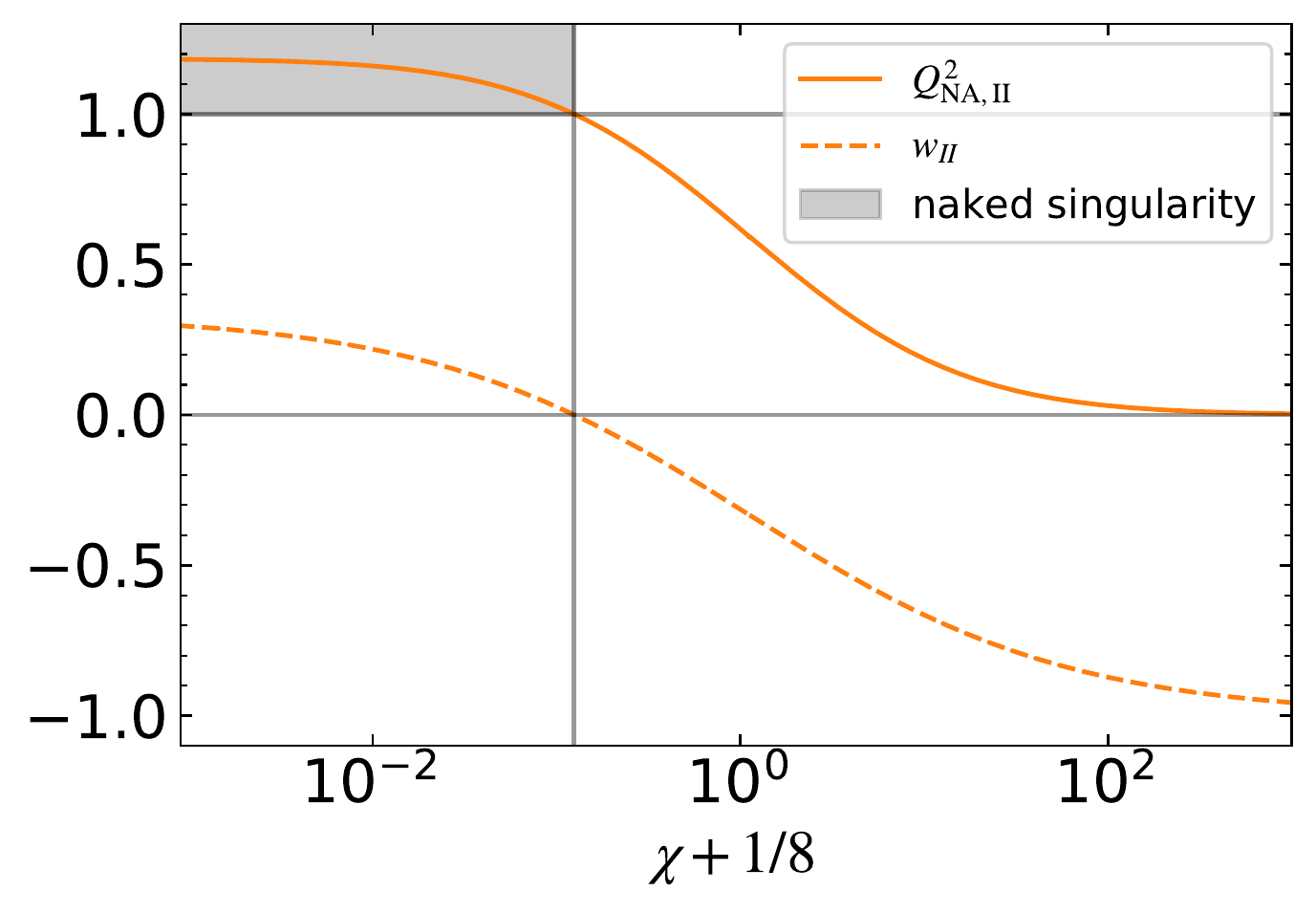}
    \caption{Values of the vector field and effective charge as a function of $\chi$ for the branch I (left panel) and for the branch II (right panel). In both cases when $\chi\to \infty$ the charge tends to zero, thus the solution becomes the Schwarzschild spacetime. If the mass of the black hole in normalized units is $M=1$, values of $Q^2_{\rm NA}>1$ represent a naked singularity. These last cases are shown as gray regions. In the branch II the charge is always real. In the branch I for $0<\chi<1$ the charge is imaginary, which implies that the energy density is negative. Notice also that the shift $\chi+1/8$ in the abscissa has been done for convenience.}
    \label{fig:Q2_ws}
\end{figure*}

Consequently, the value of the effective charge $Q^2_{\rm NA}$ is given by,
\begin{equation}
    Q^2_{\rm NA, I,II}=\frac{1-4\chi(5+2\chi)\mp (1+8\chi)^{3/2}}{2(1-\chi)^3}\label{eqn:charges}\, ,
\end{equation}
where the $-$ sign  corresponds to the branch I, and the + sign to the branch II. The dependence of both quantities on the coupling constant is displayed in Figure \ref{fig:Q2_ws}.

There exist solutions for the interval $-1/8<\chi<1 \cup \chi>1$. When $\chi = 0$ the branch I solution corresponds to the Schwarzschild solution, and the branch II solution corresponds to the EYM charged solution with $Q^2=1$. When $\chi \to 1$ the branch I has a divergence, in contrast, in the same limit, the branch II is finite and tends to $w_{\rm II}\to -1/3$, $Q^2_{\rm NA,II}\to 16/27$.

On the hand, if we assume that the total mass of the black hole in normalized units is $M=1$, we can find regions where $Q^2_{\rm NA}>1$, corresponding to naked singularities. These regions are given by $-1/8<\chi<(11 - 5\sqrt{5})/2 \cup 1 < \chi < (11 + 5\sqrt{5})/2$, for the branch I, and $-1/8<\chi<0$, for the branch II (see the gray regions in Figure \ref{fig:Q2_ws}). Notice that we conjecture the existence of a naked singularity based solely on the vanishing of the event horizon. However, this criterion alone is insufficient to ensure the formation of such a spacetime. Therefore, a formal stability analysis  must be carried out to determine whether  a naked singularity can indeed form in this theory\footnote{There are in fact physical reasons to believe that the formation of a naked singularity, at least in a Kerr BH spacetime, is highly unlikely in any realistic astrophysical collapse scenario \cite{Visser:2007fj}}..

Finally, it is worthwhile to mention that the energy density associated with the vector fields is $\rho = Q^2_{\rm NA}/(8\pi r^4)$. Thus, when the non-Abelian charge is imaginary the energy density is negative, and this happens only in the branch I in the interval $0<\chi<1$.
\begin{figure*}
\centering
\includegraphics[width=0.47\hsize,clip]{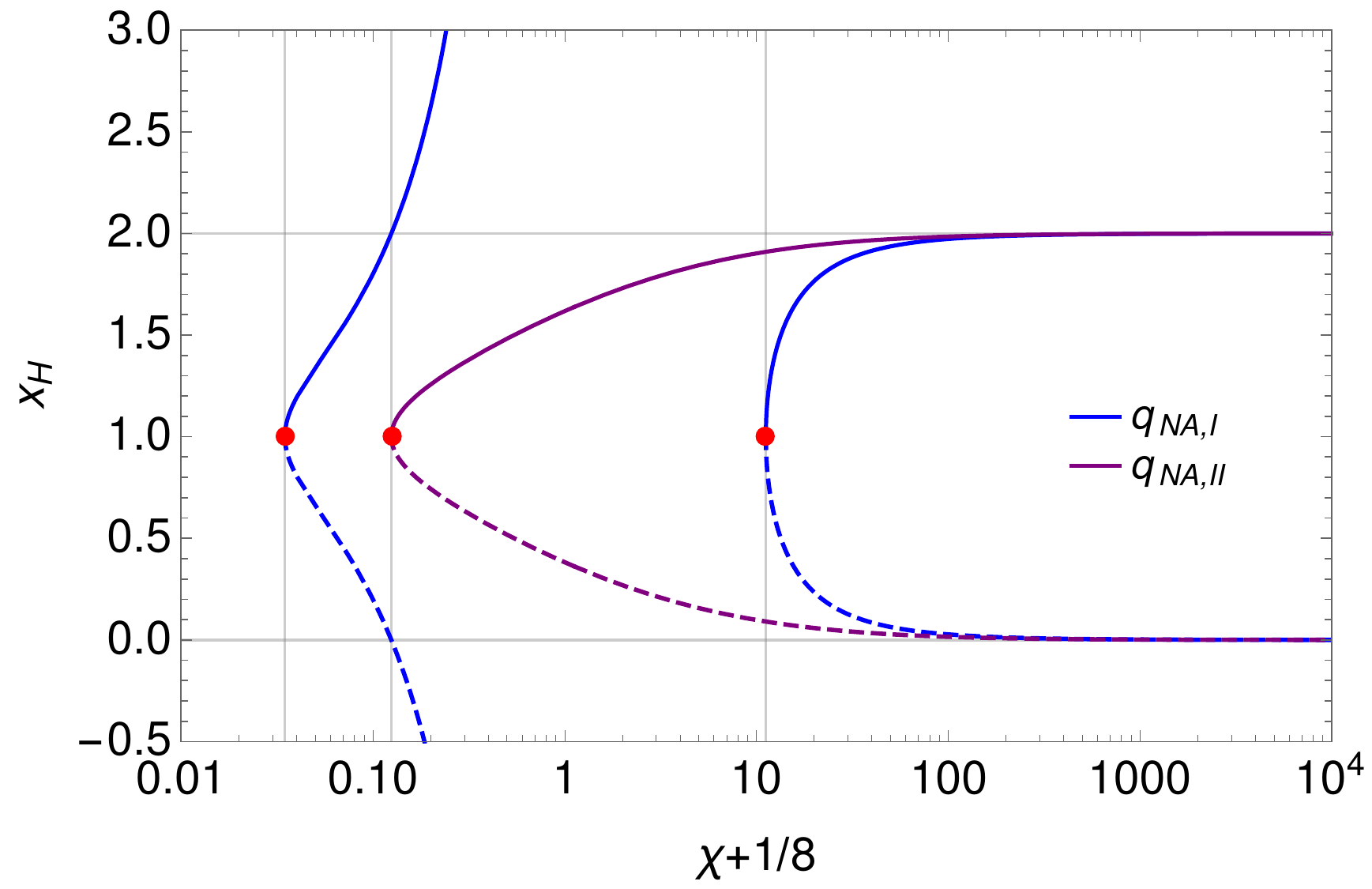}
\includegraphics[width=0.47\hsize,clip]{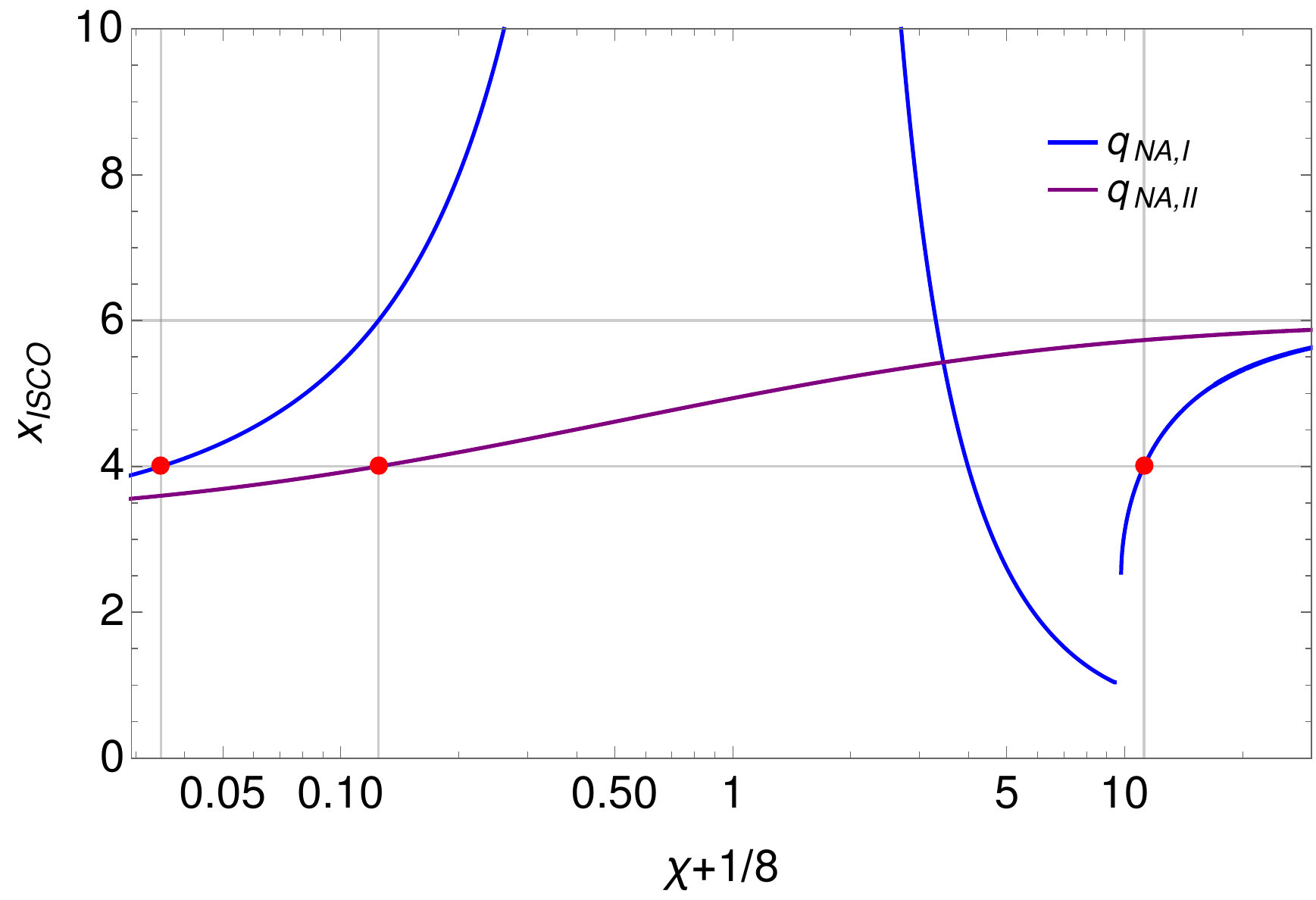}
\caption{\textit{Left panel}: position of the event horizon $x_{H}$ as a function of the coupling constant for both charges as specified by the legend. \textit{Right panel}: innermost stable circular orbit $x_{\rm ISCO}$ as a function of the coupling parameter $\chi$ as well for both non-Abelian charges. In both plots red points indicate the match with the extremal RN solution. The two-fold degeneracy of $q_{\rm NA,I}$ and the convergence to the Schwarzschild solutions when $\chi\to \infty$ for both charges is evident here. Note that the parameter $\chi$ explicitly appears in the expression for $q_{NA}$ as given by Eq. (\ref{eqn:charges}). Therefore, the event horizon and ISCO radii, which are determined by Eqs. (\ref{eqn:horizons}) and (\ref{eqn:isco}), respectively, are dependent on $\chi$ as well.} \label{fig:isco}
\end{figure*}
\subsection{Event horizon}
In the stationary and spherically symmetric case, the vanishing of the metric function $g_{tt}$ defines unequivocally the horizon (see e.g. \cite{1968JMP.....9.1319V}). When multiple solutions exist the greatest positive solution is identified with the event horizon of the black hole $r_{+}=r_{H}$. In particular, the metric function associated with the Reissner-Nordstr\"{o}m solution has two distinct roots
\begin{equation}
    r\pm =M\pm \sqrt{M^{2}-Q_{\rm NA}^{2}}.
\label{eqn:horizons}
\end{equation}
The internal solution  $r_{-}$ is an apparent horizon and the external solution corresponds to the event horizon. Hence, any observer outside the black hole (in asymptotically flat spacetimes), or on the event horizon itself, cannot see any singularity because they are protected by an event horizon. Otherwise it is said to posses a naked singularity at $r=0$. We do not mention here all the minor details and conditions about the precise formulation of what is called the \textit{weak cosmic censorship conjecture}\footnote{We refer reader to \cite{Wald:106274} for a general and robust formulation of the cosmic censor conjecture.}. %Violation of this can be possible, for instance, 

The structure of the RN solution has been studied extensively whereby we do not pretend to make here a detailed examination on this subject. Nevertheless, an intriguing query arises when one asks about the implications of the coupling constant $\chi$ on the charge and, therefore, on the event horizon. In particular, we are interested in finding which values of the coupling constant account for the convergence to both the Schwarzschild and the extremal RN black holes, as limit cases of our solution.

When performing numerical analysis and presenting the corresponding general discussion, we shall work with dimensionless variables by normalizing all physical quantities by the black hole mass $M$, unless otherwise said. Accordingly, we introduce, as usual, the charge to mass ratio $q_{\rm NA}=Q_{\rm NA}/M$ and the dimensionless radial coordinate $x=r/M$.

The event horizon is depicted in left panel of Figure \ref{fig:isco} as a function of the coupling constant for both charges\footnote{When plotting, the shift $\chi+1/8$ in the abscissa is done for convenience.}. For $q_{\rm NA,II}$ (purple solid line) the event horizon is a well-behaved function of the coupling constant. Indeed, it covers continually the full range $\chi\in(0,\infty)$ where the finite extreme value corresponds to the extremal RN case, in which case both solutions of Eq.~(\ref{eqn:horizons}) meet at $x_{H}=1$ (red point on the purple curve); while large values lead to the uncharged solution where $x_{H}=2$ and the apparent horizon (dashed curve) coincides with the singularity. This is hence a quite normal behavior that reproduces plainly the standard RN solution. 

On the contrary, the event horizon for the case $q_{\rm NA,I}$ (blue solid curves) exhibits a peculiar structure: there is a two-fold degeneracy with respect to the constant coupling.
It means that the limit cases, that is, the Schwarzschild solution and the extremal RN black hole, can be described in two distinct regions of the parameter space. The same degeneracy is also presented for the apparent horizon (blue dashed curves). This is clearly appreciated in the regions defined by the ranges $\chi\in(-0.0901,0)$ and $\chi\in(11.0902,\infty)$ of the left panel of Figure \ref{fig:isco}. There is also a special region of the parameter space $\chi\in (0, 1)$, where the square of the effective charge becomes negative and gravity repulsive. Out of the mentioned ranges results in $q_{\rm NA}^{2}>1$, a case that describes a naked singularity.

%%%%%%%%%%%%%%%%%%%%%%%%%%%%%%%%%%%%%

%%%%%%%%%%%%%%%%%%%%%%%%%%%%%%%%

\subsection{Inner most stable circular orbit}
%On the other hand, considering that the spacetime
%around this black hole is static and symmetric, and 
Next, we turn our attention to the motion of massive test particles in the equatorial plane, as usual, to find the position of the so-called \textit{innermost stable circular orbits} (ISCO). 
Following the standard procedure (see e.g. \cite{Pugliese:2010ps}), one finds that the geodesic equation can be written as
\begin{equation}
\left( \frac{dr}{ds} \right)^2 = \left[ E^2 - N \left( 1 + \frac{L^2}{r^2} \right) \right],
\end{equation}
where $E$ and $L$ are identified, respectively, as the energy and angular momentum  which correspond to conserved quantities as in the classical Keplerian motion. From the above equation, one can also identify the effective potential  
\begin{equation}
    V_{\rm eff} =N \left(1+\frac{L^{2}}{r^{2}}\right).
\end{equation}
We remind the reader that $N=1-2m/r$, the mass function $m(r)$ has the form given by Eq.~(\ref{eqn:metricsol}), and the (squared) angular momentum is defined as
\begin{equation}
L^{2} = \frac{N^{\prime}(r) r^{3}} {2 N(r) - r N^{\prime}(r)}.
\end{equation}
In marginally stable circular orbits, the local extremum of
the effective potential, i.e. $V_{\rm eff}^{\prime\prime}=0$, determines the position of the ISCO. It yields
\begin{widetext}
\begin{equation}
   r_{\rm ISCO} = 2M+\frac{4M^{4}-3M^{2} Q_{\rm NA}^{2} +
 \left( 8M^{6}-9M^{4}Q_{\rm NA}^{2}+2M^{2}Q_{\rm NA}^{4}+\sqrt{5M^{8} Q_{\rm NA}^{4}-9 M^{6} Q_{\rm NA}^{6} +4 M^{4} Q_{\rm NA}^{8}}\right)^{2/3}}{M\left( 8M^{6}-9M^{4}Q_{\rm NA}^{2}+2M^{2}Q_{\rm NA}^{4}+\sqrt{5M^{8} Q_{\rm NA}^{4}-9 M^{6} Q_{\rm NA}^{6} +4 M^{4} Q_{\rm NA}^{8}}\right)^{1/3}}.\label{eqn:isco}
\end{equation}
\end{widetext}
This expression of course equals the corresponding standard RN case \cite{Pugliese:2010ps}. At this point we are able to compute the location of the ISCO in terms of the coupling constant $\chi$ with the aid of Eq.~(\ref{eqn:charges}). The result is shown in right panel of Figure \ref{fig:isco}. Clearly the effect of the coupling constant is replicated on the ISCO structure in the same fashion as it is on the event horizon. Hence the same parameter space region establishes the corresponding Schwarzschild solution $x_{\rm ISCO}=6$, and the extremal RN case $x_{\rm ISCO}=4$ (red point on all curves) for both positive square of the non-Abelian charges as can be read from the plot. Thus, all the discussed properties for the event horizon are preserved for the ISCO location. It is interesting to notice, however, that for the charge $q_{\rm NA,I}$, there appears another region where gravity is repulsive in the interval $\chi\in(1,11.092)$, in comparison with the ones seen for the event horizon. For this charge it is admissible then to have stable circular orbits for almost all the parameter space with the exception of its extreme values even though there does not exist an event horizon. This particular behavior is reminiscent of having stable accretion flows onto a RN black hole, even thought there is not a physical event horizon. Hence, in a naked singularity situation it is (mathematically) possible to have stable circular orbits. We will not discuss this point in detail, however, since it is not of physical interest for the present work. 

\subsection{Photon sphere and shadow}
\begin{figure*}
    \centering
    \includegraphics[width=0.49\textwidth]{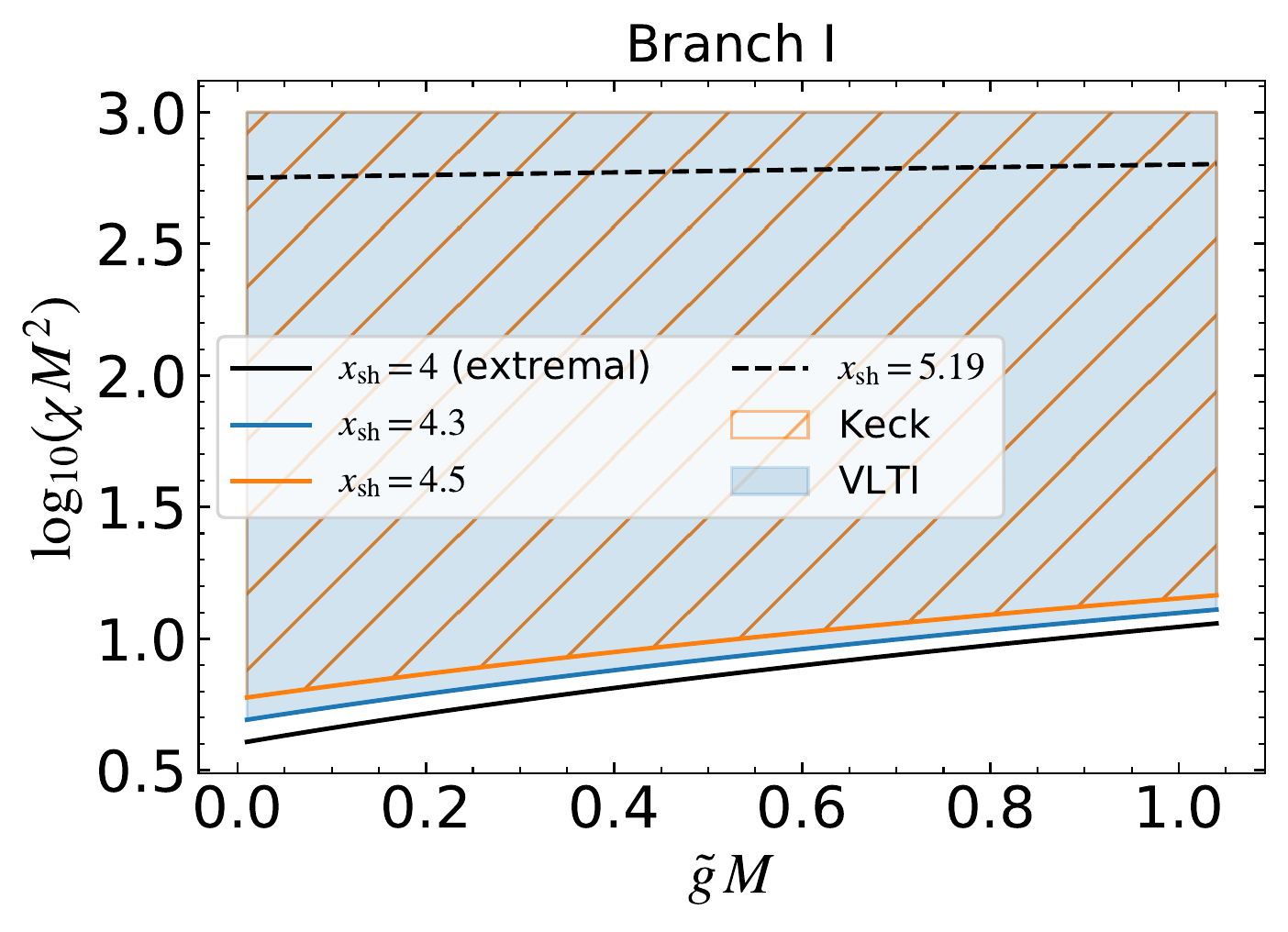}
    \includegraphics[width=0.49\textwidth]{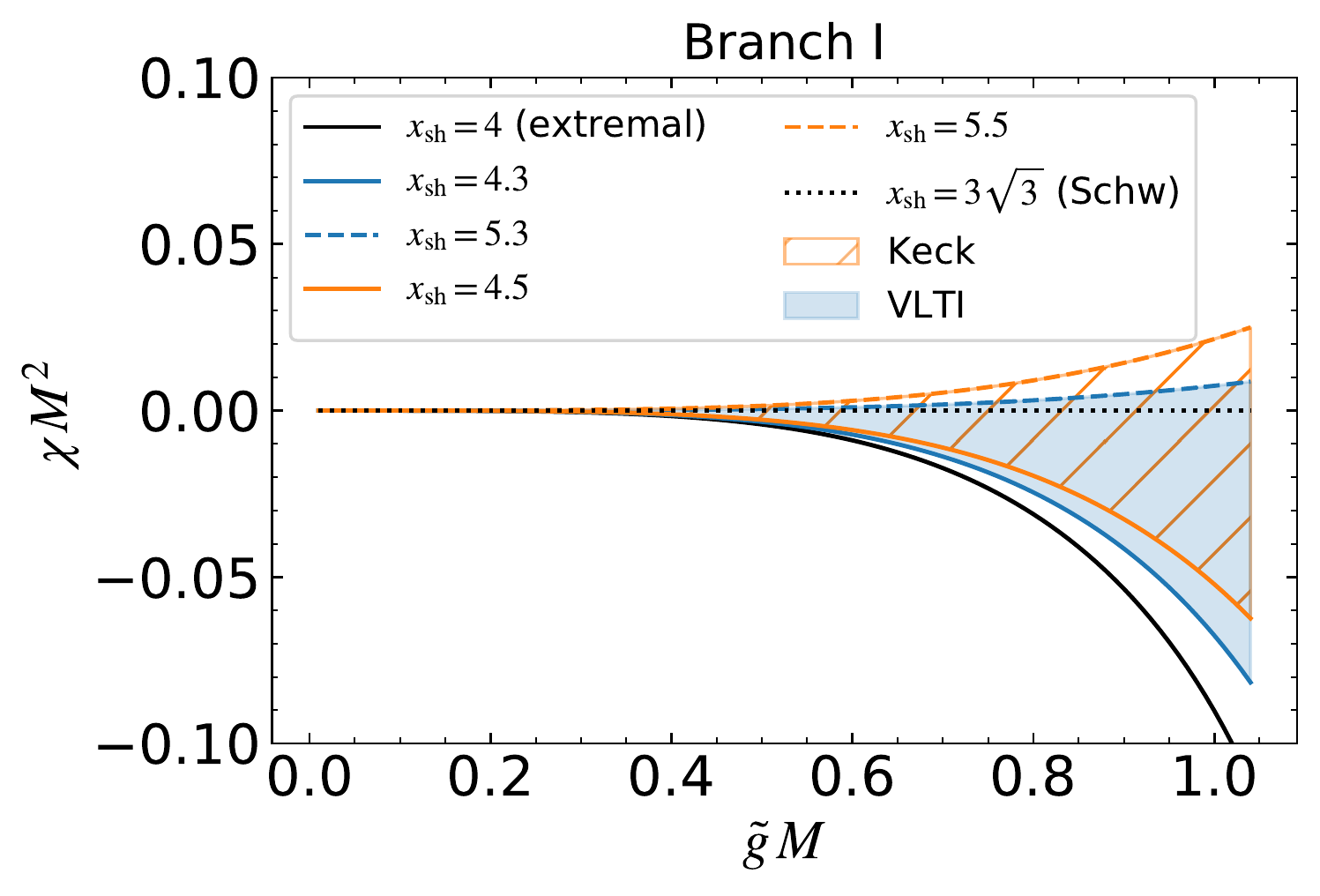}
    \includegraphics[width=0.49\textwidth]{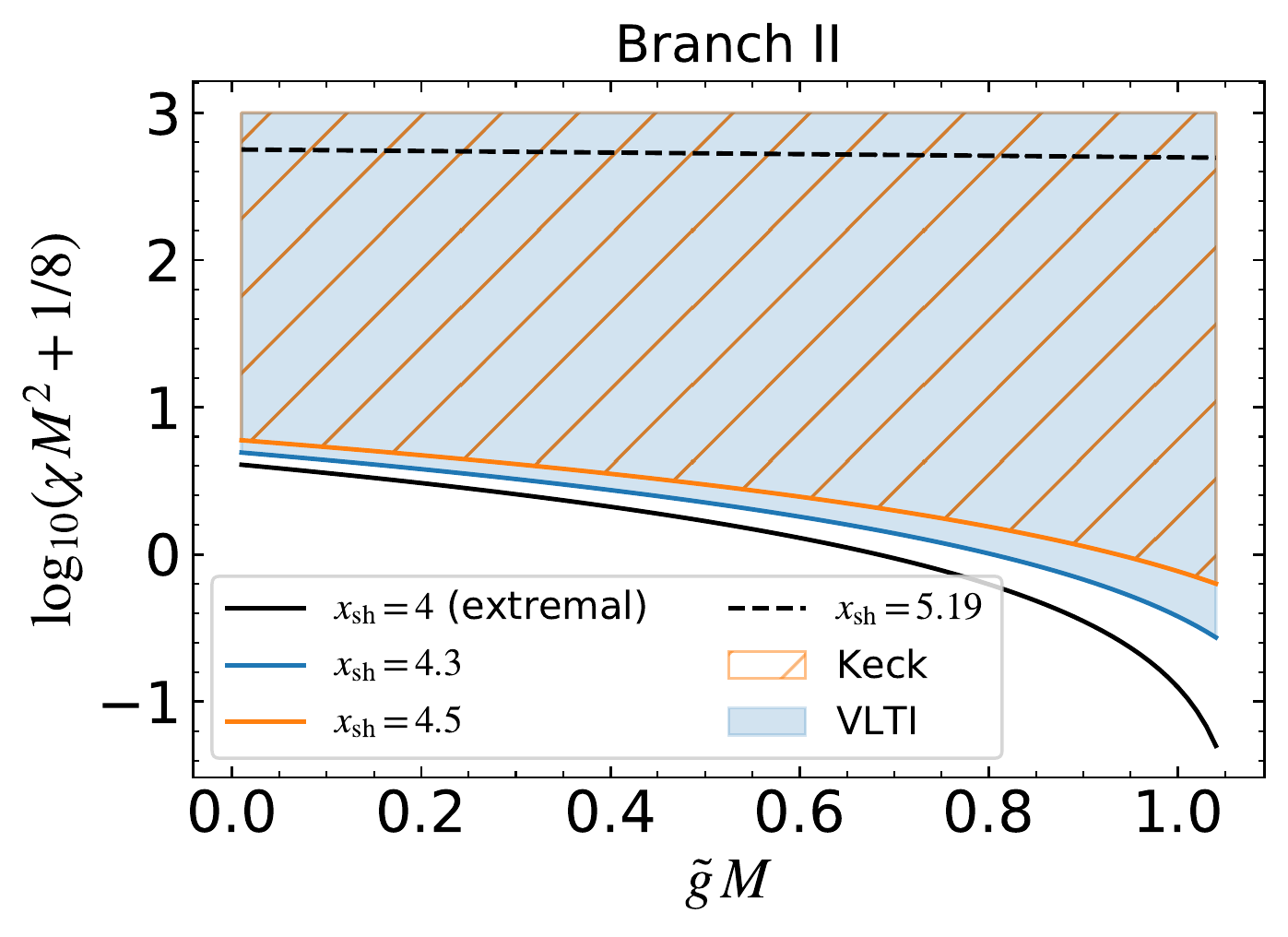}
    \caption{Constraints on the parameters $\tilde{g}$, and $\chi$ obtained from the observation of the shadow of the object located at the Galactic Center of the Milky Way, Sagittarius A$^\star$. The upper plots correspond to the branch I which has been split into two ranges of $\chi$ for better representation, and the lower plot corresponds to the branch II. The different colored curves show the possible values of $g_c$ and $\chi$ for the lower and upper limits on $x_{\rm sh}$ given by \eqref{eqn:keck} and \eqref{eqn:VLTI} as indicated by the legend. For comparison, we have added the curves for $x_{\rm sh}=4,5.19,3\sqrt{3}$. The regions with slanted orange lines represent the constraints obtained from the mass-to-distance ratio with Keck, and the blue regions represent the constraints with VLTI. It can be seen that a naked singularity is not consistent with the observational data.}
    \label{fig:shadow-constr}
\end{figure*}

A black hole has a central dark area called the shadow. This shadow is not delimited by the event horizon, but by the photon sphere, which is made of circular photon orbits. The radius of the photon sphere $r_{\rm ph}$ is given by solving,
\begin{equation}
    r_{\rm ph} g_{tt}'(r_{\rm ph})-2g_{tt}(r_{\rm ph})=0.
\end{equation}
However, the observed shadow radius, $r_{\rm sh}$ is given by the lensed image of this surface \cite{2008PhRvD..77f4006P},
\begin{equation}
    r_{\rm sh} = \frac{r_{\rm ph}}{\sqrt{g_{tt}(r_{\rm ph})}}.
\end{equation}
For the Reissner-Nordstr\"om solution the shadow radius is,
\begin{equation}
    \frac{r_{\rm sh}}{M} = x_{\rm sh}=\frac{\sqrt{2} \left(\sqrt{9-8 q_{\rm NA}^2}+3\right)}{\sqrt{\frac{4 q_{\rm NA}^2+\sqrt{9-8 q_{\rm NA}^2}-3}{q_{\rm NA}^2}}}\, .
\end{equation}
This last equation can be inverted such that, for a given observed shadow radius, we can constrain the charge $q_{\rm NA}$. From this value we can constrain in turn the free parameters of the theory $\tilde{g}$ and $\chi$. The observations of Sagittarius A$^*$ made by the EHT collaboration give the following constraints on the size of the shadow, which depend on the mass-to-distance ratio \cite{2022ApJ...930L..17E},
\begin{equation}
4.5 \lesssim x_{\rm sh} \lesssim 5.5, \label{eqn:keck}
\end{equation}
with Keck and
\begin{equation}
  4.3 \lesssim x_{\rm sh} \lesssim 5.3, \label{eqn:VLTI}  
\end{equation}
with VLTI. With this information, we can place  constraints on the  parameters $\tilde{g}$ and $\chi$, shown in Figure \ref{fig:shadow-constr}. Since \eqref{eqn:charges} is normalized but depends physically on the $\tilde{g}$ and $\chi$, only in this part of the analysis do we put back the units. The first conclusion is that an extremal RN black hole is not consistent with the current observations.

If the non-Abelian charge is real, the maximum value of the shadow is $ r_{\rm sh} = 3\sqrt{3} M$, obtained when $Q_{\rm NA}=0$, i.e., when the solution is the Schwarzschild spacetime. This last case corresponds to the limit $\chi\to\infty$ in both branches I and II. The shadow of a Schwarzschild black hole is inside the intervals \eqref{eqn:keck} and \eqref{eqn:VLTI}. 
Since the charge is always real for \emph{branch II}, we can give only lower limits on $\chi M^2$ for this branch (see the blue and orange curves in the lower plot of Figure \ref{fig:shadow-constr}). For example, if $\tilde{g}M = 1.01$, the minimum values of $\chi M^2$ are $0.608985$ and $0.226688$, for Keck and VLTI, respectively.

In contrast, the \emph{branch I} exhibits imaginary values of the charge, allowing a greater shadow than the one of a Schwarzschild black hole. Also, in this branch the allowed region of values for $\tilde{g}M$ and $\chi M^2$ are greater than in branch II. For instance, when $\tilde{g} M = 0.9$ the minimum values of $\chi M^2$ are $-0.0326136$ and $-0.0414562$, for mass-to-distance ratio with Keck and VLTI, respectively. This implies a less stringent constraint in the parameters than in the branch discussed above. More exactly, if $\tilde{g}M=1.01$, the constraints are $(-0.0545704 \lesssim\chi M^2\lesssim 0.0223263)\, \cup \, 14.3271\lesssim \chi M^2$, for Keck and $(-0.0708403\lesssim\chi M^2\lesssim0.00774184 )\,\cup\, 12.6237\lesssim\chi M^2$, for VLTI, see the upper plots in \ref{fig:shadow-constr}. More cases for branch I can be inferred from the Figure \ref{fig:shadow-constr}.

The main purpose now is to figure out how the structure of the non-Abelian RN black hole can impact the transonic properties of accretion flows. After formulating the basic hydrodynamics equations, this subject will be investigated by employing both numerical and analytical treatments in the subsequent sections.

%%%%%%%%%%%%%%%%%%%%%%%%%%%%
\section{Spherical steady accretion flows in a spherically symmetric spacetime}\label{sec:3}

Bondi accretion process in a spherically symmetric spacetime is briefly described in this section following the general prescription presented in Ref.~ \cite{Bauer:2021atk}. Accordingly,  we consider a spacetime with line element given by \eqref{eqn:ds2-0}, with $m$ and $Q^2_{\rm NA}$ given, respectively, by \eqref{eqn:metricsol} and \eqref{eqn:charges}, describing a black hole of mass $M$ and non-Abelian charge $Q_{\rm NA}^{2}$ in Schwarzschild coordinates. Although the non-Abelian RN solution  Eq.~(\ref{eqn:metricsol}) is, in principle, globally indistinguishable for the standard RN BH solution, it was found, formally speaking, in the framework of modified gravity. So, we keep as much as possible the generality in the description\footnote{This general treatment serves also as a starting point to other (non-analytical) BH solutions found in the GSU2P theory \cite{Martinez:2022wsy}.}. On the other hand, we consider a steady fluid with total density $\rho$, mass density $\rho_{0}$ and internal energy density $\epsilon$, such that $\rho=\rho_{0}+\epsilon$. For isentropic fluids the pressure can be defined as $P=k\rho^{\gamma}$ where $k$ is a constant and $\gamma$ is the adiabatic index. For perfect fluids, the stress energy momentum tensor is given by
\begin{equation}
   T^{\mu\nu}= (\rho+P)u^{\mu}u^{\nu}+P g^{\mu\nu},\label{eqn:streetensor} 
\end{equation}
where $u^{\mu}=(u^{t},u^{r},0,0)$ is the four velocity of the fluid characterized by infall radial flow. The normalization condition allows ones to obtain the relation between the components $u^{t}=\sqrt{\frac{g_{rr}(u^{2}+g_{tt})}{g_{tt}}}$, where we have defined for abbreviation $u \equiv u^{r}$. From the baryon conservation and energy momentum conservation
\begin{align}
    \nabla_{\mu}(\rho_{0}u^{\mu})&=0,\\
    \nabla_{\mu}T^{\mu\nu}&=0,
\end{align}
one obtains two master equations, respectively
\begin{align}
    \frac{\rho^\prime_{0}}{\rho_{0}}+\frac{u^\prime}{u}+&\Sigma=0,\label{eqn:densityeqn}\\
    u u^\prime + \frac{g_{tt}\prime}{2 g_{rr} g_{tt}} (1 + g_{rr} u^{2}) +\frac{g_{rr}^\prime}{2g_{rr}} u^{2}&+   \frac{c_{s}^{2}}{g_{rr}}(1+ g_{rr} u^{2}) \frac{\rho^\prime_{0}}{\rho_{0}}=0,\label{eqn:velocityeqn}
\end{align}
where prime denotes radial derivative and the quantity $\Sigma\equiv \frac{(\sqrt{-g})^\prime}{\sqrt{-g}}$ has been introduced for brevity as in \cite{Bauer:2021atk}. These equations reduce to the Schwarzschild case when $-g_{tt}=\frac{1}{g_{rr}}=1-\frac{2M}{r}$. In finding Eq.~(\ref{eqn:velocityeqn}), we have used the definition of the sound speed of a medium at constant entropy $c_{s}^{2}\equiv \frac{dP}{d\rho}$, and the useful relation $P^{\prime}=\frac{(\rho+P)}{\rho_{0}}c_{s}^{2}\;\rho_{0}^{\prime}$ derived from the first law of thermodynamics in the form $\frac{d\rho}{d\rho_{0}}=\frac{\rho+P}{\rho_{0}}$. Integration of Eqs.~(\ref{eqn:densityeqn}) and (\ref{eqn:velocityeqn}) 
gives, respectively, the mass accretion rate
\begin{equation}
    \dot{M}=4\pi r^{2} u \rho_{0},\label{eqn:accretion0}
\end{equation}
and, after some algebraic manipulations, the relativistic version of the Bernoulli equation (see Ref.~ \cite{Bauer:2021atk}. for more details)
\begin{equation}
    g_{tt}(1 + g_{rr} u^{2})\left(\frac{\rho+P}{\rho_{0}}\right) = C,\label{Bernoullieqn}
\end{equation}
where $C$ is an integration constant. Once $C$ is defined by the boundary conditions at infinity for instance, and the metric functions are specified, the inward radial velocity can be computed for a given equation of state $P=P(\rho_{0})$.  Our primary concern is to solve this equation to determine, in turn, the accretion rate given by Eq.~(\ref{eqn:accretion0}). Before computing this, we find critical values at which accretion flow is regular and causality is guaranteed.

%%%%%%%%%%%%%%%%%%%%
\section{Accretion  in a non-Abelian Reissner-Nordstr\"{o}m black hole}\label{sec:4}

%%%%%%%%%%%%%%%%%%%%
\subsection{Critical accretion}

We study in this part general conditions under which transonic flow can take place in the vicinity of black holes. Next, we shall describe both the spacetime geometry and the fluid nature. Let us first write  Eqs.~(\ref{eqn:densityeqn}) and (\ref{eqn:velocityeqn}) in the more convenient form
\begin{align}
\frac{u^{\prime}}{u}&=   \frac{g_{tt}  (2 c_{s}^{2} (1 +  u^{2} g_{rr}) \Sigma - 
    u^{2} g_{rr}^{\prime}) - (1 + u^{2} g_{rr})
   g_{tt}^{\prime}}{2 g_{rr} g_{tt}(u^{2}-c_{s}^{2}(g^{rr}u^{2}))},\\
 \frac{\rho_{0}^{\prime}}{\rho_{0}} &= -\frac{u^{2} g_{tt} (-2 g_{rr} \Sigma + g_{rr}^{\prime}) + (1 + 
     u^{2}g_{rr}) g_{tt}^{\prime}}{2 g_{rr} g_{tt}(u^{2}-c_{s}^{2}(g^{rr}u^{2}))}.
\end{align}
Imposing regular condition in both equations, implies that both numerators must vanish simultaneously at some critical point $r_{c}$, resulting in
\begin{align}
    u_{c}^{2} &= -\frac{g_{tt}^{\prime}}{
 g_{tt}g_{rr}^{\prime}+ g_{rr}(-2 g_{tt} \Sigma + g_{tt}^{\prime})},\label{eqn:critvelrad}\\
 c_{s,c}^{2} &= \frac{g_{rr} g_{tt}^{\prime}}{2 g_{rr}g_{tt}\Sigma - g_{tt} g_{rr}^{\prime}}.\label{eqn:critsoundvel}
\end{align}
Causality constraint $c_{s}^{2}<1$ in the flow sets a special point ($r_{c}$) by which the flow must pass. This physical requirement leads to the relation
\begin{equation}
    u_{c}^{2} = -\frac{c_{s,c}^{2}\; g_{rr}}{-1 + c_{s,c}^{2}},\label{eqn:criticalvel}
\end{equation}
between the radial velocity and the sound speed. At large radius, the flow is in the subsonic regime $u^{2}<c_{s}^{2}$. So far these results are general in the sense that can be applied to any spherically symmetric black hole solution. At this point we must specify necessarily the metric functions to obtain the exact forms for the critical velocity and sound speed Eqs.~(\ref{eqn:critvelrad}) and (\ref{eqn:critsoundvel}), respectively. Thus, considering the non-Abelian Reissner Nordstr\"{o}m BH solution Eq.~(\ref{eqn:metricsol}), leads to the critical values
\begin{equation}
    u_{c}^{2}= -\frac{Q_{\rm NA}^2 - M r}{2 r^{2}},\;  
  c_{s,c}^{2}=\frac{-Q_{\rm NA}^2 + M r}{Q_{NA}^{2} + r (-3 M + 2 r)}.\label{eqn:criticalvalues}
\end{equation}
%
%At this point in the discussion it is suggestive to ask which region of the parameter space leads to the fulfillment of the causality condition in the flow.

From the transonic condition that $u_{c}^{2}=c_{s,c}^{2}$ at the critical radius, and considering Eq.~(\ref{eqn:criticalvel}), one obtains unequivocally the critical radius 
\begin{equation}
    r_{c}=\frac{M + 3 c_{s,c}^{2} M \pm \sqrt{(M + 3 c_{s,c}^{2} M)^{2} - 
  8 c_{s,c}^{2} (1 + c_{s,c}^{2}) Q_{\rm NA}^2}}{4c_{s,c}^{2}},\label{eqn:criticalradius}
\end{equation}
where the non-trivial contribution of the non-Abelian charge to the Schwarzschild solution, $r_{c,\rm Sch}=M(1 + 3 c_{s,c}^{2})/2 c_{s,c}^{2}$ and $u_{c,\rm Sch}^{2}=M/2r_{c}$, is clearly manifested. From Eq.~(\ref{eqn:criticalradius}) we can see that there exist, mathematically speaking, two distinct critical points but the positive branch corresponds only to the physical solution which resides outside the event horizon, and therefore, it is the one to which we shall pay our attention. Of course, the negative branch is in a region observationally inaccessible since it is delimited by the event horizon. Existence of the critical radius demands
\begin{equation}
\frac{Q_{\rm NA}^{2}}{M^{2}} < \frac{1 + 6 c_{s,c}^{2} + 9 c_{s,c}^{4}}{8 c_{s,c}^{2} + 8 c_{s,c}^{4}},\label{eqn:chargecond}
\end{equation}
which in turn puts constraints on the coupling parameter for a given non-Abelian charge. Nevertheless, the resulting expressions are very lengthy (and not illuminating) to be reported here. We shall illustrate below this aspect numerically.

In order to understand the striking structure of the critical radius we must necessarily specify
the non-Abelian charge and the nature of the fluid that characterizes the sound speed. For simplicity in the former analysis, we assume an isothermal fluid\footnote{Another possibility is to take a polytropic fluid but it introduces an extra parameter that overclouds the present analysis about the structure of the critical radius. Nevertheless, it will be successfully addressed in the next part for the sake of completeness when looking for the critical mass accretion rate.} so that the sound speed equals its equation of state. 
This simple choice will give us, however, a profound insight about the behavior of the critical radius in terms of the model parameters. So the present case suffices to prove the rich structure of the critical radius due to the non-Abelian charge. It should be noticed, on the other hand, that for the (physical) critical radius there are two solutions due to the existence of the non-Abelian charges  $Q_{\rm NA, I,II}$.

%In what follows, we work with dimensionless variables by normalizing all physical quantities, as appropriate, by the black hole mass $M$. For instance: $x_{c}=r_{c}/M$ and $q_{\rm NA}=Q_{\rm NA}/M$. 

%Let's first explore the effect of changing the sound speed on the critical radius for some specific values of the coupling constant that cover mostly the (physical) BH solutions of interest. This is shown shown in Fig.~\ref{fig:critradius1}. The behavior of the critical radius depends however on the non-Abelian charge chosen. For instance, for $q_{\rm NA,I}$ (left panel) and a given coupling parameters $\chi = -0.0901$,  $\chi =0$, $\chi=20$ ($q_{\rm NA,I}=0.4112$) and $\chi=11.022$ ($q_{\rm NA, I}=1.01$), each of these correspond respectively to the extremal case, Schwarzschild solution,  charged solution and a naked singularity. Taking now $q_{\rm NA, II}$ (right panel) for the same values of the coupling parameter, leads to a different physical situation that, effectively, changes the behavior of the critical radius. Thus, the solutions $\chi=-0.0901$, $\chi=0$, $\chi=20$ and $\chi=11.0222$ correspond, respectively, to a naked singularity ($q_{\rm NA, II}=1.11352$), a extremal case and charge BH cases $q_{\rm NA, II}=0.113438$ and  $q_{\rm NA, II}=0.17345$. The exact Schwarzschild solution (green curve) in the right panel is shown as asymptotic limit of large values of the coupling constant.

\begin{figure*}
\centering
\includegraphics[width=0.47\hsize,clip]{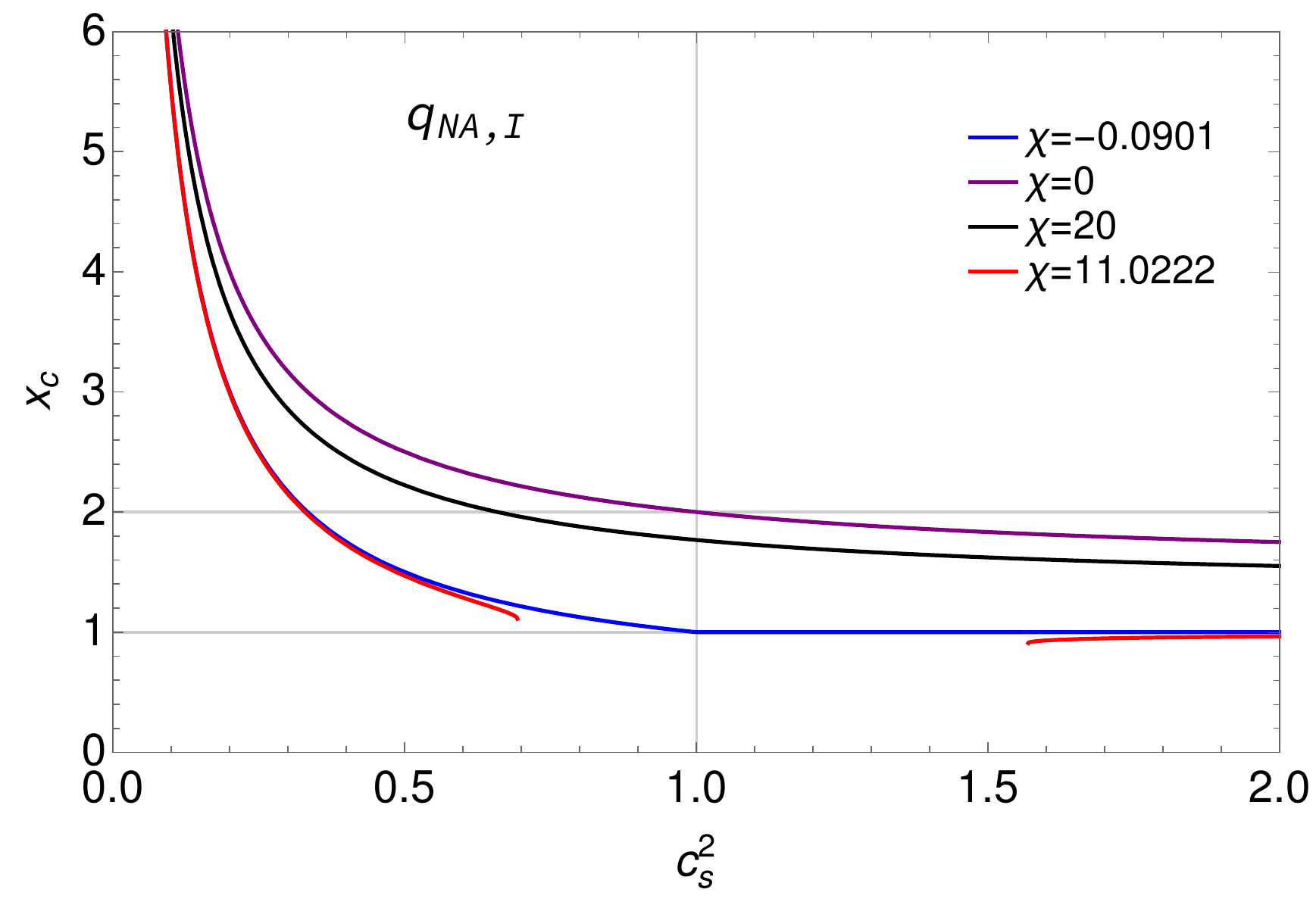}
\includegraphics[width=0.47\hsize,clip]{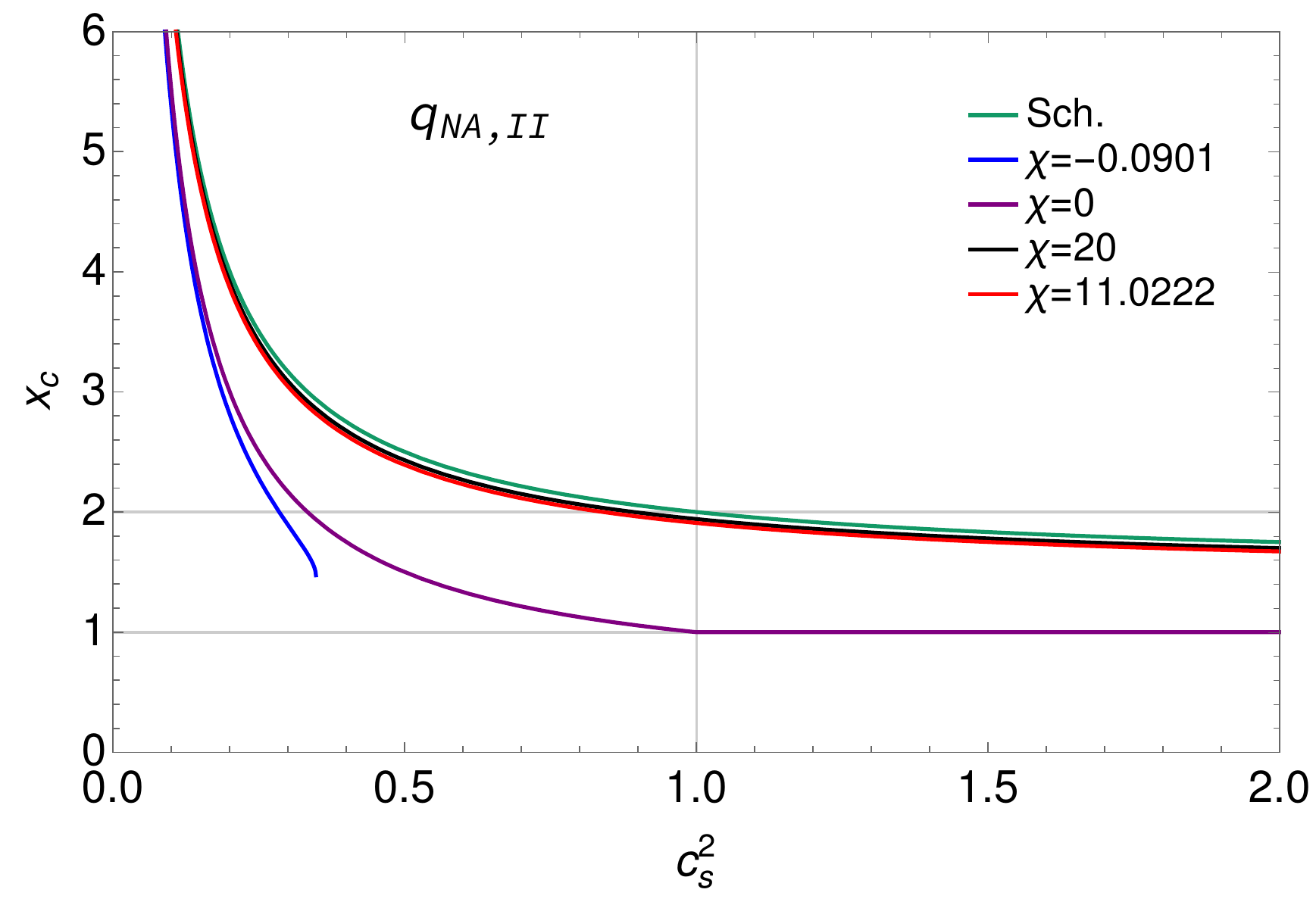}
\caption{\textit{Left panel}: critical radius $x_c \equiv r_c/M$ \eqref{eqn:criticalradius}  for the non-abelian charge $q_{\rm NA,I}$ for a given coupling parameters that correspond to the extremal case $\chi = -0.0901$, Schwarzschild solution $\chi =0$, charged solution $\chi=20$ ($q_{\rm NA,I}=0.4112$) and a naked singularity $\chi=11.0222$ ($q_{\rm NA, I}=1.01$). \textit{Right panel}: critical radius for the non-abelian charge $q_{\rm NA, II}$ for the same values of parameters as left panel. The exact Schwarzschild solution (green curve) has been included here for comparison. Notice that, on the contrary, the solutions $\chi=-0.0901$, $\chi=0$, $\chi=20$ and $\chi=11.0222$ correspond, respectively, to a naked singularity ($q_{\rm NA, II}=1.11352$), a extremal case and charge BH cases $q_{\rm NA, II}=0.113438$ and $q_{\rm NA, II}=0.17345$.} \label{fig:critradius1}
\end{figure*}

Let us first explore the effect of changing the sound speed on the critical radius for some specific values of the coupling constant that cover mostly the (physical) BH solutions of interest. The behavior of the critical radius depends, however, on the non-Abelian charge chosen. This is shown in Fig.~\ref{fig:critradius1}, where the left and right panels correspond, respectively, to the non-Abelian charges $q_{\rm NA,I}$ and $q_{\rm NA,II}$. In particular, for $q_{\rm NA,I}$ and $\chi=0$, the critical radius matches exactly the Schwarzschild solution, whereas for $q_{\rm NA,II}$ the convergence is possible provided that $\chi\to \infty$. This is the reason why we have displayed the exact Schwarzschild solution (green curve) in right panel as an asymptotic limit of large coupling constant values. Accordingly, for the larger case shown, $\chi=20$, the solution is barely distinguishable from the uncharged case but physically different since the former is endowed with a charge $q_{\rm NA,I}=0.113438$. Hence, despite taking the same values of $\chi$ for both non-Abelian charges, the physical implications do not hold for the critical radius as it also happens for the event horizon and ISCO. 

To better illustrate  the role of $\chi$ in the critical radius, we take the same values of the coupling constant, as described below, for both charges. For instance, taking $q_{\rm NA,I}$ (see left panel) and a given coupling parameter lead to the following physical situations: the extremal case $\chi = -0.0901$, the Schwarzschild solution $\chi =0$, the charged solution $\chi=20$ ($q_{\rm NA,I}=0.4112$) and a naked singularity $\chi=11.0222$, that affect in a way different the behavior of the critical radius. Now, taking the same values for $\chi$ as before but for $q_{\rm NA,II}$, the previous physical meaning is lost. Naked singularities can take place, for instance, for $\chi=-0.0901$ which in the $q_{\rm NA,I}$ case occurs for $\chi=11.0222$ . Notice that for the former case accretion is possible only for (subsonic) sound speeds $c_{s,c}^{2}<0.3478$ while for the latter it occurs out of the range $0.6930<c_{s,c}^{2}<1.5677$. This explains the discontinuity of those curves. Hence accretion may be possible even though the event horizon vanishes. An interesting discussion about how to distinguish a BH from a naked singularity spacetime by using the image of  thin accretion disks is addressed in \cite{Shaikh:2019hbm}.

As a general trend, for subsonic sound speeds the critical radius can be located far away from the event horizon,
and for supersonic sound speeds it can be accommodated, on the contrary, between the event horizon and the apparent horizon. For $c_{s}^2=1$ all critical points coincide with their respective position of the event horizon as can be easily verified. Another interesting feature is that for the Schwarzschild case and subsonic speeds $c_{s,c}^{2}<1$
the critical radius is outside the event horizon while for $c_{s,c}^{2}>1$ the critical radius is located behind it. See purple and green curves in the left and right panels, respectively. This general discussion is in agreement with former studies about the accretion of perfect fluids in the RN metric \cite{Babichev:2008jb}. 

\begin{figure*}
\centering
\includegraphics[width=0.47\hsize,clip]{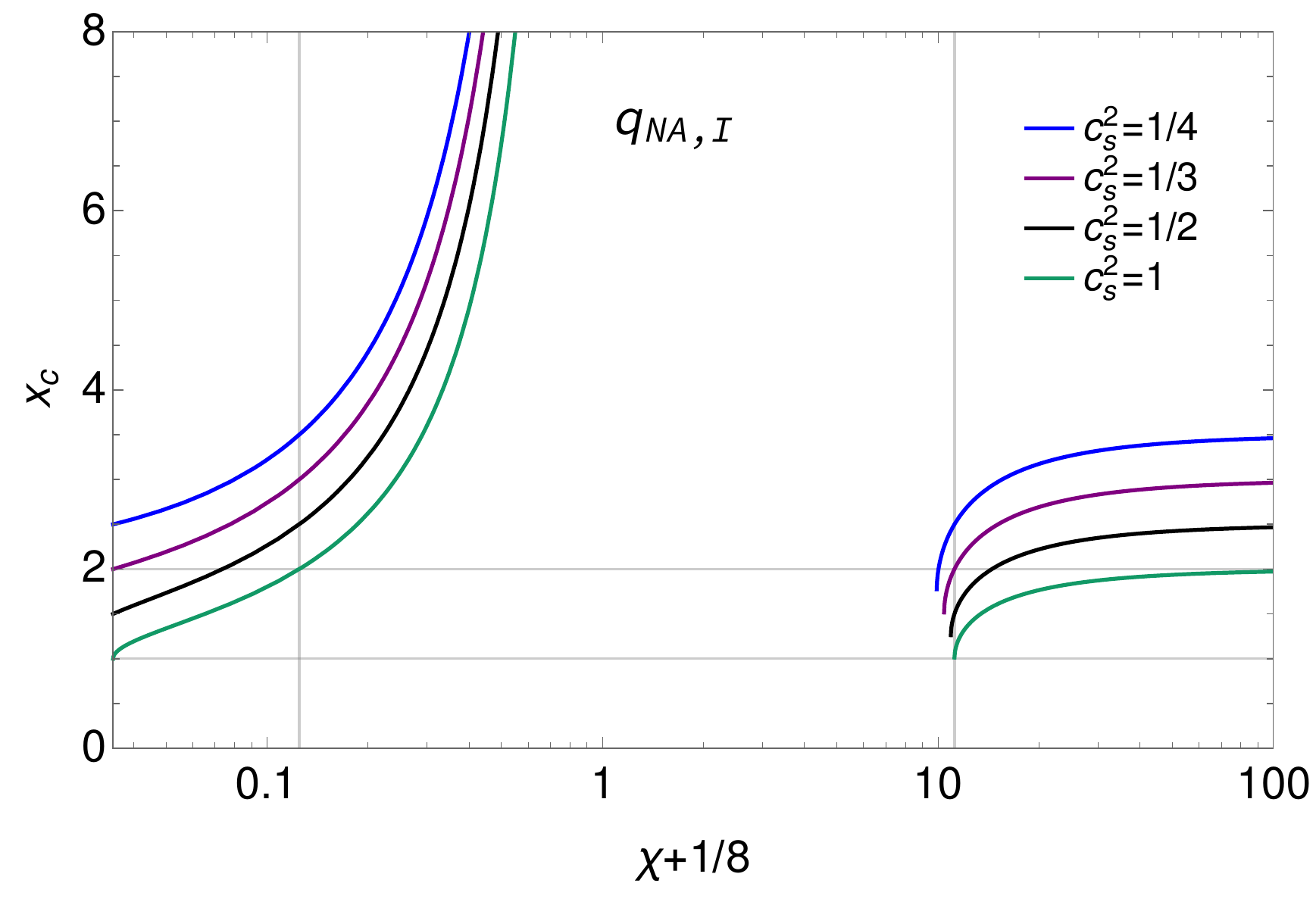}
\includegraphics[width=0.47\hsize,clip]{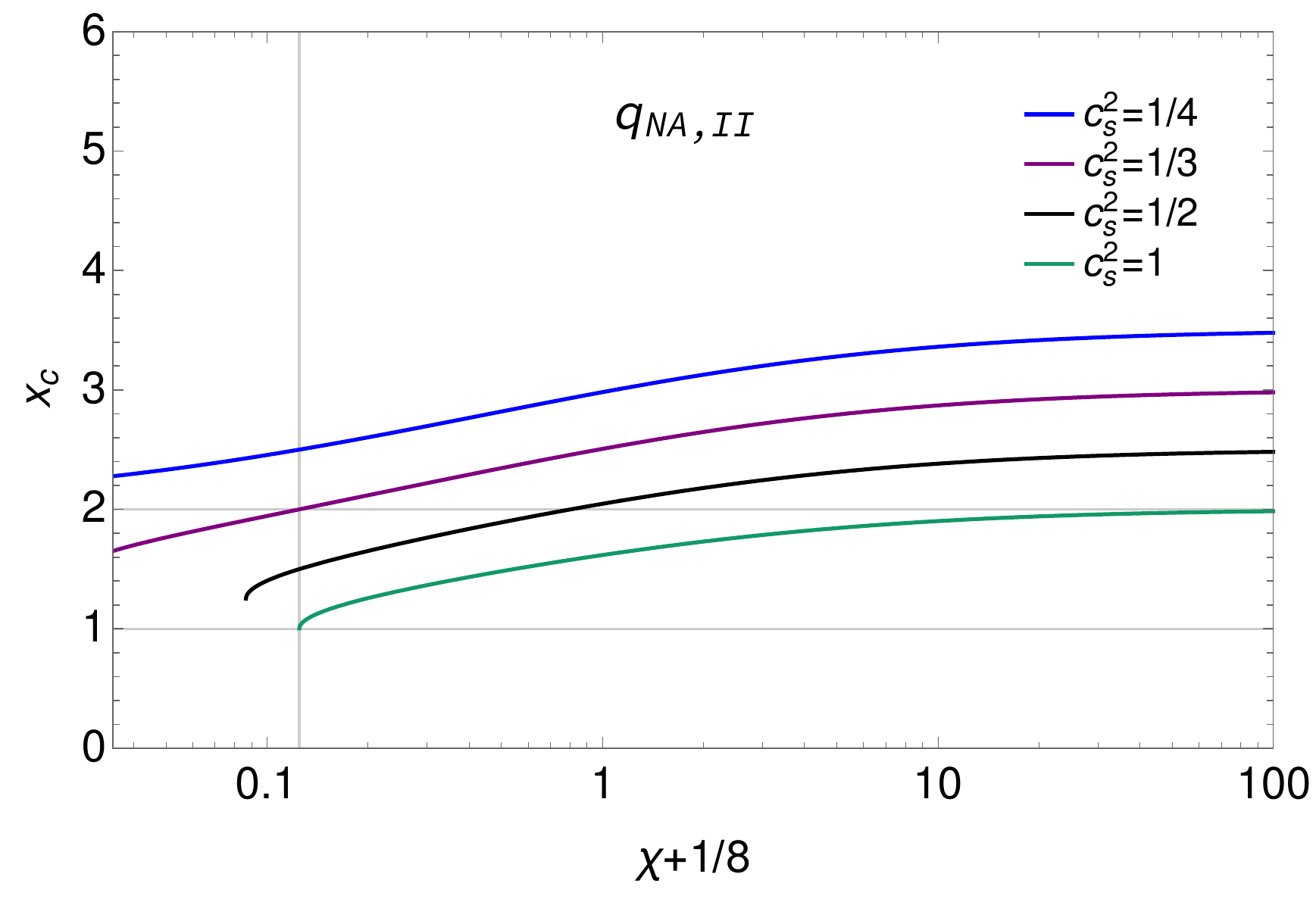}
\caption{\textit{Left panel}: critical radius for the non-abelian charge $q_{\rm NA,I}$ for certain critical sound speeds, as described in the legend,  as a function of the coupling parameter. \textit{Right panel}: critical radius for the non-abelian charge $q_{\rm NA,II}$ for the same parameters as left panel.} \label{fig:critradius2}
\end{figure*}

In contrast, in Fig.~\ref{fig:critradius2}, the coupling parameter is fixed while the sound speed varies. Notice that for the case $q_{\rm NA,I}$ the critical radius exhibits a discontinuity as it was also perceived for the event horizon structure.  It should be noticed that as $c_{s,c}^{2}$ decreases (see for instance $c_{s,c}^{2}$=1/4) the critical radius can be in a region where naked singularity takes place. As to the case $q_{\rm NA,II}$, the critical radius is also a monotonically increasing function of $\chi$, but it is well-behaved until the condition Eq.~(\ref{eqn:chargecond}) is broken for small values of $\chi$ and given squared sound speeds. This  discussion provides, in complement to the one made around Fig.~\ref{fig:critradius1}, a full picture of the general conditions under which the critical radius exists in terms of the coupling constant through Eq.~(\ref{eqn:chargecond}).

Now, we are in a more grounded position to investigate the effect of changing the coupling parameter on the accretion properties, concretely on the radial velocity and on the mass density around this class of black hole. This will be carried out first numerically for an isothermal fluid, and later with the aid of some analytical treatments for polytropic fluids, to allow a more robust and complete exploration of the involved parameters. These two cases will be then treated separately in the next sections.

%

%%%%%%%%%%%%%%%%%%%%%%%%
\subsection{Isothermal test fluid}

Before describing accretion flows for a more general fluid, let us first consider a simplistic but useful isothermal test fluid. It will provide us some physical insights on how the coupling constant influences the behavior of the infall radial velocity and mass density. This inquiry is complementary to the discussion on the critical point realized previously. 

We focus for comparison reasons in a range of the coupling constant that resembles the RN BH and the Schwarzschild solution and leave the (allowed) range that provides $q_{\rm NA,I}^{2}<0$ out of this analysis. It implies that the corresponding parameter space of $\chi$ for a given non-Abelian charge, provides the same physics whereby we focus on $q_{\rm NA,I}$. To illustrate this point we show only a limit case for $q_{\rm NA,II}$ to see the convergence. The full range of $\chi$ will be, however, considered in the calculation of the critical accretion rate for a polytropic fluid. The above  are also advantageous for numerical facilities since we have many variables involved. 

Accordingly, the equation of state is of the form $P= \kappa\rho$, with $\kappa$ being a constant, from which the simple relation for the sound speed $c_{s,c}^{2}=\kappa$ is derived.

\begin{figure*}
\centering
\includegraphics[width=0.47\hsize,clip]{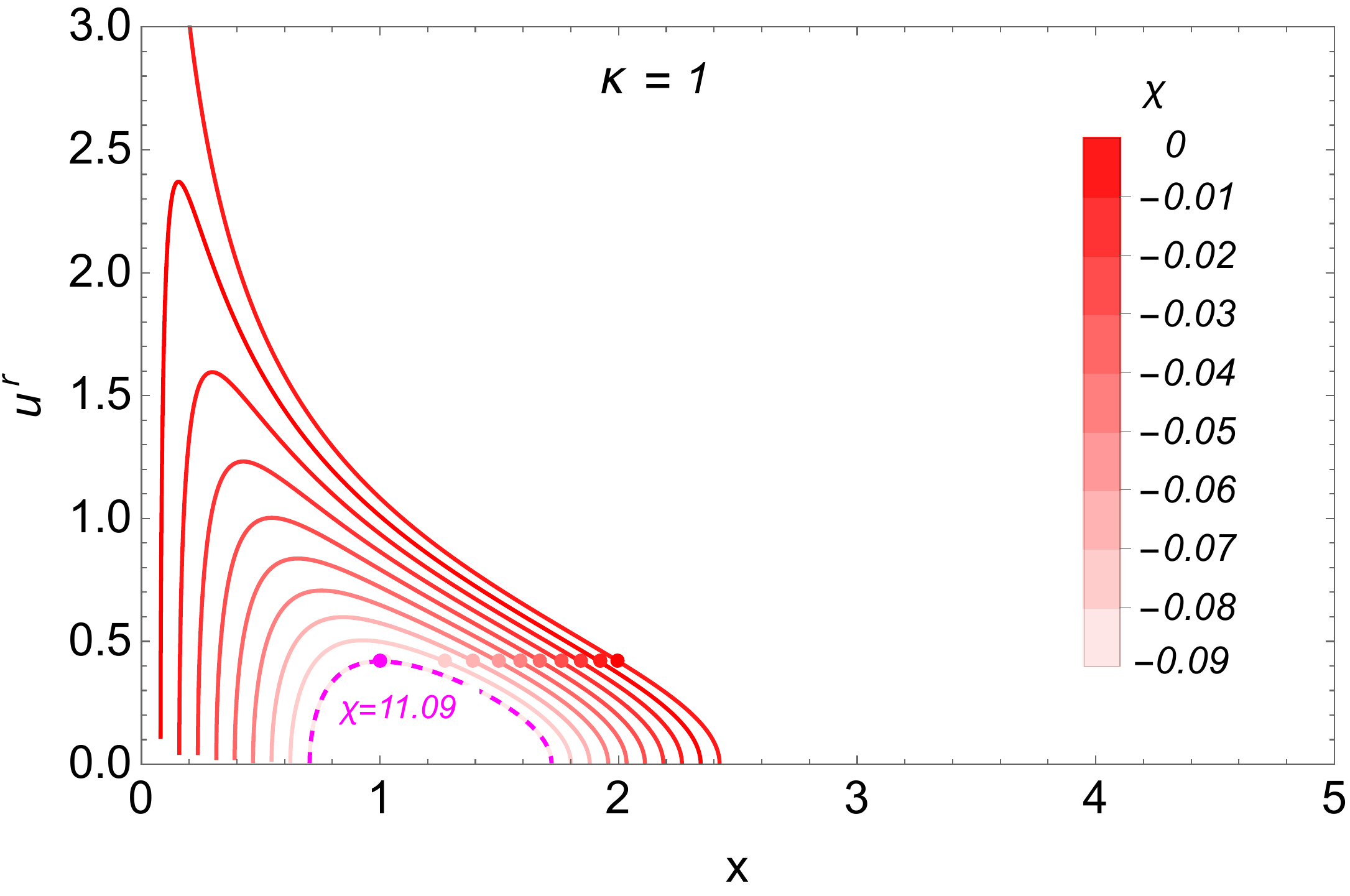}
\includegraphics[width=0.47\hsize,clip]{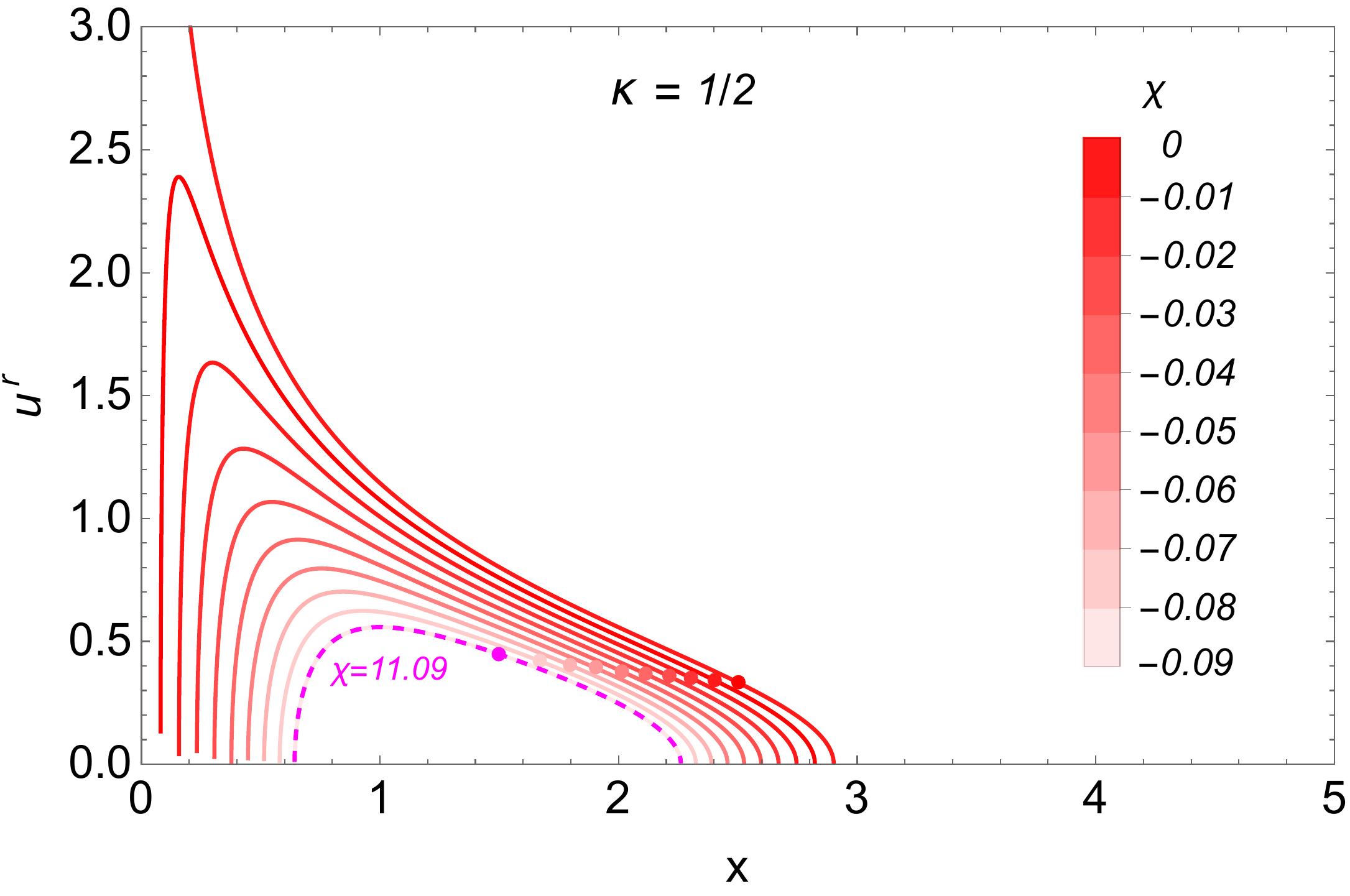}
\includegraphics[width=0.47\hsize,clip]{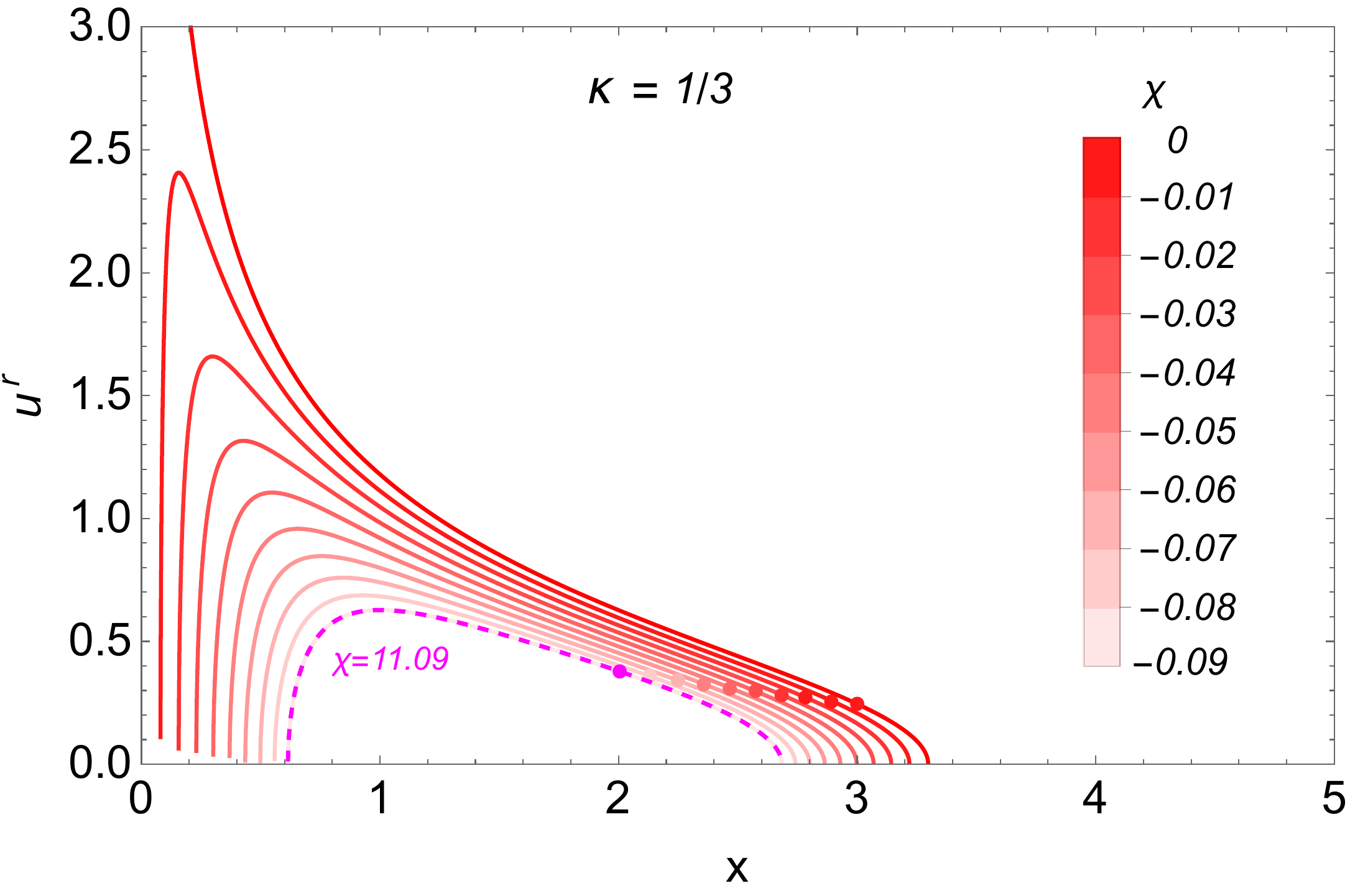}
\includegraphics[width=0.47\hsize,clip]{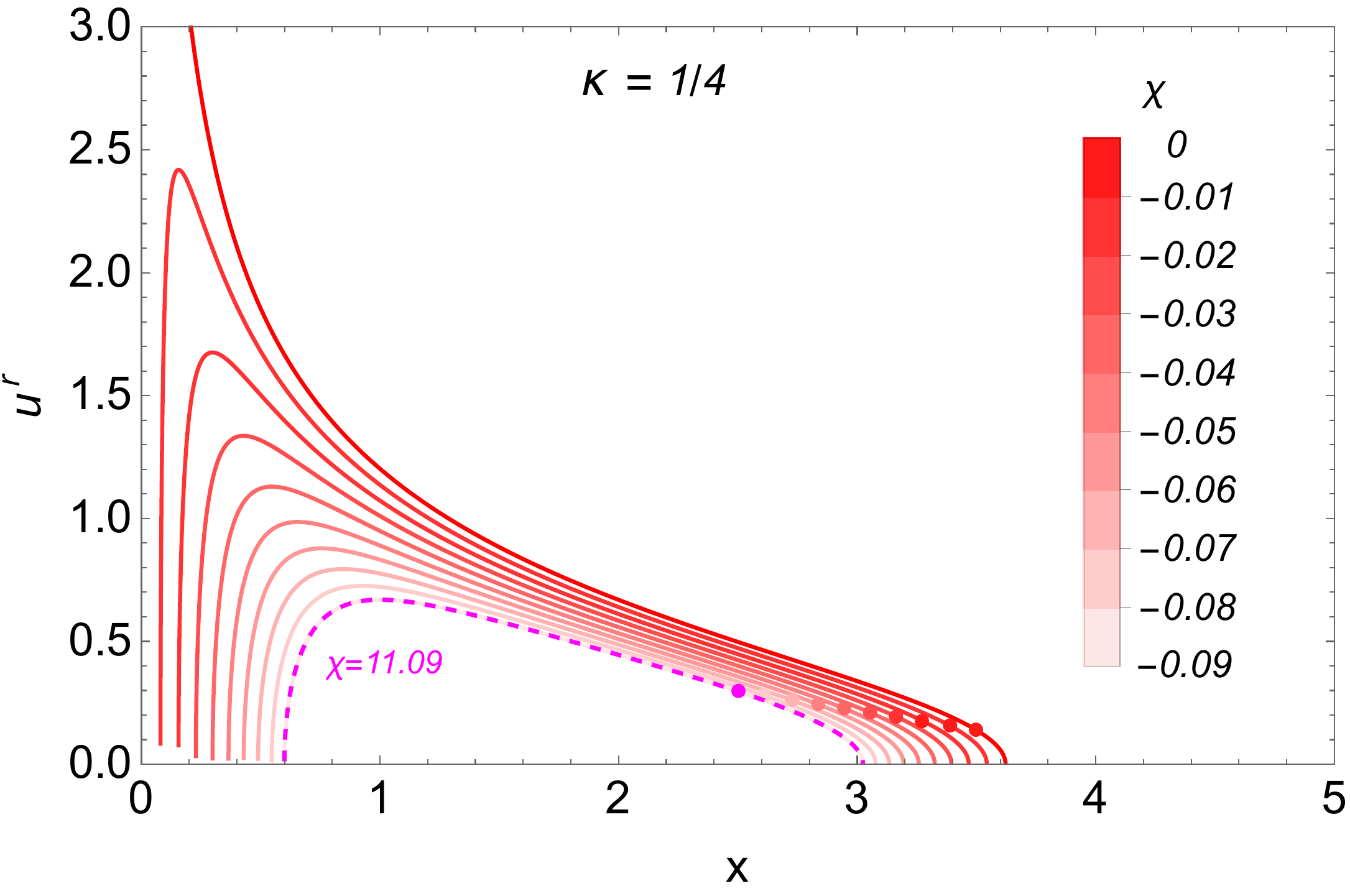}
\caption{Infall radial velocity for some specific values of $\kappa$ describing a stiff fluid ($\kappa=1$), ultra-relativistic fluid ($\kappa=1/2$), radiation fluid ($\kappa=1/3$) and  sub-relativistic fluid ($\kappa=1/4$) while the coupling parameter $\chi$ spans the allowed range as depicted by the bar legend. The dotted magenta curve that delimits all the possible transonic solutions from below corresponds to the extremal case $\chi=11.09$ for $q_{\rm NA,I}$ (or $\chi=-0.09$ for $q_{\rm NA,II}$), while $\chi\rightarrow0$ resembles the Schwarzschild solution from above. It is worth noting that the event horizon is always located in the interval $x \in[1,2]$, with the lower and upper limits corresponding to the extremal Reissner-Nordstr\"{o}m and Schwarzschild solutions, respectively. In the specific case of stiff fluid $\kappa=1$ (top left panel), the position of the critical points coincide with the location of the event horizons. However, for smaller and larger values of $\kappa$ and $\chi$, all critical points shift toward larger values of the corresponding event horizons. Note also that the divergences observed are not physical, but they are related, instead, to the coordinate singularity problem associated with the Schwarzschild coordinates. This singularity can be removed by moving to non-singular coordinates at the event horizon as the Eddington-Finkelstein coordinates.} \label{fig:radialvel}
\end{figure*}

Let us start our analysis by considering a stiff fluid $\kappa=1$. As we already discussed, for this case ($c_{s,c}^{2}=1$) critical points coincide with the event horizons no matter the value of the coupling constant. The latter spans the allowed region of the parameter space $\chi\in(-0.09,0)$ for the charge $q_{\rm NA,I}$, as can be seen in the bar legend of Figure.~\ref{fig:radialvel}. We are not considering the other possible range of values $\chi\in(11.0902,\infty)$ because the same physical properties are replicated in the already shown range. For the other cases describing an ultra-relativistic fluid $\kappa=1/2$, a radiation fluid $\kappa=1/3$ and a sub-relativistic fluid $\kappa=1/4$; all critical points move out of the BH as $\kappa$ decreases according to Eq.~(\ref{eqn:criticalradius}). As a general trend, all transonic solutions are delimited from below to the extremal RN solution $\chi=-0.0901$ and from above to the Schwarzschild case. This latter case is attained whenever the coupling constant increases until it reaches the maximum value showed here ($\chi=0$). We have also included the solution $\chi=11.09$ (dashed magenta curve) for the charge $q_{\rm NA,II}$ which matches the RN solution of the $q_{\rm NA,I}$ case. What is of physical interest here is that all infall radial velocities pass through the corresponding critical points marked as points on the curves and computed from Eq.~(\ref{eqn:criticalradius}), guaranteeing thus the transonic flow of all solutions, otherwise a stellar wind is generated. This selects a unique solution with constant inward mass flux that connects the subsonic with the supersonic regimes, as expected in Bondi-type accretion \cite{Bondi:1952ni,1972Ap&SS..15..153M}. Far beyond the BH influence, that is, in the non-relativistic regime, the radial velocities of the particles are too low such that the inflow rate is decreased compared with particles with high-speed velocities. 

\begin{figure*}
\centering
\includegraphics[width=0.47\hsize,clip]{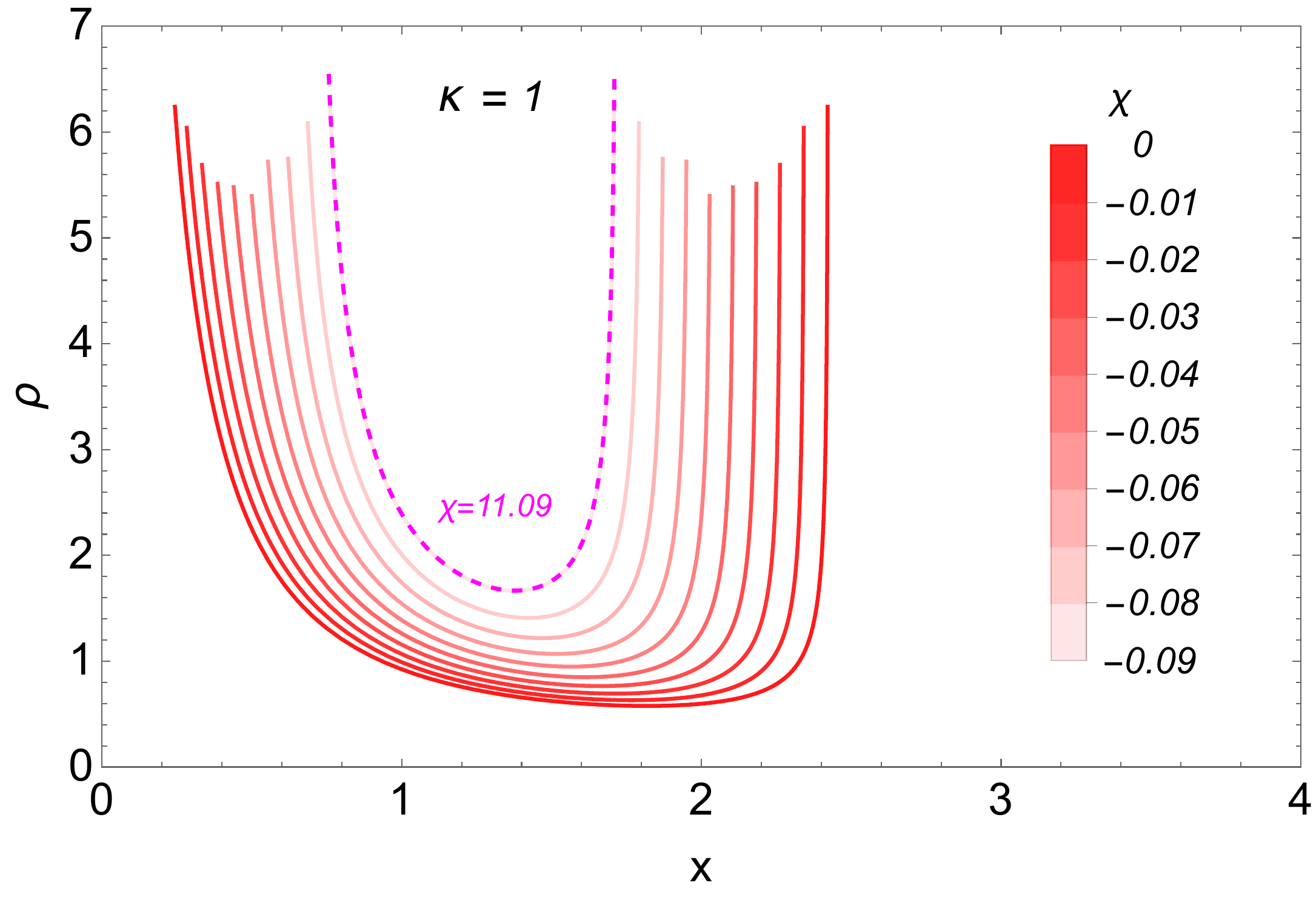}
\includegraphics[width=0.47\hsize,clip]{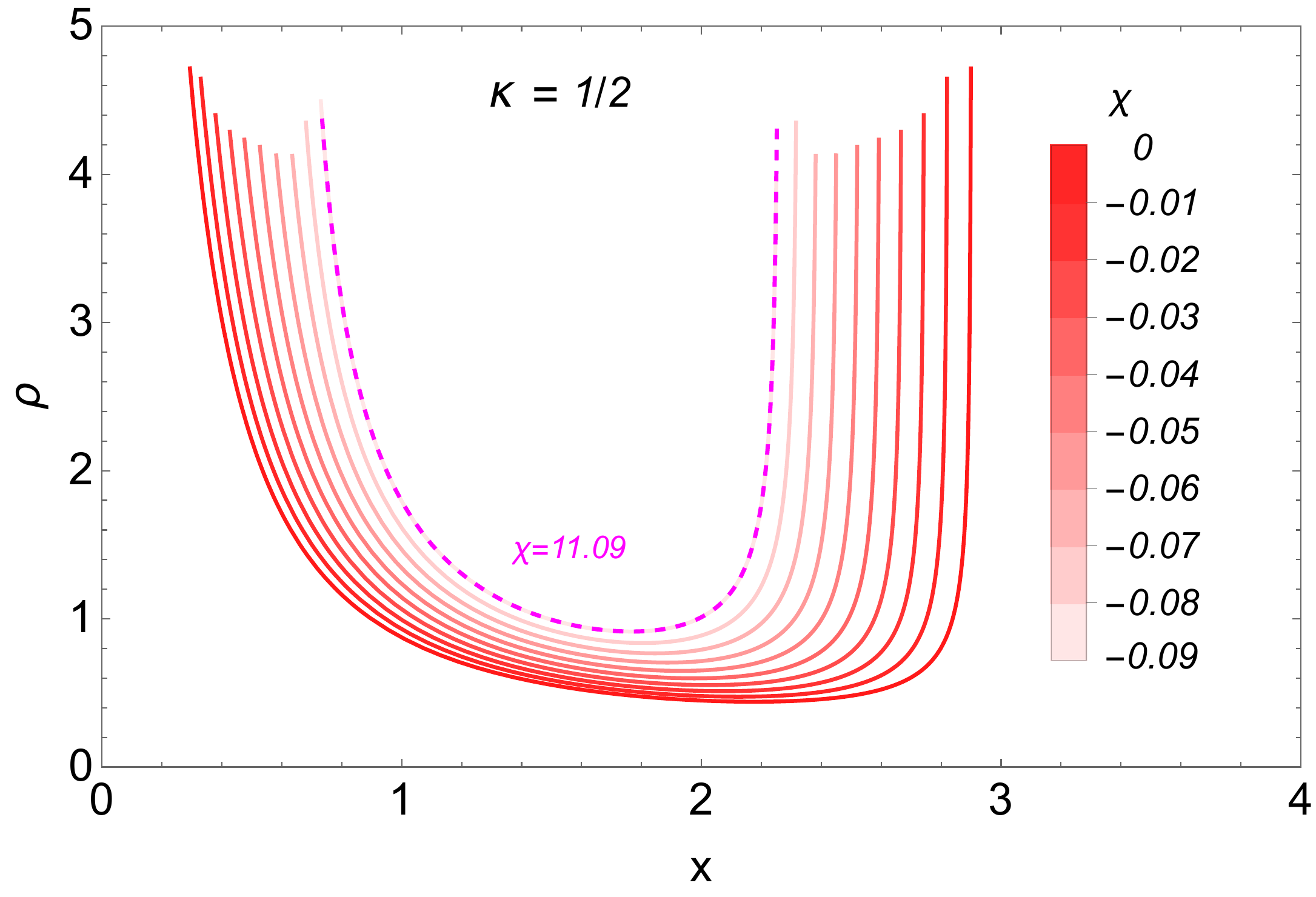}
\includegraphics[width=0.47\hsize,clip]{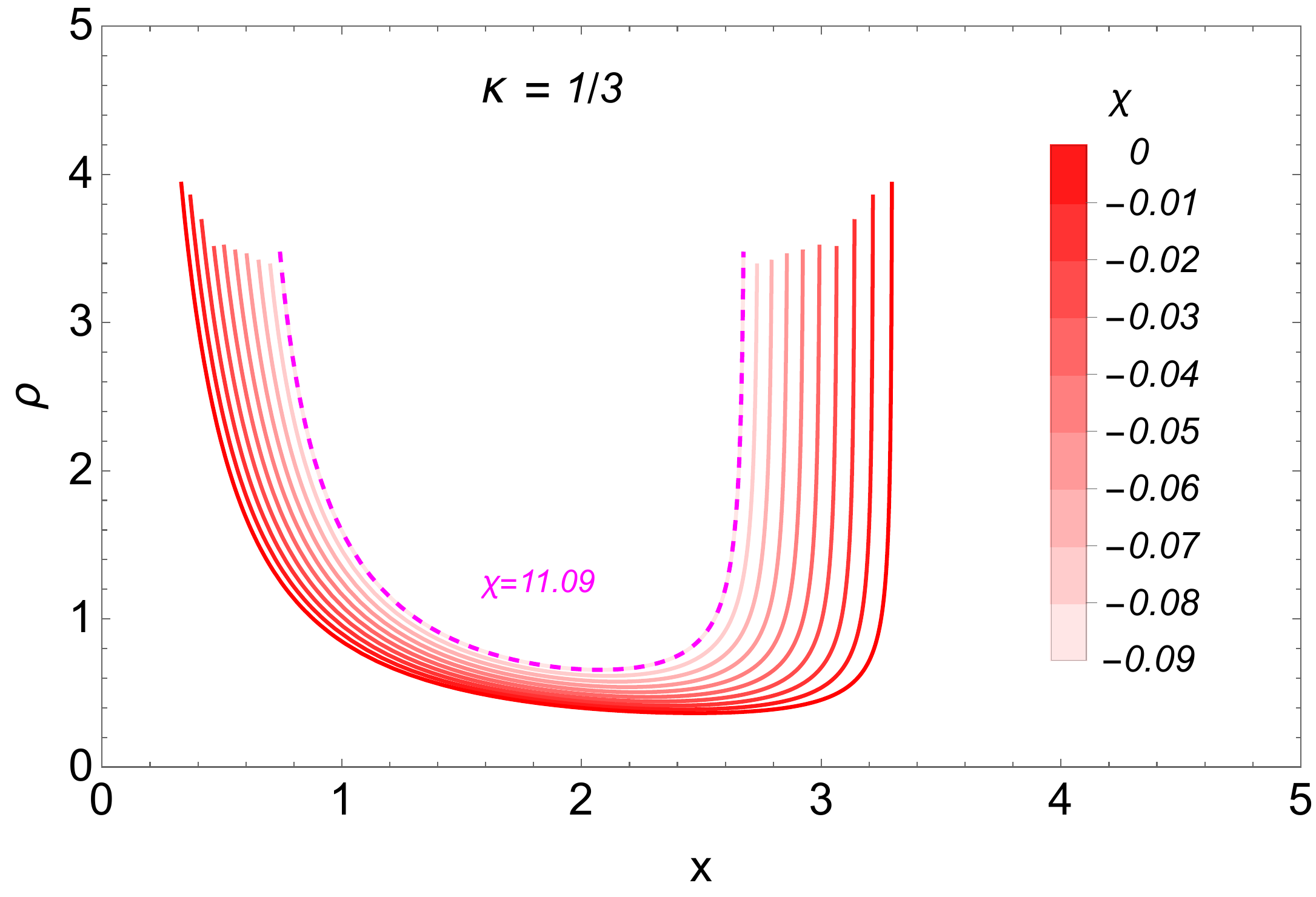}
\includegraphics[width=0.47\hsize,clip]{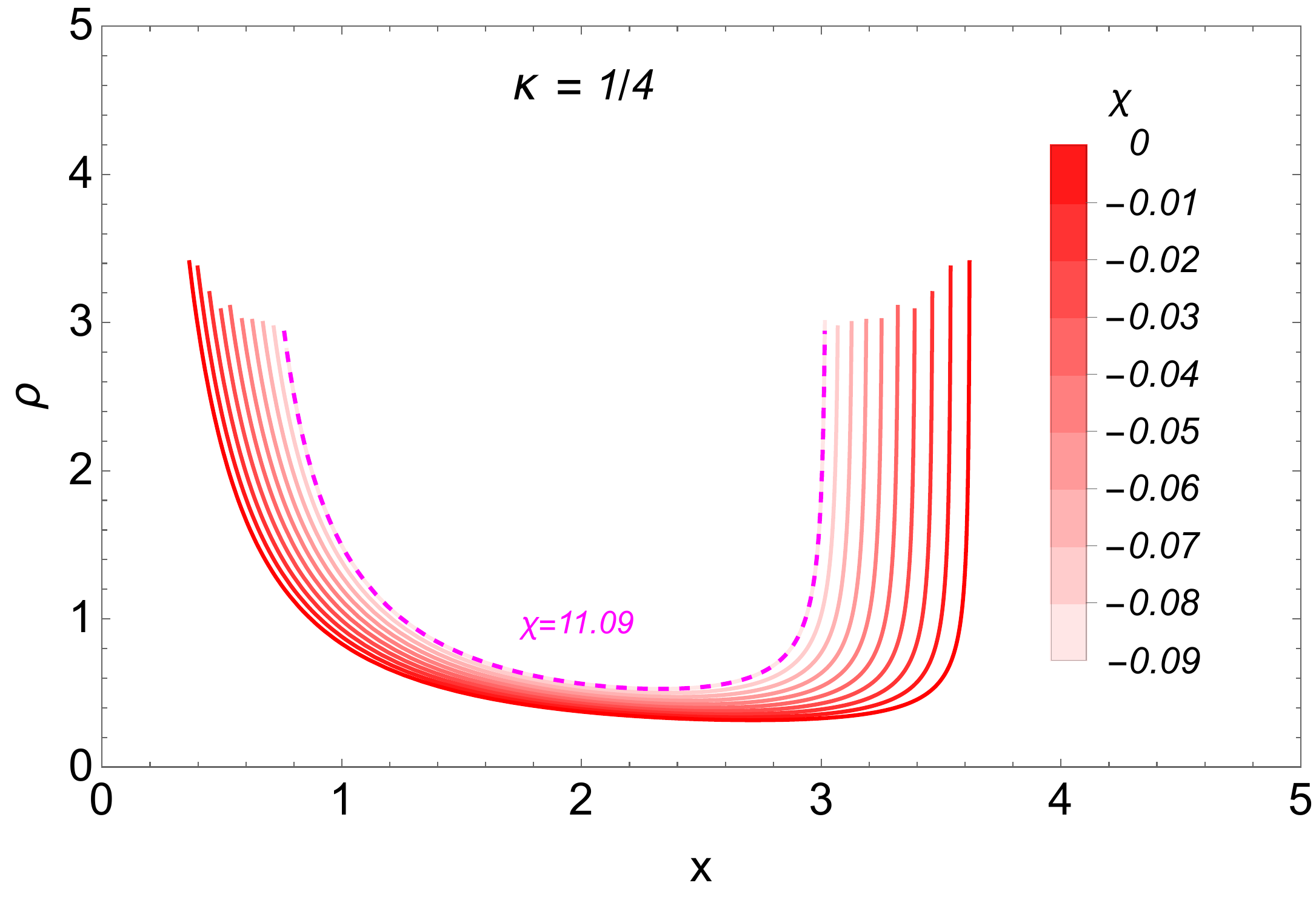}
\caption{Mass density distribution for some specific values of $\kappa$ describing a stiff fluid ($\kappa=1$), ultra-relativistic fluid ($\kappa=1/2$), radiation fluid ($\kappa=1/3$) and sub-relativistic fluid ($\kappa=1/4$) while the coupling parameter $\chi$ spans the allowed range as depicted by the bar legend. The dotted magenta curve that delimits all the possible transonic solutions from below corresponds to the extremal case $\chi=11.09$ for $q_{\rm NA,I}$ (or $\chi=-0.09$ for $q_{\rm NA,II}$), while $\chi\rightarrow0$ (or $\chi\rightarrow\infty$) resembles the Schwarzschild solution from above.} \label{fig:massden}
\end{figure*}

For the mass density distribution due to the BH gravitational potential, the RN solution now delimits all solutions from above, as seen in  Figure.~\ref{fig:massden}. As in the radial velocity case, $\chi$ covers the same range of values for each value of the constant $\kappa$ considered. We can observe that as $\kappa$ reduces, the mass density is more spread out along the radial coordinate.

So far we have obtained the expected behavior for the radial velocity and mass density within the steady-state, spherical accretion scenario for the discussed range of $\chi$-values. As in previous sections, this has been very useful to understand the role of the coupling constant on the transonic flows and how our solutions approach to the extremal RN and Schwarzschild solutions. Making a more robust description of the fluid, in next section we will perform analytical calculations of the critical accretion rate in the fully relativistic regime along with numerical computations for the entire range of values of $\chi$ and both non-Abelian charges.

%

%%%%%%%%%%%%%%%%%%%%%%%%
\subsection{Polytropic fluid}

Once the properties of the steady flows are known at the sonic point, this is, radial velocity, sound speed and critical radius, we can proceed to express such quantities in terms of the boundary conditions, with the help of the Bernoulli equation, as commonly done. The purpose of doing so is to calculate the accretion rate explicitly. It is necessary also to adopt an equation of state for the gas. So we study a non-relativistic baryonic gas with a polytropic equation
\begin{equation}
    P = K n^{\gamma},\label{sub32:eqn1}
\end{equation}
where $\gamma$ is the adiabatic index and $K$ is a constant. With this and from the first of law of thermodynamics one can get \cite{Shapiro:1983du}
\begin{equation}
    \rho = m n + \frac{K}{\gamma-1} n^{\gamma}.\label{sub32:eqn2}
\end{equation}
%

%%%%%%%%%%%%%%%%%%%%%%%%%%%%%%%%%%%%%%%%%%
\subsubsection{Relativistic regime}

Surprisingly, the fully relativistic accretion rate for the Reissner Nordstr\"{o}m solution has not been thoroughly treated in the literature, where most of existing works have focused on the non-relativistic limit. It leaves undoubtedly an incomplete comprehension of the full picture. We first derive some analytical expressions and show some numerical examples to illustrate better the role of the coupling constant on the mass accretion rate. We follow closely Ref.~\cite{Richards:2021zbr}, where Bondi accretion of steady spherical gas flow onto a Schwarzschild black hole has been studied. We extend this work to the charged case.

With the aid of the polytropic equation, it is possible to relate the sound speed with the mass density 
\begin{equation}
    c_{s}^{2}=\frac{\gamma k \rho_{0}^{\gamma-1}}{1+\gamma k\rho_{0}^{\gamma-1}/(\gamma-1)}.\label{soundspeed2}
\end{equation}
This expression can be evaluated at the critical point and in the asymptotic region to provide the useful relation
\begin{equation}
    \rho_{0,s} = \rho_{0,\infty} \left( \frac{c_{s,c}^{2}}{c_{s,\infty}^{2}}\right)^{\frac{1}{\gamma-1}} \left( \frac{\gamma -1 - c_{s,\infty}^{2}}{\gamma -1 - c_{s,c}^{2}}\right)^{\frac{1}{\gamma-1}}.\label{criticaldensity2} 
\end{equation}
This closed expression requires the knowledge of the critical sound speed that can be extracted from  the relativistic Bernoulli equation
\begin{equation}
    (1+3c_{s,c}^{2})\left(1-\frac{c_{s,c}^{2}}{\gamma-1}\right)^{2} = \left(1-\frac{c_{s,\infty}^{2}}{\gamma-1}\right)^{2}\label{Bernoullicriticaleqn}.
\end{equation}
So, once the sound speed at infinity is specified, the  sound speed and the mass density at the critical point are uniquely determined. The Bernoulli equation is actually a cubic equation for $c_{s,c}^{2}$ with one real solution for the range $1<\gamma<5/3$. There are some procedures to solve analytically this equations as, for instance, a standard root-finding schema which we implement to.

\begin{figure*}
\centering
\includegraphics[width=0.47\hsize,clip]{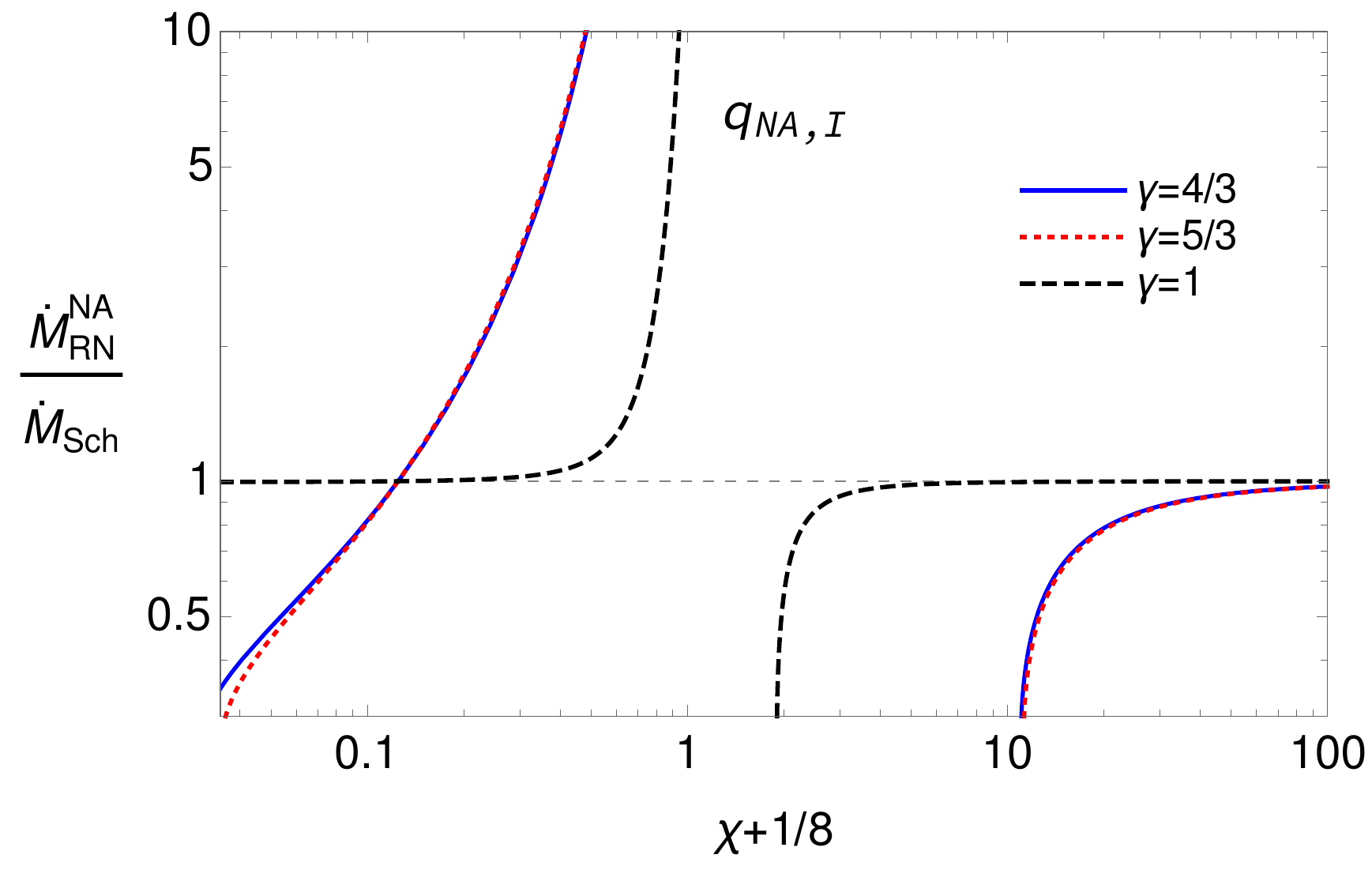}
\includegraphics[width=0.47\hsize,clip]{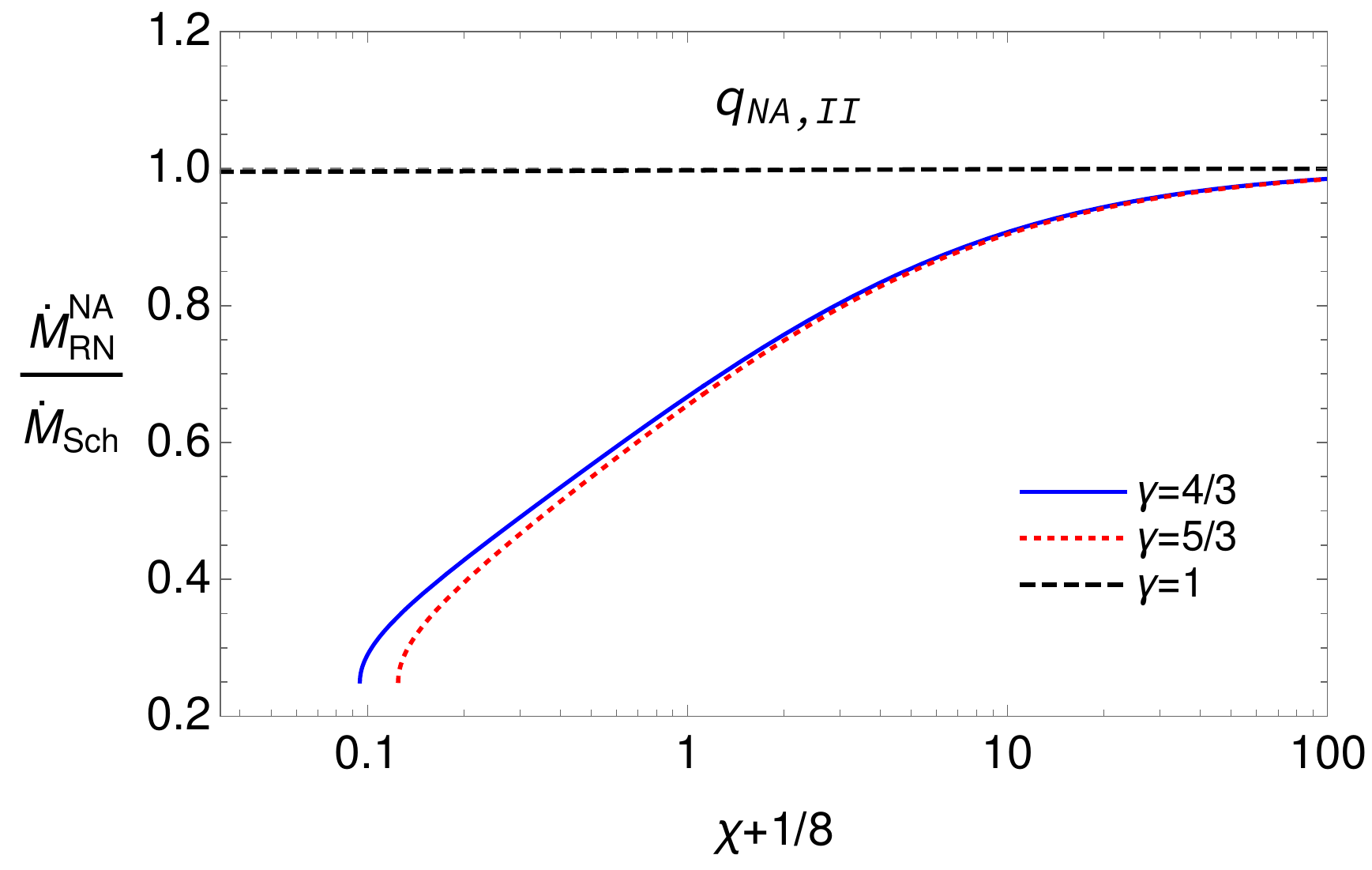}
\includegraphics[width=0.47\hsize,clip]{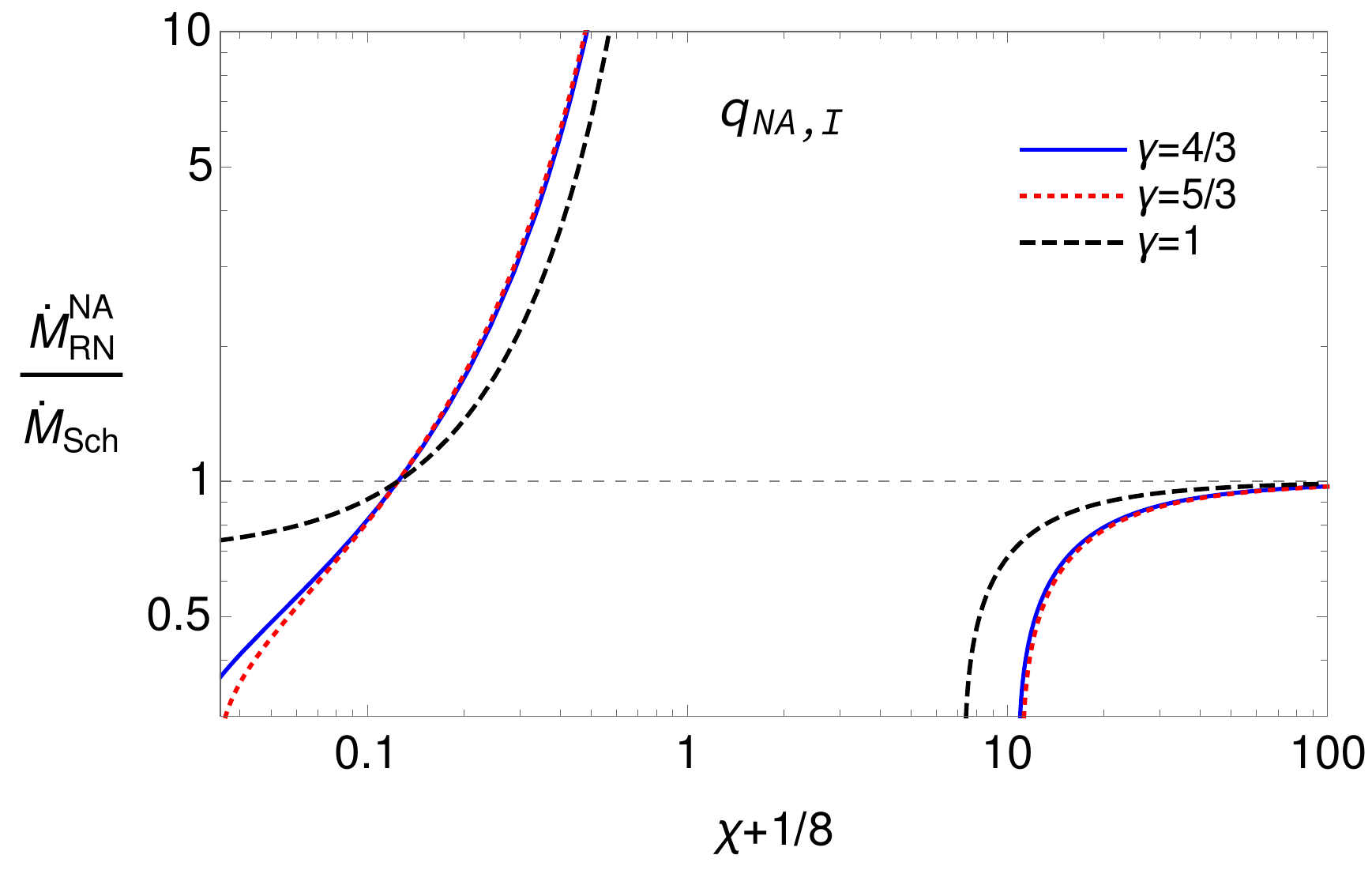}
\includegraphics[width=0.47\hsize,clip]{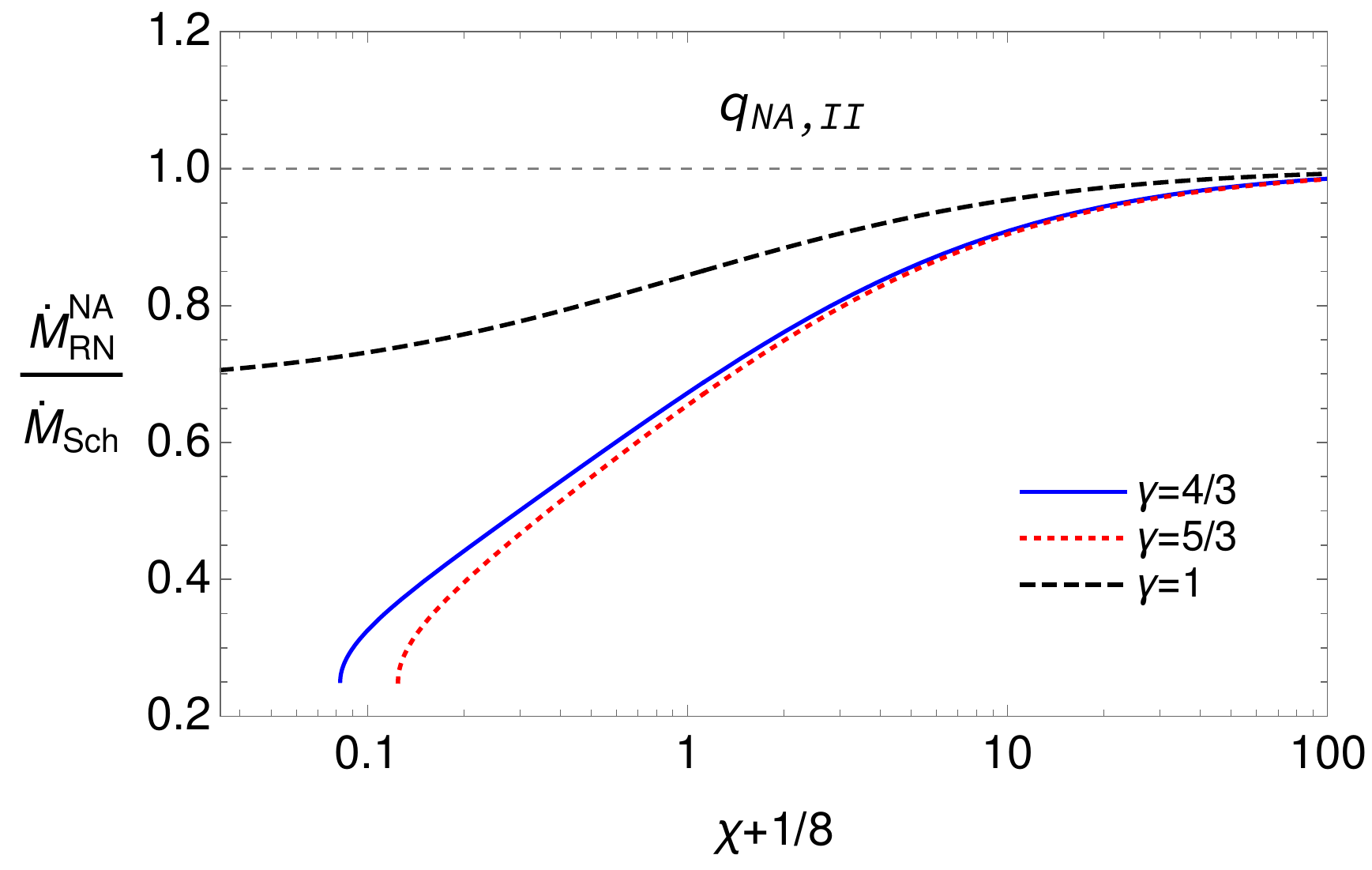}
\includegraphics[width=0.47\hsize,clip]{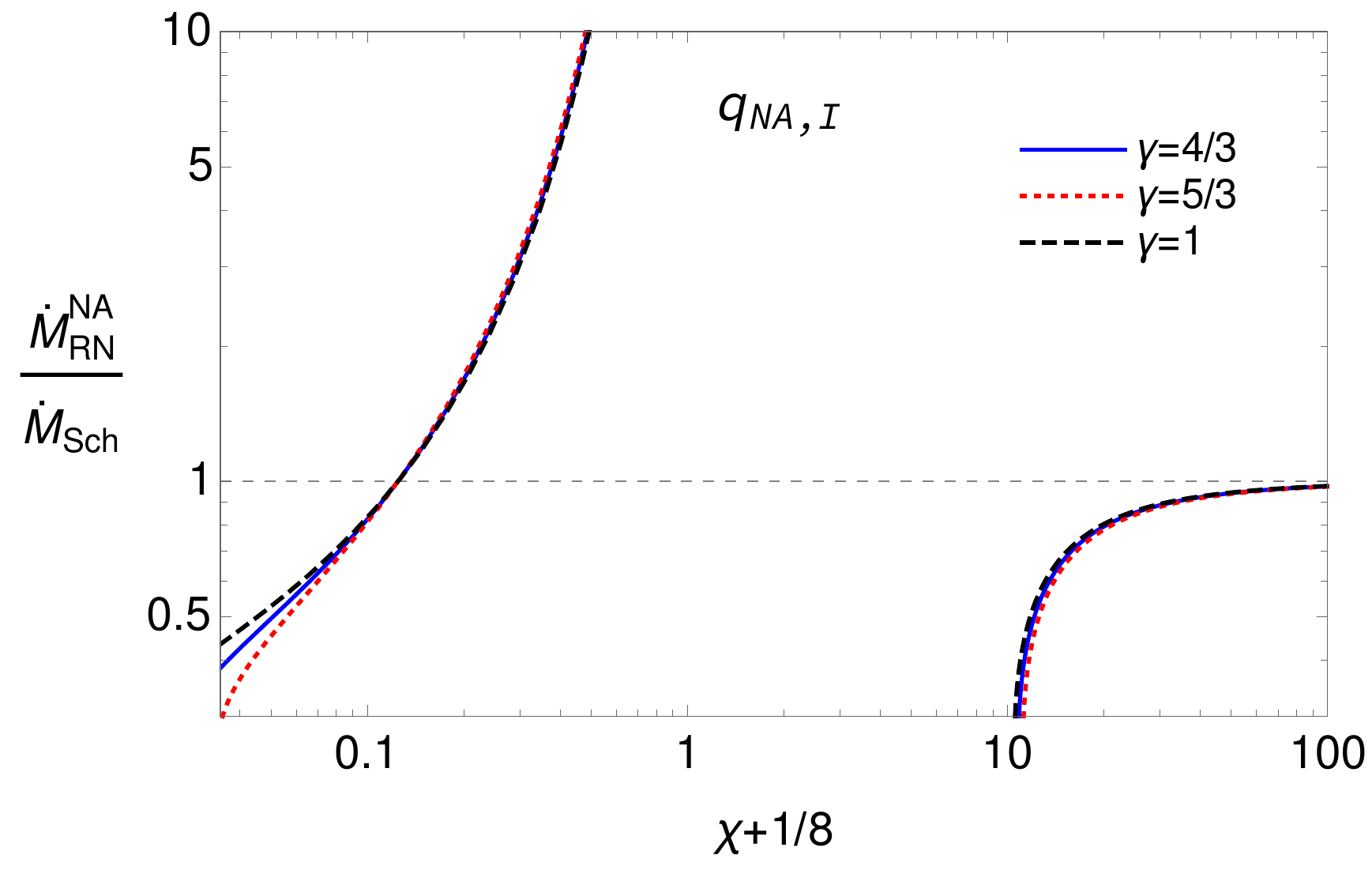}
\includegraphics[width=0.47\hsize,clip]{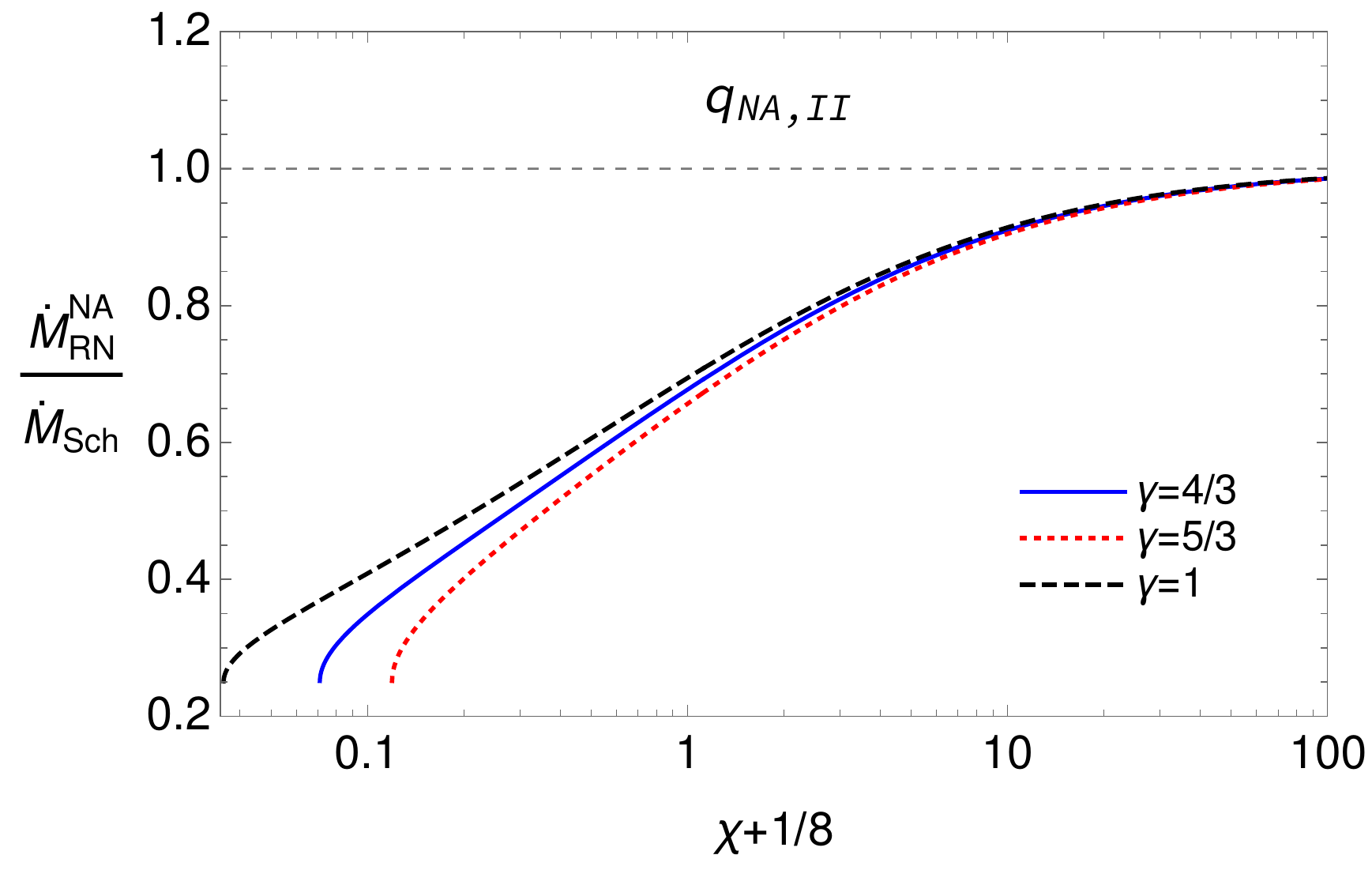}
\caption{Ratio of mass accretion rate in the non-Abelian RN BH to the accretion rate in the Schwarzschild  BH for different adiabatic indices as indicated in the legend. Left and right panels correspond to the cases $q_{\rm NA}=q_{\rm NA,I}$ and $q_{\rm NA}=q_{\rm NA,II}$, respectively. In tops panels the boundary condition approaching to the non-relativistic regime $c_{s,\infty}^{2}=0.001$ has been taken, whereas in middle and bottom panels a relativistic boundary sound speed $c_{s,\infty}^{2}=0.1$ and $c_{s,\infty}^{2}=0.5$ have been chosen, respectively, in contrasting. In all cases the condition $u_{c}>c_{s,\infty}$ is guaranteed, ensuring thus the transonic flow.} \label{accretion}
\end{figure*}

Having expressed all quantities at the critical radius in terms of the boundary conditions, the critical accretion rate
\begin{equation}
    \dot{M}=4\pi \rho_{0,s} u_{s} r_{s}^{2},\label{accretion2}
\end{equation}
can be computed easily
\begin{equation}
    \dot{M}=4\pi \left(\frac{M}{c_{s,\infty}^{2}}\right)^{2} c_{s,\infty}^{2}\;\rho_{0,\infty}\;\lambda_{\rm RN}^{\rm NA},\label{accretionRN}
    \end{equation}
with the accretion rate eigenvalue
\begin{equation}
   \lambda_{\rm RN}^{\rm NA} \equiv   \left( \frac{c_{s,c}^{2}}{c_{s,\infty}^{2}}  \right)^{\frac{5-3\gamma}{\gamma-1}} \left( \frac{\gamma-1-c_{s,\infty}^{2}}{\gamma-1-c_{s,c}^{2}} \right)^{\frac{1}{\gamma-1}} \frac{(1+3c_{s,c}^{2})^{3/2}}{4}\beta,\label{eigenRN}
\end{equation}
and the $\beta$ factor, containing information of the non-Abelian charge, is
\begin{equation}
    \beta = \frac{1}{4}\left[ 1+ \sqrt{1-\frac{8 c_{s,c}^{2}(1+c_{s,c}^{2})q_{\rm NA}^{2}}{(1+3c_{s,c}^{2})^{2}}}\right]^{2}.\label{betafactor}
\end{equation}
This quantity clearly accounts for the deviation from the Schwarzschild case. In what follows, we quantify such a deviation by computing the ratio of both accretion rates
\begin{equation}
   \frac{\dot{M}_{\rm RN}^{\rm NA}}{\dot{M}_{\rm Sch}} = \frac{\lambda_{\rm RN}^{\rm NA}}{\lambda_{\rm Sch}}=\beta. \label{accretionrations}
\end{equation}
As it is known, the electric charge of the RN black hole reduces the accretion rate compared to the Schwarzschild black hole. Our case may be, however, different if the imaginary charge of the black hole is allowed, i.e. when $q_{NA}^{2}=q_{NA,I}^{2}<0$ and $\chi\in(0,1)$. Under this choice, $\dot{M}_{\rm RN}^{\rm NA}>\dot{M}_{\rm Sch}$ as can be verified in all left panels of Figure \ref{accretion} where the accretion rate has been plotted as a function of the coupling constant $\chi$ for different adiabatic indices as indicated. Notice that all curves meet in the corresponding position of the event horizon $\chi\to0,\infty$. Out of the mentioned range, the expected behavior of the Reissner Nordstr\"{o}m black hole is displayed just as the case $q_{\rm NA}=q_{\rm NA,II}$ (right panels). Even though the non-Abelian charge case is effectively distinguishable from its electric counterpart in the small range $\chi\in(0,1)$ for $q_{\rm NA,I}$, these numerical computations allow ones, in addition, to understand better the multiplicity of the non-Abelian Reissner Nordstr\"{o}m black solution and its implications in the accretion rate, in particular for the range of values of $\chi$ that leads to a significant enhancement of the accretion flow.

As a last remark about the boundary conditions, the larger the boundary sound speed, the shorter the accretion rate variations are among the polytropic fluid considered. The stiff case $\gamma=1$ is more sensible to the increase of the boundary sound speed: notice how the dashed black curves in right panels meet the other curves for low $\chi$ or, which is equivalent, in the non-Abelian extremal case $q_{\rm NA}\to1$. 

%

%%%%%%%%%%%%%%%%%%%%%%%%%%%%%%%%%%%%%%
\subsubsection{Asymptotic limit}

We do not describe here in detail all derivations concerning the asymptotic limit of the mass accretion rate since it can be found, for instance, in reference \cite{Shapiro:1983du}. The purpose of this part is to check the consistency with the well-known Newtonian limit and the corresponding dependence on the non-Abelian charge. From the Bernoulli equation, one can derive the useful relation
\begin{equation}
   c_{s,c}^2 \approx \frac{2 c_{s,\infty}^{2}}{(5-3\gamma)} ,\label{sub32:eqn3}
\end{equation}
under the non-relativistic condition $c_{s,c}\ll 1$ which holds for reasonable large radius, $r\gg r_{c}$, far away from the BH gravitational influence. The same condition leads to the simple relation 
\begin{equation}
   c_{s,c}^2 \approx K \gamma\; \rho_{0,c}^{\gamma-1} ,\label{sub32:eqn4}
\end{equation}
between the sound speed and the mass density at the critical point. This implies that the mass density can be expressed in terms of the sound speed at the infinity in view of Eq.~(\ref{sub32:eqn3}) to yield
\begin{equation}
   \rho_{0,c} \approx \rho_{0,\infty}\left(\frac{c_{s,c}^2}{c_{s,\infty}^2}\right)^{\frac{1}{\gamma-1}}\approx \left( \frac{2}{5-3\gamma}\right)^{\frac{1}{\gamma-1}}.\label{sub32:eqn5}
\end{equation}
As in the uncharged case, physical solutions require $\gamma<5/3$ in this. 
At this point, all of the above are standard and these quantities do not receive contributions from the effective charge $Q_{\rm NA}$ at lowest order in $c_{s,c}$. This is not the case however for the critical radius Eq.~(\ref{eqn:criticalradius}) where, at leading order, $Q_{\rm NA}$ already appears explicitly as second order power
%
%\begin{widetext}
\begin{equation}
   r_{c} \approx \frac{M}{2 c_{s,c}^{2}} + \frac{3 M^{2} - 2 Q_{\rm NA}^{2}}{2 M} + \frac{2 (M^{2} Q_{\rm NA}^{2} - Q_{\rm NA}^{4})} {M^{3}} c_{s,c}^{2} + \mathcal{O}(c_{s,c}^{4}).\label{sub32:eqn6}
\end{equation}
%\end{widetext}
%
Keeping only higher order contributions in $c_{s,c}$ and using Eq.~(\ref{sub32:eqn3}), the critical radius can be approximated to
\begin{equation}
   r_{c} \approx \frac{(5-3\gamma)}{4c_{s,\infty}^{2}} M\eta(Q_{\rm NA}),\label{sub32:eqn7}
\end{equation}
where the dimensionless correction factor due to the effective charge has been defined as
\begin{equation}
   \eta(Q_{\rm NA}) = 1 + \frac{c_{s,\infty}^{2}}{(5-3\gamma)}\left(6-4\frac{Q_{\rm NA}^{2}}{M^2}\right).\label{sub32:eqn8}
\end{equation}
With all this, the critical accretion mass rate can be written in the familiar form
\begin{equation}
   \dot{M} \approx 4\pi \rho_{0,\infty} M^{2} c_{s,\infty}^{-3} \left(\frac{1}{2}\right)^{\frac{\gamma+1}{2(\gamma-1)}} \left(\frac{5-3\gamma}{4}\right)^{\frac{3\gamma-5}{2(\gamma-1)}}\eta(Q_{\rm NA})^{2},\label{sub32:eqn9}
\end{equation}
which agrees with \cite{FreitasPacheco:2011yme}. Finally, the uncharged Newtonian limit is straightforwardly achieved in the limit $Q_{\rm NA}\to 0$ and, appropriately, $c_{c,\infty}^{2}\ll1$. %since charge corrections to the accretion rate were already taken away in Eq.~(\ref{sub32:eqn7}).

%%%%%%%%%%%%%%%%%%%%%%%%%%%%%%%%%%%%%%%%
%%%%%%%%%%%%%%%%%%%%%%%%%%%%%%%%%%%%%
\section{Discussion and Conclusions}\label{sec:5}

Within the framework of modified theories of gravity, in particular in the context of  vector-tensor theories that follow the same spirit of Horndeski’s theory, a kind of RN black hole solution with two different non-Abelian effective charges has been found in terms of the coupling constants of the involved Lagrangian pieces of the generalized SU(2) Proca theory.  Such new solutions correspond to genuine non-Abelian RN BH solutions in the sense they were derived from a theory where the vector fields belong to the Lie algebra of the SU(2) group. These objects then carry a dark charge because the magnetic charge is not of electromagnetic origin. There do exist other solutions coming from other more involved Lagrangians of the theory \cite{GallegoCadavid:2020dho}, but they do not possess effective charges in the asymptotic limit, which means that they converge, formally speaking, to the Schwarzschild solution. This will be reported in a separate paper.

Even though these solutions retain the main global properties of the standard RN BH derived from the Einstein-Maxwell theory, the solutions found exhibit an appealing structure due to the non-trivial dependence on the coupling constants. Interestingly, the solution with $q_{\rm NA,I}$ is characterized by having a negative square of the effective charge for a certain range of values of the coupling constant $\chi$, similar to the tidal charge of brane BHs \cite{Dadhich:2000am,Zakharov:2021gbg,Neves:2020doc} and BH solutions in Horndeski theory \cite{Babichev:2017guv}. Preventing the naked singularity from forming in both non-Abelian RN BHs leads to discard a small region of the parameter space. This happens particularly for the solutions with $q_{\rm NA,I}$ since the solution with $q_{\rm NA,II}$ is always real and well-behaved in the sense that no divergences are present in the event horizon and ISCO structures. Studying the RN solution within the framework of the GSU2P theory allowed us to infer constraints on the free parameters of the model. This finding is of particular physical interest as it can be compared with future astrophysical constraints to check the consistency of the theory at the relevant scales.

Some phenomenological implications  of the BH solutions were also investigated in the astrophysical setting in order to constrain the parameter space in a joint way with the aforementioned theoretical considerations. They are summarized as follows.

%Studying the event horizon structure and the ISCO of massive test particles, allowed us to assess 
%
\begin{itemize}
    \item Observations of the EHT's first images of Sagittarius A$^{\star}$ of our Galaxy, along with Keck telescope results, set the first serious constraint on the free parameters of the theory ($\tilde{g},\chi$), leaving almost all the available parameter space of the non-Abelian RN BHs basically unconstrained. For a given $\tilde{g}$, a lower limit on $\chi$ is determined as can be inferred from the parameter space Fig.~ \ref{fig:shadow-constr}. As in the electric RN BH case, these observational constraints also rule out regions of the parameter space of the BH solutions associated with naked singularities and with the extremal BH. On the contrary, the corresponding region of the parameter space for which $q_{\rm NA,I}^{2}<0$, i.e., imaginary non-Abelian charge, is allowed.

     \item As a first step toward a more realistic and elaborate description of accretion processes, a fully relativistic treatment of spherical accretion of isothermal and polytropic fluids onto this class of BHs has been performed to quantify the effect of the non-Abelian charge and, therefore, of the coupling constant, on the critical accretion rate. Interestingly, we have found some dissimilarities in the accretion process with respect to the standard electric RN case that can serve as a potential observational signature to test the theory. Concretely, the critical accretion rate efficiency can be noticeably improved compared to the Schwarzschild case (and also to the electric RN case), that is, $\dot{M}_{\rm RN}^{\rm NA}>\dot{M}_{\rm Sch}$, provided that $\chi \in (0,1)$ and  $q_{\rm NA}=q_{\rm NA,I}^{2}<0$ which is, as discussed above, admitted form the observational side. In this regard, we have examined carefully, with the aid of numerical computations for different adiabatic indices of a polytropic fluid, the role of the coupling parameter on the transonic properties of steady flows.

\item  As a way of probing the consistency of the non-Abelian BH solutions, the Schwarzschild solution and the extremal RN BH solution are recovered in our solutions, as limit cases of the theory, for certain values of $\chi$. This is a probe of concept of how the BH solutions found behave in extreme regimes of the parameter space.

\end{itemize}

 An immediate theoretical extension of this work is to implement the Newman-Janis algorithm to find rotating non-Abelian charge BH solutions. Effectively, a Kerr-(non-Abelian) Newman BH solution is naturally expected. Although the applicability of this algorithm must be taken with great care \cite{Hansen:2013owa}, the absence of direct couplings of the gauge fields to curvature terms guarantee the viability of this future work. Then, we plan to study the main properties of the image of the resulting BHs, such as the shadows and photon rings, surrounded by an optically and geometrically thin accretion disk and the subsequent comparison with current observations. In this regard, it is imperative to use observational constraints from the shadow of the supermassive BH galaxy M87$^{\star}$, as was recently done for the tidal charge \cite{Neves:2020doc}. 
 Gravitational and electromagnetic waveforms for charged black hole binaries can be used to estimate the charges of BHs in current and future gravitational wave experiments as has been discussed recently \cite{
Christiansen:2020pnv,Benavides-Gallego:2022dpn,Bozzola:2020mjx,Liu:2020vsy,Wang:2021vmi}. This is another interesting way to assess the effect of the coupling constants in the strong-field regime in the vicinity of BHs. Hence, gravitational wave observations have also the potential of putting constraints on the coupling constants of the theory.

\section*{Acknowledgments}

G. G. acknowledges financial support from Agencia Nacional de Investigaci\'on y Desarrollo (ANID) through the FONDECYT postdoctoral Grant No. 3210417.
J.F.R. is thankful for financial support from the Patrimonio Aut\'onomo - Fondo Nacional de Financiamiento para la Ciencia, la
Tecnolog\'ia y la Innovaci\'on Francisco Jos\'e de Caldas (MINCIENCIAS - COLOMBIA) under the grant No. 110685269447
RC-80740--465--2020, project 69553.

\bibliography{biblio}

\begin{thebibliography}{93}%
\makeatletter
\providecommand \@ifxundefined [1]{%
 \@ifx{#1\undefined}
}%
\providecommand \@ifnum [1]{%
 \ifnum #1\expandafter \@firstoftwo
 \else \expandafter \@secondoftwo
 \fi
}%
\providecommand \@ifx [1]{%
 \ifx #1\expandafter \@firstoftwo
 \else \expandafter \@secondoftwo
 \fi
}%
\providecommand \natexlab [1]{#1}%
\providecommand \enquote  [1]{``#1''}%
\providecommand \bibnamefont  [1]{#1}%
\providecommand \bibfnamefont [1]{#1}%
\providecommand \citenamefont [1]{#1}%
\providecommand \href@noop [0]{\@secondoftwo}%
\providecommand \href [0]{\begingroup \@sanitize@url \@href}%
\providecommand \@href[1]{\@@startlink{#1}\@@href}%
\providecommand \@@href[1]{\endgroup#1\@@endlink}%
\providecommand \@sanitize@url [0]{\catcode `\\12\catcode `\$12\catcode
  `\&12\catcode `\#12\catcode `\^12\catcode `\_12\catcode `\%12\relax}%
\providecommand \@@startlink[1]{}%
\providecommand \@@endlink[0]{}%
\providecommand \url  [0]{\begingroup\@sanitize@url \@url }%
\providecommand \@url [1]{\endgroup\@href {#1}{\urlprefix }}%
\providecommand \urlprefix  [0]{URL }%
\providecommand \Eprint [0]{\href }%
\providecommand \doibase [0]{http://dx.doi.org/}%
\providecommand \selectlanguage [0]{\@gobble}%
\providecommand \bibinfo  [0]{\@secondoftwo}%
\providecommand \bibfield  [0]{\@secondoftwo}%
\providecommand \translation [1]{[#1]}%
\providecommand \BibitemOpen [0]{}%
\providecommand \bibitemStop [0]{}%
\providecommand \bibitemNoStop [0]{.\EOS\space}%
\providecommand \EOS [0]{\spacefactor3000\relax}%
\providecommand \BibitemShut  [1]{\csname bibitem#1\endcsname}%
\let\auto@bib@innerbib\@empty
%</preamble>
\bibitem [{\citenamefont {Poisson}(2009)}]{Poisson:2009pwt}%
  \BibitemOpen
  \bibfield  {author} {\bibinfo {author} {\bibfnamefont {E.}~\bibnamefont
  {Poisson}},\ }\href {\doibase 10.1017/CBO9780511606601} {\emph {\bibinfo
  {title} {{A Relativist's Toolkit: The Mathematics of Black-Hole
  Mechanics}}}}\ (\bibinfo  {publisher} {Cambridge University Press},\ \bibinfo
  {year} {2009})\BibitemShut {NoStop}%
\bibitem [{\citenamefont {Shapiro}\ and\ \citenamefont
  {Teukolsky}(1983)}]{Shapiro:1983du}%
  \BibitemOpen
  \bibfield  {author} {\bibinfo {author} {\bibfnamefont {S.~L.}\ \bibnamefont
  {Shapiro}}\ and\ \bibinfo {author} {\bibfnamefont {S.~A.}\ \bibnamefont
  {Teukolsky}},\ }\href@noop {} {\emph {\bibinfo {title} {{Black holes, white
  dwarfs, and neutron stars: The physics of compact objects}}}}\ (\bibinfo
  {year} {1983})\BibitemShut {NoStop}%
\bibitem [{\citenamefont {Akiyama}\ \emph {et~al.}(2019)\citenamefont {Akiyama}
  \emph {et~al.}}]{EventHorizonTelescope:2019dse}%
  \BibitemOpen
  \bibfield  {author} {\bibinfo {author} {\bibfnamefont {K.}~\bibnamefont
  {Akiyama}} \emph {et~al.} (\bibinfo {collaboration} {Event Horizon
  Telescope}),\ }\href {\doibase 10.3847/2041-8213/ab0ec7} {\bibfield
  {journal} {\bibinfo  {journal} {Astrophys. J. Lett.}\ }\textbf {\bibinfo
  {volume} {875}},\ \bibinfo {pages} {L1} (\bibinfo {year} {2019})},\ \Eprint
  {http://arxiv.org/abs/1906.11238} {arXiv:1906.11238 [astro-ph.GA]}
  \BibitemShut {NoStop}%
\bibitem [{\citenamefont {Akiyama}\ \emph {et~al.}(2022)\citenamefont {Akiyama}
  \emph {et~al.}}]{EventHorizonTelescope:2022wkp}%
  \BibitemOpen
  \bibfield  {author} {\bibinfo {author} {\bibfnamefont {K.}~\bibnamefont
  {Akiyama}} \emph {et~al.} (\bibinfo {collaboration} {Event Horizon
  Telescope}),\ }\href {\doibase 10.3847/2041-8213/ac6674} {\bibfield
  {journal} {\bibinfo  {journal} {Astrophys. J. Lett.}\ }\textbf {\bibinfo
  {volume} {930}},\ \bibinfo {pages} {L12} (\bibinfo {year}
  {2022})}\BibitemShut {NoStop}%
\bibitem [{\citenamefont {Abuter}\ \emph {et~al.}(2020)\citenamefont {Abuter}
  \emph {et~al.}}]{GRAVITY:2020gka}%
  \BibitemOpen
  \bibfield  {author} {\bibinfo {author} {\bibfnamefont {R.}~\bibnamefont
  {Abuter}} \emph {et~al.} (\bibinfo {collaboration} {GRAVITY}),\ }\href
  {\doibase 10.1051/0004-6361/202037813} {\bibfield  {journal} {\bibinfo
  {journal} {Astron. Astrophys.}\ }\textbf {\bibinfo {volume} {636}},\ \bibinfo
  {pages} {L5} (\bibinfo {year} {2020})},\ \Eprint
  {http://arxiv.org/abs/2004.07187} {arXiv:2004.07187 [astro-ph.GA]}
  \BibitemShut {NoStop}%
\bibitem [{\citenamefont {Abbott}\ \emph
  {et~al.}(2017{\natexlab{a}})\citenamefont {Abbott} \emph
  {et~al.}}]{LIGOScientific:2017ync}%
  \BibitemOpen
  \bibfield  {author} {\bibinfo {author} {\bibfnamefont {B.~P.}\ \bibnamefont
  {Abbott}} \emph {et~al.} (\bibinfo {collaboration} {LIGO Scientific, Virgo,
  Fermi GBM, INTEGRAL, IceCube, AstroSat Cadmium Zinc Telluride Imager Team,
  IPN, Insight-Hxmt, ANTARES, Swift, AGILE Team, 1M2H Team, Dark Energy Camera
  GW-EM, DES, DLT40, GRAWITA, Fermi-LAT, ATCA, ASKAP, Las Cumbres Observatory
  Group, OzGrav, DWF (Deeper Wider Faster Program), AST3, CAASTRO, VINROUGE,
  MASTER, J-GEM, GROWTH, JAGWAR, CaltechNRAO, TTU-NRAO, NuSTAR, Pan-STARRS,
  MAXI Team, TZAC Consortium, KU, Nordic Optical Telescope, ePESSTO, GROND,
  Texas Tech University, SALT Group, TOROS, BOOTES, MWA, CALET, IKI-GW
  Follow-up, H.E.S.S., LOFAR, LWA, HAWC, Pierre Auger, ALMA, Euro VLBI Team, Pi
  of Sky, Chandra Team at McGill University, DFN, ATLAS Telescopes, High Time
  Resolution Universe Survey, RIMAS, RATIR, SKA South Africa/MeerKAT}),\ }\href
  {\doibase 10.3847/2041-8213/aa91c9} {\bibfield  {journal} {\bibinfo
  {journal} {Astrophys. J. Lett.}\ }\textbf {\bibinfo {volume} {848}},\
  \bibinfo {pages} {L12} (\bibinfo {year} {2017}{\natexlab{a}})},\ \Eprint
  {http://arxiv.org/abs/1710.05833} {arXiv:1710.05833 [astro-ph.HE]}
  \BibitemShut {NoStop}%
\bibitem [{\citenamefont {Abbott}\ \emph
  {et~al.}(2017{\natexlab{b}})\citenamefont {Abbott} \emph
  {et~al.}}]{LIGOScientific:2017vwq}%
  \BibitemOpen
  \bibfield  {author} {\bibinfo {author} {\bibfnamefont {B.~P.}\ \bibnamefont
  {Abbott}} \emph {et~al.} (\bibinfo {collaboration} {LIGO Scientific,
  Virgo}),\ }\href {\doibase 10.1103/PhysRevLett.119.161101} {\bibfield
  {journal} {\bibinfo  {journal} {Phys. Rev. Lett.}\ }\textbf {\bibinfo
  {volume} {119}},\ \bibinfo {pages} {161101} (\bibinfo {year}
  {2017}{\natexlab{b}})},\ \Eprint {http://arxiv.org/abs/1710.05832}
  {arXiv:1710.05832 [gr-qc]} \BibitemShut {NoStop}%
\bibitem [{\citenamefont {Will}(2014)}]{Will:2014kxa}%
  \BibitemOpen
  \bibfield  {author} {\bibinfo {author} {\bibfnamefont {C.~M.}\ \bibnamefont
  {Will}},\ }\href {\doibase 10.12942/lrr-2014-4} {\bibfield  {journal}
  {\bibinfo  {journal} {Living Rev. Rel.}\ }\textbf {\bibinfo {volume} {17}},\
  \bibinfo {pages} {4} (\bibinfo {year} {2014})},\ \Eprint
  {http://arxiv.org/abs/1403.7377} {arXiv:1403.7377 [gr-qc]} \BibitemShut
  {NoStop}%
\bibitem [{\citenamefont {Abbott}\ \emph {et~al.}(2021)\citenamefont {Abbott}
  \emph {et~al.}}]{LIGOScientific:2020tif}%
  \BibitemOpen
  \bibfield  {author} {\bibinfo {author} {\bibfnamefont {R.}~\bibnamefont
  {Abbott}} \emph {et~al.} (\bibinfo {collaboration} {LIGO Scientific,
  Virgo}),\ }\href {\doibase 10.1103/PhysRevD.103.122002} {\bibfield  {journal}
  {\bibinfo  {journal} {Phys. Rev. D}\ }\textbf {\bibinfo {volume} {103}},\
  \bibinfo {pages} {122002} (\bibinfo {year} {2021})},\ \Eprint
  {http://arxiv.org/abs/2010.14529} {arXiv:2010.14529 [gr-qc]} \BibitemShut
  {NoStop}%
\bibitem [{\citenamefont {Cardoso}\ and\ \citenamefont
  {Pani}(2019)}]{Cardoso:2019rvt}%
  \BibitemOpen
  \bibfield  {author} {\bibinfo {author} {\bibfnamefont {V.}~\bibnamefont
  {Cardoso}}\ and\ \bibinfo {author} {\bibfnamefont {P.}~\bibnamefont {Pani}},\
  }\href {\doibase 10.1007/s41114-019-0020-4} {\bibfield  {journal} {\bibinfo
  {journal} {Living Rev. Rel.}\ }\textbf {\bibinfo {volume} {22}},\ \bibinfo
  {pages} {4} (\bibinfo {year} {2019})},\ \Eprint
  {http://arxiv.org/abs/1904.05363} {arXiv:1904.05363 [gr-qc]} \BibitemShut
  {NoStop}%
\bibitem [{\citenamefont {Psaltis}\ \emph {et~al.}(2020)\citenamefont {Psaltis}
  \emph {et~al.}}]{EventHorizonTelescope:2020qrl}%
  \BibitemOpen
  \bibfield  {author} {\bibinfo {author} {\bibfnamefont {D.}~\bibnamefont
  {Psaltis}} \emph {et~al.} (\bibinfo {collaboration} {Event Horizon
  Telescope}),\ }\href {\doibase 10.1103/PhysRevLett.125.141104} {\bibfield
  {journal} {\bibinfo  {journal} {Phys. Rev. Lett.}\ }\textbf {\bibinfo
  {volume} {125}},\ \bibinfo {pages} {141104} (\bibinfo {year} {2020})},\
  \Eprint {http://arxiv.org/abs/2010.01055} {arXiv:2010.01055 [gr-qc]}
  \BibitemShut {NoStop}%
\bibitem [{\citenamefont {Vagnozzi}\ \emph {et~al.}(2022)\citenamefont
  {Vagnozzi} \emph {et~al.}}]{Vagnozzi:2022moj}%
  \BibitemOpen
  \bibfield  {author} {\bibinfo {author} {\bibfnamefont {S.}~\bibnamefont
  {Vagnozzi}} \emph {et~al.},\ }\href@noop {} {\  (\bibinfo {year} {2022})},\
  \Eprint {http://arxiv.org/abs/2205.07787} {arXiv:2205.07787 [gr-qc]}
  \BibitemShut {NoStop}%
\bibitem [{\citenamefont {{Reissner}}(1916)}]{1916AnP...355..106R}%
  \BibitemOpen
  \bibfield  {author} {\bibinfo {author} {\bibfnamefont {H.}~\bibnamefont
  {{Reissner}}},\ }\href {\doibase 10.1002/andp.19163550905} {\bibfield
  {journal} {\bibinfo  {journal} {Annalen der Physik}\ }\textbf {\bibinfo
  {volume} {355}},\ \bibinfo {pages} {106} (\bibinfo {year}
  {1916})}\BibitemShut {NoStop}%
\bibitem [{\citenamefont {{Nordstr{\"o}m}}(1918)}]{1918KNAB...20.1238N}%
  \BibitemOpen
  \bibfield  {author} {\bibinfo {author} {\bibfnamefont {G.}~\bibnamefont
  {{Nordstr{\"o}m}}},\ }\href@noop {} {\bibfield  {journal} {\bibinfo
  {journal} {Koninklijke Nederlandse Akademie van Wetenschappen Proceedings
  Series B Physical Sciences}\ }\textbf {\bibinfo {volume} {20}},\ \bibinfo
  {pages} {1238} (\bibinfo {year} {1918})}\BibitemShut {NoStop}%
\bibitem [{\citenamefont {De~Rujula}\ \emph {et~al.}(1990)\citenamefont
  {De~Rujula}, \citenamefont {Glashow},\ and\ \citenamefont
  {Sarid}}]{DeRujula:1989fe}%
  \BibitemOpen
  \bibfield  {author} {\bibinfo {author} {\bibfnamefont {A.}~\bibnamefont
  {De~Rujula}}, \bibinfo {author} {\bibfnamefont {S.~L.}\ \bibnamefont
  {Glashow}}, \ and\ \bibinfo {author} {\bibfnamefont {U.}~\bibnamefont
  {Sarid}},\ }\href {\doibase 10.1016/0550-3213(90)90227-5} {\bibfield
  {journal} {\bibinfo  {journal} {Nucl. Phys. B}\ }\textbf {\bibinfo {volume}
  {333}},\ \bibinfo {pages} {173} (\bibinfo {year} {1990})}\BibitemShut
  {NoStop}%
\bibitem [{\citenamefont {Cardoso}\ \emph {et~al.}(2016)\citenamefont
  {Cardoso}, \citenamefont {Macedo}, \citenamefont {Pani},\ and\ \citenamefont
  {Ferrari}}]{Cardoso:2016olt}%
  \BibitemOpen
  \bibfield  {author} {\bibinfo {author} {\bibfnamefont {V.}~\bibnamefont
  {Cardoso}}, \bibinfo {author} {\bibfnamefont {C.~F.~B.}\ \bibnamefont
  {Macedo}}, \bibinfo {author} {\bibfnamefont {P.}~\bibnamefont {Pani}}, \ and\
  \bibinfo {author} {\bibfnamefont {V.}~\bibnamefont {Ferrari}},\ }\href
  {\doibase 10.1088/1475-7516/2016/05/054} {\bibfield  {journal} {\bibinfo
  {journal} {JCAP}\ }\textbf {\bibinfo {volume} {05}},\ \bibinfo {pages} {054}
  (\bibinfo {year} {2016})},\ \bibinfo {note} {[Erratum: JCAP 04, E01
  (2020)]},\ \Eprint {http://arxiv.org/abs/1604.07845} {arXiv:1604.07845
  [hep-ph]} \BibitemShut {NoStop}%
\bibitem [{\citenamefont {Zaja\v{c}ek}\ and\ \citenamefont
  {Tursunov}(2019)}]{Zajacek:2019kla}%
  \BibitemOpen
  \bibfield  {author} {\bibinfo {author} {\bibfnamefont {M.}~\bibnamefont
  {Zaja\v{c}ek}}\ and\ \bibinfo {author} {\bibfnamefont {A.}~\bibnamefont
  {Tursunov}},\ }\href@noop {} {\  (\bibinfo {year} {2019})},\ \Eprint
  {http://arxiv.org/abs/1904.04654} {arXiv:1904.04654 [astro-ph.GA]}
  \BibitemShut {NoStop}%
\bibitem [{\citenamefont {Kocherlakota}\ \emph {et~al.}(2021)\citenamefont
  {Kocherlakota} \emph {et~al.}}]{EventHorizonTelescope:2021dqv}%
  \BibitemOpen
  \bibfield  {author} {\bibinfo {author} {\bibfnamefont {P.}~\bibnamefont
  {Kocherlakota}} \emph {et~al.} (\bibinfo {collaboration} {Event Horizon
  Telescope}),\ }\href {\doibase 10.1103/PhysRevD.103.104047} {\bibfield
  {journal} {\bibinfo  {journal} {Phys. Rev. D}\ }\textbf {\bibinfo {volume}
  {103}},\ \bibinfo {pages} {104047} (\bibinfo {year} {2021})},\ \Eprint
  {http://arxiv.org/abs/2105.09343} {arXiv:2105.09343 [gr-qc]} \BibitemShut
  {NoStop}%
\bibitem [{\citenamefont {Bozzola}\ and\ \citenamefont
  {Paschalidis}(2021)}]{Bozzola:2020mjx}%
  \BibitemOpen
  \bibfield  {author} {\bibinfo {author} {\bibfnamefont {G.}~\bibnamefont
  {Bozzola}}\ and\ \bibinfo {author} {\bibfnamefont {V.}~\bibnamefont
  {Paschalidis}},\ }\href {\doibase 10.1103/PhysRevLett.126.041103} {\bibfield
  {journal} {\bibinfo  {journal} {Phys. Rev. Lett.}\ }\textbf {\bibinfo
  {volume} {126}},\ \bibinfo {pages} {041103} (\bibinfo {year} {2021})},\
  \Eprint {http://arxiv.org/abs/2006.15764} {arXiv:2006.15764 [gr-qc]}
  \BibitemShut {NoStop}%
\bibitem [{\citenamefont {Liu}\ \emph {et~al.}(2020)\citenamefont {Liu},
  \citenamefont {Christiansen}, \citenamefont {Guo}, \citenamefont {Cai},\ and\
  \citenamefont {Kim}}]{Liu:2020vsy}%
  \BibitemOpen
  \bibfield  {author} {\bibinfo {author} {\bibfnamefont {L.}~\bibnamefont
  {Liu}}, \bibinfo {author} {\bibfnamefont {O.}~\bibnamefont {Christiansen}},
  \bibinfo {author} {\bibfnamefont {Z.-K.}\ \bibnamefont {Guo}}, \bibinfo
  {author} {\bibfnamefont {R.-G.}\ \bibnamefont {Cai}}, \ and\ \bibinfo
  {author} {\bibfnamefont {S.~P.}\ \bibnamefont {Kim}},\ }\href {\doibase
  10.1103/PhysRevD.102.103520} {\bibfield  {journal} {\bibinfo  {journal}
  {Phys. Rev. D}\ }\textbf {\bibinfo {volume} {102}},\ \bibinfo {pages}
  {103520} (\bibinfo {year} {2020})},\ \Eprint
  {http://arxiv.org/abs/2008.02326} {arXiv:2008.02326 [gr-qc]} \BibitemShut
  {NoStop}%
\bibitem [{\citenamefont {Christiansen}\ \emph {et~al.}(2021)\citenamefont
  {Christiansen}, \citenamefont {Beltr\'an~Jim\'enez},\ and\ \citenamefont
  {Mota}}]{Christiansen:2020pnv}%
  \BibitemOpen
  \bibfield  {author} {\bibinfo {author} {\bibfnamefont {O.}~\bibnamefont
  {Christiansen}}, \bibinfo {author} {\bibfnamefont {J.}~\bibnamefont
  {Beltr\'an~Jim\'enez}}, \ and\ \bibinfo {author} {\bibfnamefont {D.~F.}\
  \bibnamefont {Mota}},\ }\href {\doibase 10.1088/1361-6382/abdaf5} {\bibfield
  {journal} {\bibinfo  {journal} {Class. Quant. Grav.}\ }\textbf {\bibinfo
  {volume} {38}},\ \bibinfo {pages} {075017} (\bibinfo {year} {2021})},\
  \Eprint {http://arxiv.org/abs/2003.11452} {arXiv:2003.11452 [gr-qc]}
  \BibitemShut {NoStop}%
\bibitem [{\citenamefont {Wang}\ \emph {et~al.}(2021)\citenamefont {Wang},
  \citenamefont {Li}, \citenamefont {Jiang}, \citenamefont {Yuan},
  \citenamefont {Hu},\ and\ \citenamefont {Fan}}]{Wang:2021vmi}%
  \BibitemOpen
  \bibfield  {author} {\bibinfo {author} {\bibfnamefont {H.-T.}\ \bibnamefont
  {Wang}}, \bibinfo {author} {\bibfnamefont {P.-C.}\ \bibnamefont {Li}},
  \bibinfo {author} {\bibfnamefont {J.-L.}\ \bibnamefont {Jiang}}, \bibinfo
  {author} {\bibfnamefont {G.-W.}\ \bibnamefont {Yuan}}, \bibinfo {author}
  {\bibfnamefont {Y.-M.}\ \bibnamefont {Hu}}, \ and\ \bibinfo {author}
  {\bibfnamefont {Y.-Z.}\ \bibnamefont {Fan}},\ }\href {\doibase
  10.1140/epjc/s10052-021-09555-1} {\bibfield  {journal} {\bibinfo  {journal}
  {Eur. Phys. J. C}\ }\textbf {\bibinfo {volume} {81}},\ \bibinfo {pages} {769}
  (\bibinfo {year} {2021})}\BibitemShut {NoStop}%
\bibitem [{\citenamefont {Benavides-Gallego}\ and\ \citenamefont
  {Han}(2022)}]{Benavides-Gallego:2022dpn}%
  \BibitemOpen
  \bibfield  {author} {\bibinfo {author} {\bibfnamefont {C.~A.}\ \bibnamefont
  {Benavides-Gallego}}\ and\ \bibinfo {author} {\bibfnamefont {W.-B.}\
  \bibnamefont {Han}},\ }\href@noop {} {\  (\bibinfo {year} {2022})},\ \Eprint
  {http://arxiv.org/abs/2209.00874} {arXiv:2209.00874 [gr-qc]} \BibitemShut
  {NoStop}%
\bibitem [{\citenamefont {Zakharov}(2014)}]{Zakharov:2014lqa}%
  \BibitemOpen
  \bibfield  {author} {\bibinfo {author} {\bibfnamefont {A.~F.}\ \bibnamefont
  {Zakharov}},\ }\href {\doibase 10.1103/PhysRevD.90.062007} {\bibfield
  {journal} {\bibinfo  {journal} {Phys. Rev. D}\ }\textbf {\bibinfo {volume}
  {90}},\ \bibinfo {pages} {062007} (\bibinfo {year} {2014})},\ \Eprint
  {http://arxiv.org/abs/1407.7457} {arXiv:1407.7457 [gr-qc]} \BibitemShut
  {NoStop}%
\bibitem [{\citenamefont {Zaja\v{c}ek}\ \emph {et~al.}(2018)\citenamefont
  {Zaja\v{c}ek}, \citenamefont {Tursunov}, \citenamefont {Eckart},\ and\
  \citenamefont {Britzen}}]{Zajacek:2018ycb}%
  \BibitemOpen
  \bibfield  {author} {\bibinfo {author} {\bibfnamefont {M.}~\bibnamefont
  {Zaja\v{c}ek}}, \bibinfo {author} {\bibfnamefont {A.}~\bibnamefont
  {Tursunov}}, \bibinfo {author} {\bibfnamefont {A.}~\bibnamefont {Eckart}}, \
  and\ \bibinfo {author} {\bibfnamefont {S.}~\bibnamefont {Britzen}},\ }\href
  {\doibase 10.1093/mnras/sty2182} {\bibfield  {journal} {\bibinfo  {journal}
  {Mon. Not. Roy. Astron. Soc.}\ }\textbf {\bibinfo {volume} {480}},\ \bibinfo
  {pages} {4408} (\bibinfo {year} {2018})},\ \Eprint
  {http://arxiv.org/abs/1808.07327} {arXiv:1808.07327 [astro-ph.GA]}
  \BibitemShut {NoStop}%
\bibitem [{\citenamefont {Done}\ \emph {et~al.}(2007)\citenamefont {Done},
  \citenamefont {Gierlinski},\ and\ \citenamefont {Kubota}}]{Done:2007nc}%
  \BibitemOpen
  \bibfield  {author} {\bibinfo {author} {\bibfnamefont {C.}~\bibnamefont
  {Done}}, \bibinfo {author} {\bibfnamefont {M.}~\bibnamefont {Gierlinski}}, \
  and\ \bibinfo {author} {\bibfnamefont {A.}~\bibnamefont {Kubota}},\ }\href
  {\doibase 10.1007/s00159-007-0006-1} {\bibfield  {journal} {\bibinfo
  {journal} {Astron. Astrophys. Rev.}\ }\textbf {\bibinfo {volume} {15}},\
  \bibinfo {pages} {1} (\bibinfo {year} {2007})},\ \Eprint
  {http://arxiv.org/abs/0708.0148} {arXiv:0708.0148 [astro-ph]} \BibitemShut
  {NoStop}%
\bibitem [{\citenamefont {Juraeva}\ \emph {et~al.}(2021)\citenamefont
  {Juraeva}, \citenamefont {Rayimbaev}, \citenamefont {Abdujabbarov},
  \citenamefont {Ahmedov},\ and\ \citenamefont {Palvanov}}]{Juraeva:2021gwb}%
  \BibitemOpen
  \bibfield  {author} {\bibinfo {author} {\bibfnamefont {N.}~\bibnamefont
  {Juraeva}}, \bibinfo {author} {\bibfnamefont {J.}~\bibnamefont {Rayimbaev}},
  \bibinfo {author} {\bibfnamefont {A.}~\bibnamefont {Abdujabbarov}}, \bibinfo
  {author} {\bibfnamefont {B.}~\bibnamefont {Ahmedov}}, \ and\ \bibinfo
  {author} {\bibfnamefont {S.}~\bibnamefont {Palvanov}},\ }\href {\doibase
  10.1140/epjc/s10052-021-08876-5} {\bibfield  {journal} {\bibinfo  {journal}
  {Eur. Phys. J. C}\ }\textbf {\bibinfo {volume} {81}},\ \bibinfo {pages} {70}
  (\bibinfo {year} {2021})}\BibitemShut {NoStop}%
\bibitem [{\citenamefont {{Fragione}}\ and\ \citenamefont
  {{Loeb}}(2020)}]{2020ApJ...901L..32F}%
  \BibitemOpen
  \bibfield  {author} {\bibinfo {author} {\bibfnamefont {G.}~\bibnamefont
  {{Fragione}}}\ and\ \bibinfo {author} {\bibfnamefont {A.}~\bibnamefont
  {{Loeb}}},\ }\href {\doibase 10.3847/2041-8213/abb9b4} {\bibfield  {journal}
  {\bibinfo  {journal} {\apjl}\ }\textbf {\bibinfo {volume} {901}},\ \bibinfo
  {eid} {L32} (\bibinfo {year} {2020})}\BibitemShut {NoStop}%
\bibitem [{\citenamefont {{Misner}}\ \emph {et~al.}(2017)\citenamefont
  {{Misner}}, \citenamefont {{Thorne}},\ and\ \citenamefont
  {{Wheeler}}}]{2017grav.book.....M}%
  \BibitemOpen
  \bibfield  {author} {\bibinfo {author} {\bibfnamefont {C.~W.}\ \bibnamefont
  {{Misner}}}, \bibinfo {author} {\bibfnamefont {K.~S.}\ \bibnamefont
  {{Thorne}}}, \ and\ \bibinfo {author} {\bibfnamefont {J.~A.}\ \bibnamefont
  {{Wheeler}}},\ }\href@noop {} {\emph {\bibinfo {title} {{Gravitation}}}}\
  (\bibinfo {year} {2017})\BibitemShut {NoStop}%
\bibitem [{\citenamefont {Bizon}(1990)}]{Bizon:1990sr}%
  \BibitemOpen
  \bibfield  {author} {\bibinfo {author} {\bibfnamefont {P.}~\bibnamefont
  {Bizon}},\ }\href {\doibase 10.1103/PhysRevLett.64.2844} {\bibfield
  {journal} {\bibinfo  {journal} {Phys. Rev. Lett.}\ }\textbf {\bibinfo
  {volume} {64}},\ \bibinfo {pages} {2844} (\bibinfo {year}
  {1990})}\BibitemShut {NoStop}%
\bibitem [{\citenamefont {Volkov}\ and\ \citenamefont
  {Galtsov}(1989)}]{Volkov:1989fi}%
  \BibitemOpen
  \bibfield  {author} {\bibinfo {author} {\bibfnamefont {M.~S.}\ \bibnamefont
  {Volkov}}\ and\ \bibinfo {author} {\bibfnamefont {D.~V.}\ \bibnamefont
  {Galtsov}},\ }\href@noop {} {\bibfield  {journal} {\bibinfo  {journal} {JETP
  Lett.}\ }\textbf {\bibinfo {volume} {50}},\ \bibinfo {pages} {346} (\bibinfo
  {year} {1989})}\BibitemShut {NoStop}%
\bibitem [{\citenamefont {Kuenzle}\ and\ \citenamefont {Masood-ul
  Alam}(1990)}]{Kuenzle:1990is}%
  \BibitemOpen
  \bibfield  {author} {\bibinfo {author} {\bibfnamefont {H.~P.}\ \bibnamefont
  {Kuenzle}}\ and\ \bibinfo {author} {\bibfnamefont {A.~K.~M.}\ \bibnamefont
  {Masood-ul Alam}},\ }\href {\doibase 10.1063/1.528773} {\bibfield  {journal}
  {\bibinfo  {journal} {J. Math. Phys.}\ }\textbf {\bibinfo {volume} {31}},\
  \bibinfo {pages} {928} (\bibinfo {year} {1990})}\BibitemShut {NoStop}%
\bibitem [{\citenamefont {Lee}\ \emph {et~al.}(1992)\citenamefont {Lee},
  \citenamefont {Nair},\ and\ \citenamefont {Weinberg}}]{Lee:1991qs}%
  \BibitemOpen
  \bibfield  {author} {\bibinfo {author} {\bibfnamefont {K.-M.}\ \bibnamefont
  {Lee}}, \bibinfo {author} {\bibfnamefont {V.~P.}\ \bibnamefont {Nair}}, \
  and\ \bibinfo {author} {\bibfnamefont {E.~J.}\ \bibnamefont {Weinberg}},\
  }\href {\doibase 10.1103/PhysRevLett.68.1100} {\bibfield  {journal} {\bibinfo
   {journal} {Phys. Rev. Lett.}\ }\textbf {\bibinfo {volume} {68}},\ \bibinfo
  {pages} {1100} (\bibinfo {year} {1992})},\ \Eprint
  {http://arxiv.org/abs/hep-th/9111045} {arXiv:hep-th/9111045} \BibitemShut
  {NoStop}%
\bibitem [{\citenamefont {Radu}\ and\ \citenamefont
  {Tchrakian}(2012)}]{Radu:2011ip}%
  \BibitemOpen
  \bibfield  {author} {\bibinfo {author} {\bibfnamefont {E.}~\bibnamefont
  {Radu}}\ and\ \bibinfo {author} {\bibfnamefont {D.~H.}\ \bibnamefont
  {Tchrakian}},\ }\href {\doibase 10.1103/PhysRevD.85.084022} {\bibfield
  {journal} {\bibinfo  {journal} {Phys. Rev. D}\ }\textbf {\bibinfo {volume}
  {85}},\ \bibinfo {pages} {084022} (\bibinfo {year} {2012})},\ \Eprint
  {http://arxiv.org/abs/1111.0418} {arXiv:1111.0418 [gr-qc]} \BibitemShut
  {NoStop}%
\bibitem [{\citenamefont {Mazharimousavi}\ and\ \citenamefont
  {Halilsoy}(2009)}]{Mazharimousavi:2009mb}%
  \BibitemOpen
  \bibfield  {author} {\bibinfo {author} {\bibfnamefont {S.~H.}\ \bibnamefont
  {Mazharimousavi}}\ and\ \bibinfo {author} {\bibfnamefont {M.}~\bibnamefont
  {Halilsoy}},\ }\href {\doibase 10.1016/j.physletb.2009.10.006} {\bibfield
  {journal} {\bibinfo  {journal} {Phys. Lett. B}\ }\textbf {\bibinfo {volume}
  {681}},\ \bibinfo {pages} {190} (\bibinfo {year} {2009})},\ \Eprint
  {http://arxiv.org/abs/0908.0308} {arXiv:0908.0308 [gr-qc]} \BibitemShut
  {NoStop}%
\bibitem [{\citenamefont {Volkov}\ and\ \citenamefont
  {Gal'tsov}(1999)}]{Volkov:1998cc}%
  \BibitemOpen
  \bibfield  {author} {\bibinfo {author} {\bibfnamefont {M.~S.}\ \bibnamefont
  {Volkov}}\ and\ \bibinfo {author} {\bibfnamefont {D.~V.}\ \bibnamefont
  {Gal'tsov}},\ }\href {\doibase 10.1016/S0370-1573(99)00010-1} {\bibfield
  {journal} {\bibinfo  {journal} {Phys. Rept.}\ }\textbf {\bibinfo {volume}
  {319}},\ \bibinfo {pages} {1} (\bibinfo {year} {1999})},\ \Eprint
  {http://arxiv.org/abs/hep-th/9810070} {arXiv:hep-th/9810070} \BibitemShut
  {NoStop}%
\bibitem [{\citenamefont {Herdeiro}\ \emph {et~al.}(2017)\citenamefont
  {Herdeiro}, \citenamefont {Paturyan}, \citenamefont {Radu},\ and\
  \citenamefont {Tchrakian}}]{Herdeiro:2017oxy}%
  \BibitemOpen
  \bibfield  {author} {\bibinfo {author} {\bibfnamefont {C.}~\bibnamefont
  {Herdeiro}}, \bibinfo {author} {\bibfnamefont {V.}~\bibnamefont {Paturyan}},
  \bibinfo {author} {\bibfnamefont {E.}~\bibnamefont {Radu}}, \ and\ \bibinfo
  {author} {\bibfnamefont {D.~H.}\ \bibnamefont {Tchrakian}},\ }\href {\doibase
  10.1016/j.physletb.2017.06.041} {\bibfield  {journal} {\bibinfo  {journal}
  {Phys. Lett. B}\ }\textbf {\bibinfo {volume} {772}},\ \bibinfo {pages} {63}
  (\bibinfo {year} {2017})},\ \Eprint {http://arxiv.org/abs/1705.07979}
  {arXiv:1705.07979 [gr-qc]} \BibitemShut {NoStop}%
\bibitem [{\citenamefont {Mazharimousavi}\ and\ \citenamefont
  {Halilsoy}(2011)}]{Mazharimousavi:2011nc}%
  \BibitemOpen
  \bibfield  {author} {\bibinfo {author} {\bibfnamefont {S.~H.}\ \bibnamefont
  {Mazharimousavi}}\ and\ \bibinfo {author} {\bibfnamefont {M.}~\bibnamefont
  {Halilsoy}},\ }\href {\doibase 10.1103/PhysRevD.84.064032} {\bibfield
  {journal} {\bibinfo  {journal} {Phys. Rev. D}\ }\textbf {\bibinfo {volume}
  {84}},\ \bibinfo {pages} {064032} (\bibinfo {year} {2011})},\ \Eprint
  {http://arxiv.org/abs/1105.3659} {arXiv:1105.3659 [gr-qc]} \BibitemShut
  {NoStop}%
\bibitem [{\citenamefont {{Harnad}}\ \emph {et~al.}(1980)\citenamefont
  {{Harnad}}, \citenamefont {{Tafel}},\ and\ \citenamefont
  {{Shnider}}}]{1980JMP....21.2236H}%
  \BibitemOpen
  \bibfield  {author} {\bibinfo {author} {\bibfnamefont {J.}~\bibnamefont
  {{Harnad}}}, \bibinfo {author} {\bibfnamefont {J.}~\bibnamefont {{Tafel}}}, \
  and\ \bibinfo {author} {\bibfnamefont {S.}~\bibnamefont {{Shnider}}},\ }\href
  {\doibase 10.1063/1.524658} {\bibfield  {journal} {\bibinfo  {journal}
  {Journal of Mathematical Physics}\ }\textbf {\bibinfo {volume} {21}},\
  \bibinfo {pages} {2236} (\bibinfo {year} {1980})}\BibitemShut {NoStop}%
\bibitem [{\citenamefont {{Bartnik}}\ and\ \citenamefont
  {{McKinnon}}(1988)}]{1988PhRvL..61..141B}%
  \BibitemOpen
  \bibfield  {author} {\bibinfo {author} {\bibfnamefont {R.}~\bibnamefont
  {{Bartnik}}}\ and\ \bibinfo {author} {\bibfnamefont {J.}~\bibnamefont
  {{McKinnon}}},\ }\href {\doibase 10.1103/PhysRevLett.61.141} {\bibfield
  {journal} {\bibinfo  {journal} {\prl}\ }\textbf {\bibinfo {volume} {61}},\
  \bibinfo {pages} {141} (\bibinfo {year} {1988})}\BibitemShut {NoStop}%
\bibitem [{\citenamefont {{Breitenlohner}}\ \emph {et~al.}(1995)\citenamefont
  {{Breitenlohner}}, \citenamefont {{Forg{\'a}cs}},\ and\ \citenamefont
  {{Maison}}}]{1995NuPhB.442..126B}%
  \BibitemOpen
  \bibfield  {author} {\bibinfo {author} {\bibfnamefont {P.}~\bibnamefont
  {{Breitenlohner}}}, \bibinfo {author} {\bibfnamefont {P.}~\bibnamefont
  {{Forg{\'a}cs}}}, \ and\ \bibinfo {author} {\bibfnamefont {D.}~\bibnamefont
  {{Maison}}},\ }\href {\doibase 10.1016/S0550-3213(95)00100-X} {\bibfield
  {journal} {\bibinfo  {journal} {Nuclear Physics B}\ }\textbf {\bibinfo
  {volume} {442}},\ \bibinfo {pages} {126} (\bibinfo {year}
  {1995})}\BibitemShut {NoStop}%
\bibitem [{\citenamefont {Penrose}(1965)}]{Penrose:1964wq}%
  \BibitemOpen
  \bibfield  {author} {\bibinfo {author} {\bibfnamefont {R.}~\bibnamefont
  {Penrose}},\ }\href {\doibase 10.1103/PhysRevLett.14.57} {\bibfield
  {journal} {\bibinfo  {journal} {Phys. Rev. Lett.}\ }\textbf {\bibinfo
  {volume} {14}},\ \bibinfo {pages} {57} (\bibinfo {year} {1965})}\BibitemShut
  {NoStop}%
\bibitem [{\citenamefont {Hawking}\ and\ \citenamefont
  {Penrose}(1970)}]{Hawking:1970zqf}%
  \BibitemOpen
  \bibfield  {author} {\bibinfo {author} {\bibfnamefont {S.~W.}\ \bibnamefont
  {Hawking}}\ and\ \bibinfo {author} {\bibfnamefont {R.}~\bibnamefont
  {Penrose}},\ }\href {\doibase 10.1098/rspa.1970.0021} {\bibfield  {journal}
  {\bibinfo  {journal} {Proc. Roy. Soc. Lond. A}\ }\textbf {\bibinfo {volume}
  {314}},\ \bibinfo {pages} {529} (\bibinfo {year} {1970})}\BibitemShut
  {NoStop}%
\bibitem [{\citenamefont {Deser}\ \emph {et~al.}(1975)\citenamefont {Deser},
  \citenamefont {Van~Nieuwenhuizen},\ and\ \citenamefont
  {Boulware}}]{Deser:1974hg}%
  \BibitemOpen
  \bibfield  {author} {\bibinfo {author} {\bibfnamefont {S.}~\bibnamefont
  {Deser}}, \bibinfo {author} {\bibfnamefont {P.}~\bibnamefont
  {Van~Nieuwenhuizen}}, \ and\ \bibinfo {author} {\bibfnamefont
  {D.}~\bibnamefont {Boulware}},\ }in\ \href@noop {} {\emph {\bibinfo
  {booktitle} {{7th International Conference on Gravitation and Relativity}}}}\
  (\bibinfo {year} {1975})\ pp.\ \bibinfo {pages} {1--18}\BibitemShut {NoStop}%
\bibitem [{\citenamefont {Donoghue}(1994)}]{Donoghue:1994dn}%
  \BibitemOpen
  \bibfield  {author} {\bibinfo {author} {\bibfnamefont {J.~F.}\ \bibnamefont
  {Donoghue}},\ }\href {\doibase 10.1103/PhysRevD.50.3874} {\bibfield
  {journal} {\bibinfo  {journal} {Phys. Rev. D}\ }\textbf {\bibinfo {volume}
  {50}},\ \bibinfo {pages} {3874} (\bibinfo {year} {1994})},\ \Eprint
  {http://arxiv.org/abs/gr-qc/9405057} {arXiv:gr-qc/9405057} \BibitemShut
  {NoStop}%
\bibitem [{\citenamefont {Burgess}(2004)}]{Burgess:2003jk}%
  \BibitemOpen
  \bibfield  {author} {\bibinfo {author} {\bibfnamefont {C.~P.}\ \bibnamefont
  {Burgess}},\ }\href {\doibase 10.12942/lrr-2004-5} {\bibfield  {journal}
  {\bibinfo  {journal} {Living Rev. Rel.}\ }\textbf {\bibinfo {volume} {7}},\
  \bibinfo {pages} {5} (\bibinfo {year} {2004})},\ \Eprint
  {http://arxiv.org/abs/gr-qc/0311082} {arXiv:gr-qc/0311082} \BibitemShut
  {NoStop}%
\bibitem [{\citenamefont {Heisenberg}(2019)}]{Heisenberg:2018vsk}%
  \BibitemOpen
  \bibfield  {author} {\bibinfo {author} {\bibfnamefont {L.}~\bibnamefont
  {Heisenberg}},\ }\href {\doibase 10.1016/j.physrep.2018.11.006} {\bibfield
  {journal} {\bibinfo  {journal} {Phys. Rept.}\ }\textbf {\bibinfo {volume}
  {796}},\ \bibinfo {pages} {1} (\bibinfo {year} {2019})},\ \Eprint
  {http://arxiv.org/abs/1807.01725} {arXiv:1807.01725 [gr-qc]} \BibitemShut
  {NoStop}%
\bibitem [{\citenamefont {Horndeski}(1974)}]{Horndeski:1974wa}%
  \BibitemOpen
  \bibfield  {author} {\bibinfo {author} {\bibfnamefont {G.~W.}\ \bibnamefont
  {Horndeski}},\ }\href {\doibase 10.1007/BF01807638} {\bibfield  {journal}
  {\bibinfo  {journal} {Int. J. Theor. Phys.}\ }\textbf {\bibinfo {volume}
  {10}},\ \bibinfo {pages} {363} (\bibinfo {year} {1974})}\BibitemShut
  {NoStop}%
\bibitem [{\citenamefont {Ostrogradsky}(1850)}]{Ostrogradsky:1850fid}%
  \BibitemOpen
  \bibfield  {author} {\bibinfo {author} {\bibfnamefont {M.}~\bibnamefont
  {Ostrogradsky}},\ }\href@noop {} {\bibfield  {journal} {\bibinfo  {journal}
  {Mem. Acad. St. Petersbourg}\ }\textbf {\bibinfo {volume} {6}},\ \bibinfo
  {pages} {385} (\bibinfo {year} {1850})}\BibitemShut {NoStop}%
%%CITATION = INSPIRE-1468685;%%
\bibitem [{\citenamefont {Heisenberg}(2014)}]{Heisenberg:2014rta}%
  \BibitemOpen
  \bibfield  {author} {\bibinfo {author} {\bibfnamefont {L.}~\bibnamefont
  {Heisenberg}},\ }\href {\doibase 10.1088/1475-7516/2014/05/015} {\bibfield
  {journal} {\bibinfo  {journal} {JCAP}\ }\textbf {\bibinfo {volume} {05}},\
  \bibinfo {pages} {015} (\bibinfo {year} {2014})},\ \Eprint
  {http://arxiv.org/abs/1402.7026} {arXiv:1402.7026 [hep-th]} \BibitemShut
  {NoStop}%
\bibitem [{\citenamefont {Allys}\ \emph
  {et~al.}(2016{\natexlab{a}})\citenamefont {Allys}, \citenamefont {Peter},\
  and\ \citenamefont {Rodriguez}}]{Allys:2015sht}%
  \BibitemOpen
  \bibfield  {author} {\bibinfo {author} {\bibfnamefont {E.}~\bibnamefont
  {Allys}}, \bibinfo {author} {\bibfnamefont {P.}~\bibnamefont {Peter}}, \ and\
  \bibinfo {author} {\bibfnamefont {Y.}~\bibnamefont {Rodriguez}},\ }\href
  {\doibase 10.1088/1475-7516/2016/02/004} {\bibfield  {journal} {\bibinfo
  {journal} {JCAP}\ }\textbf {\bibinfo {volume} {02}},\ \bibinfo {pages} {004}
  (\bibinfo {year} {2016}{\natexlab{a}})},\ \Eprint
  {http://arxiv.org/abs/1511.03101} {arXiv:1511.03101 [hep-th]} \BibitemShut
  {NoStop}%
\bibitem [{\citenamefont {Allys}\ \emph
  {et~al.}(2016{\natexlab{b}})\citenamefont {Allys}, \citenamefont
  {Beltran~Almeida}, \citenamefont {Peter},\ and\ \citenamefont
  {Rodr\'\i{}guez}}]{Allys:2016jaq}%
  \BibitemOpen
  \bibfield  {author} {\bibinfo {author} {\bibfnamefont {E.}~\bibnamefont
  {Allys}}, \bibinfo {author} {\bibfnamefont {J.~P.}\ \bibnamefont
  {Beltran~Almeida}}, \bibinfo {author} {\bibfnamefont {P.}~\bibnamefont
  {Peter}}, \ and\ \bibinfo {author} {\bibfnamefont {Y.}~\bibnamefont
  {Rodr\'\i{}guez}},\ }\href {\doibase 10.1088/1475-7516/2016/09/026}
  {\bibfield  {journal} {\bibinfo  {journal} {JCAP}\ }\textbf {\bibinfo
  {volume} {09}},\ \bibinfo {pages} {026} (\bibinfo {year}
  {2016}{\natexlab{b}})},\ \Eprint {http://arxiv.org/abs/1605.08355}
  {arXiv:1605.08355 [hep-th]} \BibitemShut {NoStop}%
\bibitem [{\citenamefont {Gallego~Cadavid}\ and\ \citenamefont
  {Rodriguez}(2019)}]{GallegoCadavid:2019zke}%
  \BibitemOpen
  \bibfield  {author} {\bibinfo {author} {\bibfnamefont {A.}~\bibnamefont
  {Gallego~Cadavid}}\ and\ \bibinfo {author} {\bibfnamefont {Y.}~\bibnamefont
  {Rodriguez}},\ }\href {\doibase 10.1016/j.physletb.2019.134958} {\bibfield
  {journal} {\bibinfo  {journal} {Phys. Lett. B}\ }\textbf {\bibinfo {volume}
  {798}},\ \bibinfo {pages} {134958} (\bibinfo {year} {2019})},\ \Eprint
  {http://arxiv.org/abs/1905.10664} {arXiv:1905.10664 [hep-th]} \BibitemShut
  {NoStop}%
\bibitem [{\citenamefont {Allys}\ \emph
  {et~al.}(2016{\natexlab{c}})\citenamefont {Allys}, \citenamefont {Peter},\
  and\ \citenamefont {Rodriguez}}]{Allys:2016kbq}%
  \BibitemOpen
  \bibfield  {author} {\bibinfo {author} {\bibfnamefont {E.}~\bibnamefont
  {Allys}}, \bibinfo {author} {\bibfnamefont {P.}~\bibnamefont {Peter}}, \ and\
  \bibinfo {author} {\bibfnamefont {Y.}~\bibnamefont {Rodriguez}},\ }\href
  {\doibase 10.1103/PhysRevD.94.084041} {\bibfield  {journal} {\bibinfo
  {journal} {Phys. Rev. D}\ }\textbf {\bibinfo {volume} {94}},\ \bibinfo
  {pages} {084041} (\bibinfo {year} {2016}{\natexlab{c}})},\ \Eprint
  {http://arxiv.org/abs/1609.05870} {arXiv:1609.05870 [hep-th]} \BibitemShut
  {NoStop}%
\bibitem [{\citenamefont {G\'omez}\ and\ \citenamefont
  {Rodr\'\i{}guez}(2019)}]{Gomez:2019tbj}%
  \BibitemOpen
  \bibfield  {author} {\bibinfo {author} {\bibfnamefont {L.~G.}\ \bibnamefont
  {G\'omez}}\ and\ \bibinfo {author} {\bibfnamefont {Y.}~\bibnamefont
  {Rodr\'\i{}guez}},\ }\href {\doibase 10.1103/PhysRevD.100.084048} {\bibfield
  {journal} {\bibinfo  {journal} {Phys. Rev. D}\ }\textbf {\bibinfo {volume}
  {100}},\ \bibinfo {pages} {084048} (\bibinfo {year} {2019})},\ \Eprint
  {http://arxiv.org/abs/1907.07961} {arXiv:1907.07961 [gr-qc]} \BibitemShut
  {NoStop}%
\bibitem [{\citenamefont {Gallego~Cadavid}\ \emph {et~al.}(2020)\citenamefont
  {Gallego~Cadavid}, \citenamefont {Rodriguez},\ and\ \citenamefont
  {G\'omez}}]{GallegoCadavid:2020dho}%
  \BibitemOpen
  \bibfield  {author} {\bibinfo {author} {\bibfnamefont {A.}~\bibnamefont
  {Gallego~Cadavid}}, \bibinfo {author} {\bibfnamefont {Y.}~\bibnamefont
  {Rodriguez}}, \ and\ \bibinfo {author} {\bibfnamefont {L.~G.}\ \bibnamefont
  {G\'omez}},\ }\href {\doibase 10.1103/PhysRevD.102.104066} {\bibfield
  {journal} {\bibinfo  {journal} {Phys. Rev. D}\ }\textbf {\bibinfo {volume}
  {102}},\ \bibinfo {pages} {104066} (\bibinfo {year} {2020})},\ \Eprint
  {http://arxiv.org/abs/2009.03241} {arXiv:2009.03241 [hep-th]} \BibitemShut
  {NoStop}%
\bibitem [{\citenamefont {Gallego~Cadavid}\ \emph {et~al.}(2022)\citenamefont
  {Gallego~Cadavid}, \citenamefont {Nieto},\ and\ \citenamefont
  {Rodriguez}}]{GallegoCadavid:2022uzn}%
  \BibitemOpen
  \bibfield  {author} {\bibinfo {author} {\bibfnamefont {A.}~\bibnamefont
  {Gallego~Cadavid}}, \bibinfo {author} {\bibfnamefont {C.~M.}\ \bibnamefont
  {Nieto}}, \ and\ \bibinfo {author} {\bibfnamefont {Y.}~\bibnamefont
  {Rodriguez}},\ }\href {\doibase 10.1103/PhysRevD.105.104051} {\bibfield
  {journal} {\bibinfo  {journal} {Phys. Rev. D}\ }\textbf {\bibinfo {volume}
  {105}},\ \bibinfo {pages} {104051} (\bibinfo {year} {2022})},\ \Eprint
  {http://arxiv.org/abs/2204.04328} {arXiv:2204.04328 [hep-th]} \BibitemShut
  {NoStop}%
\bibitem [{\citenamefont {Garnica}\ \emph {et~al.}(2022)\citenamefont
  {Garnica}, \citenamefont {Gomez}, \citenamefont {Navarro},\ and\
  \citenamefont {Rodriguez}}]{Garnica:2021fuu}%
  \BibitemOpen
  \bibfield  {author} {\bibinfo {author} {\bibfnamefont {J.~C.}\ \bibnamefont
  {Garnica}}, \bibinfo {author} {\bibfnamefont {L.~G.}\ \bibnamefont {Gomez}},
  \bibinfo {author} {\bibfnamefont {A.~A.}\ \bibnamefont {Navarro}}, \ and\
  \bibinfo {author} {\bibfnamefont {Y.}~\bibnamefont {Rodriguez}},\ }\href
  {\doibase 10.1002/andp.202100453} {\bibfield  {journal} {\bibinfo  {journal}
  {Annalen Phys.}\ }\textbf {\bibinfo {volume} {534}},\ \bibinfo {pages}
  {2100453} (\bibinfo {year} {2022})},\ \Eprint
  {http://arxiv.org/abs/2109.10154} {arXiv:2109.10154 [gr-qc]} \BibitemShut
  {NoStop}%
\bibitem [{\citenamefont {Martinez}\ \emph {et~al.}(2023)\citenamefont
  {Martinez}, \citenamefont {Rodriguez}, \citenamefont {Rodriguez},\ and\
  \citenamefont {Gomez}}]{Martinez:2022wsy}%
  \BibitemOpen
  \bibfield  {author} {\bibinfo {author} {\bibfnamefont {J.~N.}\ \bibnamefont
  {Martinez}}, \bibinfo {author} {\bibfnamefont {J.~F.}\ \bibnamefont
  {Rodriguez}}, \bibinfo {author} {\bibfnamefont {Y.}~\bibnamefont
  {Rodriguez}}, \ and\ \bibinfo {author} {\bibfnamefont {G.}~\bibnamefont
  {Gomez}},\ }\href {\doibase 10.1088/1475-7516/2023/04/032} {\bibfield
  {journal} {\bibinfo  {journal} {JCAP}\ }\textbf {\bibinfo {volume} {04}},\
  \bibinfo {pages} {032} (\bibinfo {year} {2023})},\ \Eprint
  {http://arxiv.org/abs/2212.13832} {arXiv:2212.13832 [gr-qc]} \BibitemShut
  {NoStop}%
\bibitem [{\citenamefont {Bondi}(1952)}]{Bondi:1952ni}%
  \BibitemOpen
  \bibfield  {author} {\bibinfo {author} {\bibfnamefont {H.}~\bibnamefont
  {Bondi}},\ }\href {\doibase 10.1093/mnras/112.2.195} {\bibfield  {journal}
  {\bibinfo  {journal} {Mon. Not. Roy. Astron. Soc.}\ }\textbf {\bibinfo
  {volume} {112}},\ \bibinfo {pages} {195} (\bibinfo {year}
  {1952})}\BibitemShut {NoStop}%
\bibitem [{\citenamefont {{Michel}}(1972)}]{1972Ap&SS..15..153M}%
  \BibitemOpen
  \bibfield  {author} {\bibinfo {author} {\bibfnamefont {F.~C.}\ \bibnamefont
  {{Michel}}},\ }\href {\doibase 10.1007/BF00649949} {\bibfield  {journal}
  {\bibinfo  {journal} {\apss}\ }\textbf {\bibinfo {volume} {15}},\ \bibinfo
  {pages} {153} (\bibinfo {year} {1972})}\BibitemShut {NoStop}%
\bibitem [{\citenamefont {Richards}\ \emph {et~al.}(2021)\citenamefont
  {Richards}, \citenamefont {Baumgarte},\ and\ \citenamefont
  {Shapiro}}]{Richards:2021zbr}%
  \BibitemOpen
  \bibfield  {author} {\bibinfo {author} {\bibfnamefont {C.~B.}\ \bibnamefont
  {Richards}}, \bibinfo {author} {\bibfnamefont {T.~W.}\ \bibnamefont
  {Baumgarte}}, \ and\ \bibinfo {author} {\bibfnamefont {S.~L.}\ \bibnamefont
  {Shapiro}},\ }\href {\doibase 10.1093/mnras/stab2069} {\bibfield  {journal}
  {\bibinfo  {journal} {Mon. Not. Roy. Astron. Soc.}\ }\textbf {\bibinfo
  {volume} {502}},\ \bibinfo {pages} {3003} (\bibinfo {year} {2021})},\
  \bibinfo {note} {[Erratum: Mon.Not.Roy.Astron.Soc. 506, 3935 (2021)]},\
  \Eprint {http://arxiv.org/abs/2101.08797} {arXiv:2101.08797 [astro-ph.HE]}
  \BibitemShut {NoStop}%
\bibitem [{\citenamefont {Aguayo-Ortiz}\ \emph {et~al.}(2021)\citenamefont
  {Aguayo-Ortiz}, \citenamefont {Tejeda}, \citenamefont {Sarbach},\ and\
  \citenamefont {L\'opez-C\'amara}}]{Aguayo-Ortiz:2021jzv}%
  \BibitemOpen
  \bibfield  {author} {\bibinfo {author} {\bibfnamefont {A.}~\bibnamefont
  {Aguayo-Ortiz}}, \bibinfo {author} {\bibfnamefont {E.}~\bibnamefont
  {Tejeda}}, \bibinfo {author} {\bibfnamefont {O.}~\bibnamefont {Sarbach}}, \
  and\ \bibinfo {author} {\bibfnamefont {D.}~\bibnamefont {L\'opez-C\'amara}},\
  }\href {\doibase 10.1093/mnras/stab1127} {\  (\bibinfo {year} {2021}),\
  10.1093/mnras/stab1127},\ \Eprint {http://arxiv.org/abs/2102.12529}
  {arXiv:2102.12529 [astro-ph.HE]} \BibitemShut {NoStop}%
\bibitem [{\citenamefont {Bauer}\ \emph {et~al.}(2022)\citenamefont {Bauer},
  \citenamefont {C\'ardenas-Avenda\~no}, \citenamefont {Gammie},\ and\
  \citenamefont {Yunes}}]{Bauer:2021atk}%
  \BibitemOpen
  \bibfield  {author} {\bibinfo {author} {\bibfnamefont {A.~M.}\ \bibnamefont
  {Bauer}}, \bibinfo {author} {\bibfnamefont {A.}~\bibnamefont
  {C\'ardenas-Avenda\~no}}, \bibinfo {author} {\bibfnamefont {C.~F.}\
  \bibnamefont {Gammie}}, \ and\ \bibinfo {author} {\bibfnamefont
  {N.}~\bibnamefont {Yunes}},\ }\href {\doibase 10.3847/1538-4357/ac3a03}
  {\bibfield  {journal} {\bibinfo  {journal} {Astrophys. J.}\ }\textbf
  {\bibinfo {volume} {925}},\ \bibinfo {pages} {119} (\bibinfo {year}
  {2022})},\ \Eprint {http://arxiv.org/abs/2111.02178} {arXiv:2111.02178
  [gr-qc]} \BibitemShut {NoStop}%
\bibitem [{\citenamefont {Salahshoor}\ and\ \citenamefont
  {Nozari}(2018)}]{Salahshoor:2018plr}%
  \BibitemOpen
  \bibfield  {author} {\bibinfo {author} {\bibfnamefont {K.}~\bibnamefont
  {Salahshoor}}\ and\ \bibinfo {author} {\bibfnamefont {K.}~\bibnamefont
  {Nozari}},\ }\href {\doibase 10.1140/epjc/s10052-018-5946-2} {\bibfield
  {journal} {\bibinfo  {journal} {Eur. Phys. J. C}\ }\textbf {\bibinfo {volume}
  {78}},\ \bibinfo {pages} {486} (\bibinfo {year} {2018})},\ \Eprint
  {http://arxiv.org/abs/1806.08949} {arXiv:1806.08949 [gr-qc]} \BibitemShut
  {NoStop}%
\bibitem [{\citenamefont {Feng}\ \emph {et~al.}(2022)\citenamefont {Feng},
  \citenamefont {Li}, \citenamefont {Liang},\ and\ \citenamefont
  {Yang}}]{Feng:2022bst}%
  \BibitemOpen
  \bibfield  {author} {\bibinfo {author} {\bibfnamefont {H.}~\bibnamefont
  {Feng}}, \bibinfo {author} {\bibfnamefont {M.}~\bibnamefont {Li}}, \bibinfo
  {author} {\bibfnamefont {G.-R.}\ \bibnamefont {Liang}}, \ and\ \bibinfo
  {author} {\bibfnamefont {R.-J.}\ \bibnamefont {Yang}},\ }\href {\doibase
  10.1088/1475-7516/2022/04/027} {\bibfield  {journal} {\bibinfo  {journal}
  {JCAP}\ }\textbf {\bibinfo {volume} {04}},\ \bibinfo {pages} {027} (\bibinfo
  {year} {2022})},\ \Eprint {http://arxiv.org/abs/2203.02924} {arXiv:2203.02924
  [gr-qc]} \BibitemShut {NoStop}%
\bibitem [{\citenamefont {Zuluaga}\ and\ \citenamefont
  {S\'anchez}(2021)}]{Zuluaga:2021vjc}%
  \BibitemOpen
  \bibfield  {author} {\bibinfo {author} {\bibfnamefont {F.~H.}\ \bibnamefont
  {Zuluaga}}\ and\ \bibinfo {author} {\bibfnamefont {L.~A.}\ \bibnamefont
  {S\'anchez}},\ }\href {\doibase 10.1140/epjc/s10052-021-09644-1} {\bibfield
  {journal} {\bibinfo  {journal} {Eur. Phys. J. C}\ }\textbf {\bibinfo {volume}
  {81}},\ \bibinfo {pages} {840} (\bibinfo {year} {2021})},\ \Eprint
  {http://arxiv.org/abs/2106.03140} {arXiv:2106.03140 [gr-qc]} \BibitemShut
  {NoStop}%
\bibitem [{\citenamefont {Ditta}\ and\ \citenamefont
  {Abbas}(2020)}]{Ditta:2020jud}%
  \BibitemOpen
  \bibfield  {author} {\bibinfo {author} {\bibfnamefont {A.}~\bibnamefont
  {Ditta}}\ and\ \bibinfo {author} {\bibfnamefont {G.}~\bibnamefont {Abbas}},\
  }\href {\doibase 10.1007/s10714-020-02724-9} {\bibfield  {journal} {\bibinfo
  {journal} {Gen. Rel. Grav.}\ }\textbf {\bibinfo {volume} {52}},\ \bibinfo
  {pages} {77} (\bibinfo {year} {2020})}\BibitemShut {NoStop}%
\bibitem [{\citenamefont {Uniyal}\ \emph {et~al.}(2022)\citenamefont {Uniyal},
  \citenamefont {Pantig},\ and\ \citenamefont {\"Ovg\"un}}]{Uniyal:2022vdu}%
  \BibitemOpen
  \bibfield  {author} {\bibinfo {author} {\bibfnamefont {A.}~\bibnamefont
  {Uniyal}}, \bibinfo {author} {\bibfnamefont {R.~C.}\ \bibnamefont {Pantig}},
  \ and\ \bibinfo {author} {\bibfnamefont {A.}~\bibnamefont {\"Ovg\"un}},\
  }\href@noop {} {\  (\bibinfo {year} {2022})},\ \Eprint
  {http://arxiv.org/abs/2205.11072} {arXiv:2205.11072 [gr-qc]} \BibitemShut
  {NoStop}%
\bibitem [{\citenamefont {Chakhchi}\ \emph {et~al.}(2022)\citenamefont
  {Chakhchi}, \citenamefont {El~Moumni},\ and\ \citenamefont
  {Masmar}}]{Chakhchi:2022fls}%
  \BibitemOpen
  \bibfield  {author} {\bibinfo {author} {\bibfnamefont {L.}~\bibnamefont
  {Chakhchi}}, \bibinfo {author} {\bibfnamefont {H.}~\bibnamefont {El~Moumni}},
  \ and\ \bibinfo {author} {\bibfnamefont {K.}~\bibnamefont {Masmar}},\ }\href
  {\doibase 10.1103/PhysRevD.105.064031} {\bibfield  {journal} {\bibinfo
  {journal} {Phys. Rev. D}\ }\textbf {\bibinfo {volume} {105}},\ \bibinfo
  {pages} {064031} (\bibinfo {year} {2022})}\BibitemShut {NoStop}%
\bibitem [{\citenamefont {John}(2019)}]{John:2019was}%
  \BibitemOpen
  \bibfield  {author} {\bibinfo {author} {\bibfnamefont {A.~J.}\ \bibnamefont
  {John}},\ }\href {\doibase 10.1093/mnras/stz2889} {\bibfield  {journal}
  {\bibinfo  {journal} {Mon. Not. Roy. Astron. Soc.}\ }\textbf {\bibinfo
  {volume} {490}},\ \bibinfo {pages} {3824} (\bibinfo {year} {2019})},\ \Eprint
  {http://arxiv.org/abs/1603.09425} {arXiv:1603.09425 [gr-qc]} \BibitemShut
  {NoStop}%
\bibitem [{\citenamefont {P\'erez}\ \emph {et~al.}(2017)\citenamefont
  {P\'erez}, \citenamefont {Lopez~Armengol},\ and\ \citenamefont
  {Romero}}]{Perez:2017spz}%
  \BibitemOpen
  \bibfield  {author} {\bibinfo {author} {\bibfnamefont {D.}~\bibnamefont
  {P\'erez}}, \bibinfo {author} {\bibfnamefont {F.~G.}\ \bibnamefont
  {Lopez~Armengol}}, \ and\ \bibinfo {author} {\bibfnamefont {G.~E.}\
  \bibnamefont {Romero}},\ }\href {\doibase 10.1103/PhysRevD.95.104047}
  {\bibfield  {journal} {\bibinfo  {journal} {Phys. Rev. D}\ }\textbf {\bibinfo
  {volume} {95}},\ \bibinfo {pages} {104047} (\bibinfo {year} {2017})},\
  \Eprint {http://arxiv.org/abs/1705.02713} {arXiv:1705.02713 [astro-ph.HE]}
  \BibitemShut {NoStop}%
\bibitem [{\citenamefont {Liu}\ \emph {et~al.}(2022)\citenamefont {Liu},
  \citenamefont {Yang}, \citenamefont {Wu},\ and\ \citenamefont
  {Zhu}}]{Liu:2021yev}%
  \BibitemOpen
  \bibfield  {author} {\bibinfo {author} {\bibfnamefont {C.}~\bibnamefont
  {Liu}}, \bibinfo {author} {\bibfnamefont {S.}~\bibnamefont {Yang}}, \bibinfo
  {author} {\bibfnamefont {Q.}~\bibnamefont {Wu}}, \ and\ \bibinfo {author}
  {\bibfnamefont {T.}~\bibnamefont {Zhu}},\ }\href {\doibase
  10.1088/1475-7516/2022/02/034} {\bibfield  {journal} {\bibinfo  {journal}
  {JCAP}\ }\textbf {\bibinfo {volume} {02}},\ \bibinfo {pages} {034} (\bibinfo
  {year} {2022})},\ \Eprint {http://arxiv.org/abs/2107.04811} {arXiv:2107.04811
  [gr-qc]} \BibitemShut {NoStop}%
\bibitem [{\citenamefont {Heydari-Fard}\ \emph {et~al.}(2021)\citenamefont
  {Heydari-Fard}, \citenamefont {Heydari-Fard},\ and\ \citenamefont
  {Sepangi}}]{Heydari-Fard:2021ljh}%
  \BibitemOpen
  \bibfield  {author} {\bibinfo {author} {\bibfnamefont {M.}~\bibnamefont
  {Heydari-Fard}}, \bibinfo {author} {\bibfnamefont {M.}~\bibnamefont
  {Heydari-Fard}}, \ and\ \bibinfo {author} {\bibfnamefont {H.~R.}\
  \bibnamefont {Sepangi}},\ }\href {\doibase 10.1140/epjc/s10052-021-09266-7}
  {\bibfield  {journal} {\bibinfo  {journal} {Eur. Phys. J. C}\ }\textbf
  {\bibinfo {volume} {81}},\ \bibinfo {pages} {473} (\bibinfo {year} {2021})},\
  \Eprint {http://arxiv.org/abs/2105.09192} {arXiv:2105.09192 [gr-qc]}
  \BibitemShut {NoStop}%
\bibitem [{\citenamefont {Stashko}\ \emph {et~al.}(2021)\citenamefont
  {Stashko}, \citenamefont {Zhdanov},\ and\ \citenamefont
  {Alexandrov}}]{Stashko:2021lad}%
  \BibitemOpen
  \bibfield  {author} {\bibinfo {author} {\bibfnamefont {O.~S.}\ \bibnamefont
  {Stashko}}, \bibinfo {author} {\bibfnamefont {V.~I.}\ \bibnamefont
  {Zhdanov}}, \ and\ \bibinfo {author} {\bibfnamefont {A.~N.}\ \bibnamefont
  {Alexandrov}},\ }\href {\doibase 10.1103/PhysRevD.104.104055} {\bibfield
  {journal} {\bibinfo  {journal} {Phys. Rev. D}\ }\textbf {\bibinfo {volume}
  {104}},\ \bibinfo {pages} {104055} (\bibinfo {year} {2021})},\ \Eprint
  {http://arxiv.org/abs/2107.05111} {arXiv:2107.05111 [gr-qc]} \BibitemShut
  {NoStop}%
\bibitem [{\citenamefont {Shaikh}\ and\ \citenamefont
  {Joshi}(2019)}]{Shaikh:2019hbm}%
  \BibitemOpen
  \bibfield  {author} {\bibinfo {author} {\bibfnamefont {R.}~\bibnamefont
  {Shaikh}}\ and\ \bibinfo {author} {\bibfnamefont {P.~S.}\ \bibnamefont
  {Joshi}},\ }\href {\doibase 10.1088/1475-7516/2019/10/064} {\bibfield
  {journal} {\bibinfo  {journal} {JCAP}\ }\textbf {\bibinfo {volume} {10}},\
  \bibinfo {pages} {064} (\bibinfo {year} {2019})},\ \Eprint
  {http://arxiv.org/abs/1909.10322} {arXiv:1909.10322 [gr-qc]} \BibitemShut
  {NoStop}%
\bibitem [{\citenamefont {Van~Aelst}\ \emph {et~al.}(2021)\citenamefont
  {Van~Aelst}, \citenamefont {Gourgoulhon},\ and\ \citenamefont
  {Vincent}}]{VanAelst:2021uem}%
  \BibitemOpen
  \bibfield  {author} {\bibinfo {author} {\bibfnamefont {K.}~\bibnamefont
  {Van~Aelst}}, \bibinfo {author} {\bibfnamefont {E.}~\bibnamefont
  {Gourgoulhon}}, \ and\ \bibinfo {author} {\bibfnamefont {F.~H.}\ \bibnamefont
  {Vincent}},\ }\href {\doibase 10.1103/PhysRevD.104.124034} {\bibfield
  {journal} {\bibinfo  {journal} {Phys. Rev. D}\ }\textbf {\bibinfo {volume}
  {104}},\ \bibinfo {pages} {124034} (\bibinfo {year} {2021})},\ \Eprint
  {http://arxiv.org/abs/2103.01827} {arXiv:2103.01827 [gr-qc]} \BibitemShut
  {NoStop}%
\bibitem [{\citenamefont {Bekenstein}(1972)}]{Bekenstein:1971hc}%
  \BibitemOpen
  \bibfield  {author} {\bibinfo {author} {\bibfnamefont {J.~D.}\ \bibnamefont
  {Bekenstein}},\ }\href {\doibase 10.1103/PhysRevD.5.1239} {\bibfield
  {journal} {\bibinfo  {journal} {Phys. Rev. D}\ }\textbf {\bibinfo {volume}
  {5}},\ \bibinfo {pages} {1239} (\bibinfo {year} {1972})}\BibitemShut
  {NoStop}%
\bibitem [{\citenamefont {{Wald}}(1984)}]{1984ucp..book.....W}%
  \BibitemOpen
  \bibfield  {author} {\bibinfo {author} {\bibfnamefont {R.~M.}\ \bibnamefont
  {{Wald}}},\ }\href@noop {} {\emph {\bibinfo {title} {{General Relativity}}}}\
  (\bibinfo {year} {1984})\BibitemShut {NoStop}%
\bibitem [{\citenamefont {{Ashtekar}}\ and\ \citenamefont
  {{Magnon-Ashtekar}}(1979)}]{1979JMP....20..793A}%
  \BibitemOpen
  \bibfield  {author} {\bibinfo {author} {\bibfnamefont {A.}~\bibnamefont
  {{Ashtekar}}}\ and\ \bibinfo {author} {\bibfnamefont {A.}~\bibnamefont
  {{Magnon-Ashtekar}}},\ }\href {\doibase 10.1063/1.524151} {\bibfield
  {journal} {\bibinfo  {journal} {Journal of Mathematical Physics}\ }\textbf
  {\bibinfo {volume} {20}},\ \bibinfo {pages} {793} (\bibinfo {year}
  {1979})}\BibitemShut {NoStop}%
\bibitem [{\citenamefont {Visser}(2007)}]{Visser:2007fj}%
  \BibitemOpen
  \bibfield  {author} {\bibinfo {author} {\bibfnamefont {M.}~\bibnamefont
  {Visser}},\ }in\ \href@noop {} {\emph {\bibinfo {booktitle} {{Kerr Fest:
  Black Holes in Astrophysics, General Relativity and Quantum Gravity}}}}\
  (\bibinfo {year} {2007})\ \Eprint {http://arxiv.org/abs/0706.0622}
  {arXiv:0706.0622 [gr-qc]} \BibitemShut {NoStop}%
\bibitem [{\citenamefont {{Vishveshwara}}(1968)}]{1968JMP.....9.1319V}%
  \BibitemOpen
  \bibfield  {author} {\bibinfo {author} {\bibfnamefont {C.~V.}\ \bibnamefont
  {{Vishveshwara}}},\ }\href {\doibase 10.1063/1.1664717} {\bibfield  {journal}
  {\bibinfo  {journal} {Journal of Mathematical Physics}\ }\textbf {\bibinfo
  {volume} {9}},\ \bibinfo {pages} {1319} (\bibinfo {year} {1968})}\BibitemShut
  {NoStop}%
\bibitem [{\citenamefont {Wald}(1984)}]{Wald:106274}%
  \BibitemOpen
  \bibfield  {author} {\bibinfo {author} {\bibfnamefont {R.~M.}\ \bibnamefont
  {Wald}},\ }\href {https://cds.cern.ch/record/106274} {\emph {\bibinfo {title}
  {{General relativity}}}}\ (\bibinfo  {publisher} {Chicago Univ. Press},\
  \bibinfo {address} {Chicago, IL},\ \bibinfo {year} {1984})\BibitemShut
  {NoStop}%
\bibitem [{\citenamefont {Pugliese}\ \emph {et~al.}(2011)\citenamefont
  {Pugliese}, \citenamefont {Quevedo},\ and\ \citenamefont
  {Ruffini}}]{Pugliese:2010ps}%
  \BibitemOpen
  \bibfield  {author} {\bibinfo {author} {\bibfnamefont {D.}~\bibnamefont
  {Pugliese}}, \bibinfo {author} {\bibfnamefont {H.}~\bibnamefont {Quevedo}}, \
  and\ \bibinfo {author} {\bibfnamefont {R.}~\bibnamefont {Ruffini}},\ }\href
  {\doibase 10.1103/PhysRevD.83.024021} {\bibfield  {journal} {\bibinfo
  {journal} {Phys. Rev. D}\ }\textbf {\bibinfo {volume} {83}},\ \bibinfo
  {pages} {024021} (\bibinfo {year} {2011})},\ \Eprint
  {http://arxiv.org/abs/1012.5411} {arXiv:1012.5411 [astro-ph.HE]} \BibitemShut
  {NoStop}%
\bibitem [{\citenamefont {{Psaltis}}(2008)}]{2008PhRvD..77f4006P}%
  \BibitemOpen
  \bibfield  {author} {\bibinfo {author} {\bibfnamefont {D.}~\bibnamefont
  {{Psaltis}}},\ }\href {\doibase 10.1103/PhysRevD.77.064006} {\bibfield
  {journal} {\bibinfo  {journal} {\prd}\ }\textbf {\bibinfo {volume} {77}},\
  \bibinfo {eid} {064006} (\bibinfo {year} {2008})},\ \Eprint
  {http://arxiv.org/abs/0704.2426} {arXiv:0704.2426 [astro-ph]} \BibitemShut
  {NoStop}%
\bibitem [{\citenamefont {{Event Horizon Telescope Collaboration}}\ \emph
  {et~al.}(2022)\citenamefont {{Event Horizon Telescope Collaboration}},
  \citenamefont {{Akiyama}}, \citenamefont {{Alberdi}}, \citenamefont {{Alef}},
  \citenamefont {{Algaba}}, \citenamefont {{Anantua}}, \citenamefont {{Asada}},
  \citenamefont {{Azulay}}, \citenamefont {{Bach}}, \citenamefont {{Baczko}},
  \citenamefont {{Ball}}, \citenamefont {{Balokovi{\'c}}},\ and\ \citenamefont
  {{Barrett}}}]{2022ApJ...930L..17E}%
  \BibitemOpen
  \bibfield  {author} {\bibinfo {author} {\bibnamefont {{Event Horizon
  Telescope Collaboration}}}, \bibinfo {author} {\bibfnamefont
  {K.}~\bibnamefont {{Akiyama}}}, \bibinfo {author} {\bibfnamefont
  {A.}~\bibnamefont {{Alberdi}}}, \bibinfo {author} {\bibfnamefont
  {W.}~\bibnamefont {{Alef}}}, \bibinfo {author} {\bibfnamefont {J.~C.}\
  \bibnamefont {{Algaba}}}, \bibinfo {author} {\bibfnamefont {R.}~\bibnamefont
  {{Anantua}}}, \bibinfo {author} {\bibfnamefont {K.}~\bibnamefont {{Asada}}},
  \bibinfo {author} {\bibfnamefont {R.}~\bibnamefont {{Azulay}}}, \bibinfo
  {author} {\bibfnamefont {U.}~\bibnamefont {{Bach}}}, \bibinfo {author}
  {\bibfnamefont {A.-K.}\ \bibnamefont {{Baczko}}}, \bibinfo {author}
  {\bibfnamefont {D.}~\bibnamefont {{Ball}}}, \bibinfo {author} {\bibfnamefont
  {M.}~\bibnamefont {{Balokovi{\'c}}}}, \ and\ \bibinfo {author} {\bibnamefont
  {{Barrett}}},\ }\href {\doibase 10.3847/2041-8213/ac6756} {\bibfield
  {journal} {\bibinfo  {journal} {\apjl}\ }\textbf {\bibinfo {volume} {930}},\
  \bibinfo {eid} {L17} (\bibinfo {year} {2022})}\BibitemShut {NoStop}%
\bibitem [{\citenamefont {Babichev}\ \emph {et~al.}(2011)\citenamefont
  {Babichev}, \citenamefont {Chernov}, \citenamefont {Dokuchaev},\ and\
  \citenamefont {Eroshenko}}]{Babichev:2008jb}%
  \BibitemOpen
  \bibfield  {author} {\bibinfo {author} {\bibfnamefont {E.}~\bibnamefont
  {Babichev}}, \bibinfo {author} {\bibfnamefont {S.}~\bibnamefont {Chernov}},
  \bibinfo {author} {\bibfnamefont {V.}~\bibnamefont {Dokuchaev}}, \ and\
  \bibinfo {author} {\bibfnamefont {Y.}~\bibnamefont {Eroshenko}},\ }\href
  {\doibase 10.1134/S1063776111040157} {\bibfield  {journal} {\bibinfo
  {journal} {J. Exp. Theor. Phys.}\ }\textbf {\bibinfo {volume} {112}},\
  \bibinfo {pages} {784} (\bibinfo {year} {2011})},\ \Eprint
  {http://arxiv.org/abs/0806.0916} {arXiv:0806.0916 [gr-qc]} \BibitemShut
  {NoStop}%
\bibitem [{\citenamefont {Freitas~Pacheco}(2011)}]{FreitasPacheco:2011yme}%
  \BibitemOpen
  \bibfield  {author} {\bibinfo {author} {\bibfnamefont {J.~A.~d.}\
  \bibnamefont {Freitas~Pacheco}},\ }\href@noop {} {\  (\bibinfo {year}
  {2011})},\ \Eprint {http://arxiv.org/abs/1109.6798} {arXiv:1109.6798 [gr-qc]}
  \BibitemShut {NoStop}%
\bibitem [{\citenamefont {Dadhich}\ \emph {et~al.}(2000)\citenamefont
  {Dadhich}, \citenamefont {Maartens}, \citenamefont {Papadopoulos},\ and\
  \citenamefont {Rezania}}]{Dadhich:2000am}%
  \BibitemOpen
  \bibfield  {author} {\bibinfo {author} {\bibfnamefont {N.}~\bibnamefont
  {Dadhich}}, \bibinfo {author} {\bibfnamefont {R.}~\bibnamefont {Maartens}},
  \bibinfo {author} {\bibfnamefont {P.}~\bibnamefont {Papadopoulos}}, \ and\
  \bibinfo {author} {\bibfnamefont {V.}~\bibnamefont {Rezania}},\ }\href
  {\doibase 10.1016/S0370-2693(00)00798-X} {\bibfield  {journal} {\bibinfo
  {journal} {Phys. Lett. B}\ }\textbf {\bibinfo {volume} {487}},\ \bibinfo
  {pages} {1} (\bibinfo {year} {2000})},\ \Eprint
  {http://arxiv.org/abs/hep-th/0003061} {arXiv:hep-th/0003061} \BibitemShut
  {NoStop}%
\bibitem [{\citenamefont {Zakharov}(2022)}]{Zakharov:2021gbg}%
  \BibitemOpen
  \bibfield  {author} {\bibinfo {author} {\bibfnamefont {A.~F.}\ \bibnamefont
  {Zakharov}},\ }\href {\doibase 10.3390/universe8030141} {\bibfield  {journal}
  {\bibinfo  {journal} {Universe}\ }\textbf {\bibinfo {volume} {8}},\ \bibinfo
  {pages} {141} (\bibinfo {year} {2022})},\ \Eprint
  {http://arxiv.org/abs/2108.01533} {arXiv:2108.01533 [gr-qc]} \BibitemShut
  {NoStop}%
\bibitem [{\citenamefont {Neves}(2020)}]{Neves:2020doc}%
  \BibitemOpen
  \bibfield  {author} {\bibinfo {author} {\bibfnamefont {J.~C.~S.}\
  \bibnamefont {Neves}},\ }\href {\doibase 10.1140/epjc/s10052-020-8321-z}
  {\bibfield  {journal} {\bibinfo  {journal} {Eur. Phys. J. C}\ }\textbf
  {\bibinfo {volume} {80}},\ \bibinfo {pages} {717} (\bibinfo {year} {2020})},\
  \Eprint {http://arxiv.org/abs/2005.00483} {arXiv:2005.00483 [gr-qc]}
  \BibitemShut {NoStop}%
\bibitem [{\citenamefont {Babichev}\ \emph {et~al.}(2017)\citenamefont
  {Babichev}, \citenamefont {Charmousis},\ and\ \citenamefont
  {Leh\'ebel}}]{Babichev:2017guv}%
  \BibitemOpen
  \bibfield  {author} {\bibinfo {author} {\bibfnamefont {E.}~\bibnamefont
  {Babichev}}, \bibinfo {author} {\bibfnamefont {C.}~\bibnamefont
  {Charmousis}}, \ and\ \bibinfo {author} {\bibfnamefont {A.}~\bibnamefont
  {Leh\'ebel}},\ }\href {\doibase 10.1088/1475-7516/2017/04/027} {\bibfield
  {journal} {\bibinfo  {journal} {JCAP}\ }\textbf {\bibinfo {volume} {04}},\
  \bibinfo {pages} {027} (\bibinfo {year} {2017})},\ \Eprint
  {http://arxiv.org/abs/1702.01938} {arXiv:1702.01938 [gr-qc]} \BibitemShut
  {NoStop}%
\bibitem [{\citenamefont {Hansen}\ and\ \citenamefont
  {Yunes}(2013)}]{Hansen:2013owa}%
  \BibitemOpen
  \bibfield  {author} {\bibinfo {author} {\bibfnamefont {D.}~\bibnamefont
  {Hansen}}\ and\ \bibinfo {author} {\bibfnamefont {N.}~\bibnamefont {Yunes}},\
  }\href {\doibase 10.1103/PhysRevD.88.104020} {\bibfield  {journal} {\bibinfo
  {journal} {Phys. Rev. D}\ }\textbf {\bibinfo {volume} {88}},\ \bibinfo
  {pages} {104020} (\bibinfo {year} {2013})},\ \Eprint
  {http://arxiv.org/abs/1308.6631} {arXiv:1308.6631 [gr-qc]} \BibitemShut
  {NoStop}%
\end{thebibliography}%

\end{document}